\documentclass[12pt]{article}

\usepackage{amssymb,amsfonts,amsthm}%,psfig}
\usepackage{graphicx}
\usepackage{subfigure}
\usepackage{amssymb}
\usepackage{amsthm}
\usepackage{amsmath}
\usepackage{bm}             %% boldmath-command: \bm{}
\usepackage[mathscr]{eucal}

\usepackage{cancel}

\usepackage{psfrag}

\usepackage{multirow}

\usepackage{wasysym}

\usepackage{color}

\def\hsp5{\hspace{5mm}}

\newcommand{\sfrac}[2]{{\textstyle{#1\over#2}}}

\title{\sc Scalar field deformations of Lambda-CDM cosmology}
\begin{document}

\author{{\sc Artur Alho}\thanks{Electronic address:
{\tt aalho@math.ist.utl.pt}} \\[1ex]
Center for Mathematical Analysis, Geometry and Dynamical Systems,\\
Instituto Superior T\'ecnico, Universidade de Lisboa, \\
Av. Rovisco Pais, 1049-001 Lisboa, Portugal
\and \\
{\sc Claes Uggla}\thanks{Electronic address:
{\tt claes.uggla@kau.se}} \\[1ex]
Department of Physics, \\
Karlstad University, S-651 88 Karlstad, Sweden \\[2ex] }

%%%%%%%%%%%%%%%%%%%%%%%%%%%%%%%%%%%%%%%%%%%%%%%%%%%%%%%%%%%%%%%%%%%
%\date{\normalsize{October 04, 2015}}
\date{}
%%%%%%%%%%%%%%%%%%%%%%%%%%%%%%%%%%%%%%%%%%%%%%%%%%%%%%%%%%%%%%%%%%%
\maketitle
%%%%%%%%%%%%%%%%%%%%%%%%%%%%%%%%%%%%%%%%%%%%%%%%%%%%%%%%%%%%%%%%%%%

%%%%%%%%%%%%%%%%%%%%%%%%%%%%%%%%%%%%%%%%%%%%%%%%%%%%%%%%%%%%%%%%%%%
\begin{abstract}
%%%%%%%%%%%%%%%%%%%%%%%%%%%%%%%%%%%%%%%%%%%%%%%%%%%%%%%%%%%%%%%%%%%

This paper treats nonrelativistic matter and a scalar field $\phi$ with a
monotonically decreasing potential minimally coupled to gravity in flat
Friedmann-Lema\^{i}tre-Robertson-Walker cosmology. The field equations are
reformulated as a three-dimensional dynamical system on an extended compact
state space, complemented with cosmographic diagrams. A dynamical systems
analysis provides global dynamical results describing possible asymptotic
behavior. It is shown that one should impose \emph{global and asymptotic}
bounds on $\lambda=-V^{-1}\,dV/d\phi$ to obtain viable cosmological models
that continuously deform $\Lambda$CDM cosmology. In particular we introduce a
regularized inverse power-law potential as a simple specific example.
%Revisiting a select set of familiar potentials, but in a generic global
%dynamical systems setting, suggests that one should impose \emph{global and
%asymptotic} bounds on $\lambda=-V^{-1}\,dV/d\phi$
%, and not just demanding a shallow tail of a
%scalar field potential $V(\phi)$, as is the case for e.g. an inverse
%power-law potential,
%in order to obtain an \emph{attracting separatrix surface} that describes the
%evolution of \emph{all} solutions with $\Omega_m$ initially close to one, set
%by, e.g., inflation.
%An attracting separatrix \emph{surface} is to be contrasted with
%individually attracting solutions, such as tracker solutions that occur for
%potentials with unbounded $\lambda$.
%Furthermore, letting $\lambda \rightarrow 0$ for all $\phi$ and
%$\Omega_m\rightarrow 1$ initially yields $\Lambda$CDM cosmology and thus it
%is possible to continuously modify $\Lambda$CDM dynamics by choosing
%potentials that continuously deform the $\Lambda$CDM attracting separatrix
%urface, thereby also providing a useful testbed for $\Lambda$CDM cosmology;
%specific examples of potentials and the associated global dynamics are given.
%To facilitate
%comparisons with $\Lambda$CDM cosmology and to complement the dynamical
%systems pictures, we provide cosmographic diagrams.

%%%%%%%%%%%%%%%%%%%%%%%%%%%%%%%%%%%%%%%%%%%%%%%%%%%%%%%%%%%%%%%%%%
\end{abstract}
%%%%%%%%%%%%%%%%%%%%%%%%%%%%%%%%%%%%%%%%%%%%%%%%%%%%%%%%%%%%%%%%%%

%\centerline{\bigskip\noindent PACS numbers: 04.20.-q, 98.80.-k, 98.80.Bp,
%98.80.Jk}

%%%%%%%%%%%%%%%%%%%%%%%%%%%%%%%%%%%%%%%%%%%%%%%%%%%%%%%%%%%%%%%%%%
\section{Introduction}
%%%%%%%%%%%%%%%%%%%%%%%%%%%%%%%%%%%%%%%%%%%%%%%%%%%%%%%%%%%%%%%%%%

Recent observations, such as those of the cosmic background radiation, show
that $\Lambda$ cold dark matter ($\Lambda$CDM) cosmology is remarkably
consistent with observations, although some tensions remain,
see~\cite{plaI15,plaXIII15} and references therein. On the other hand, the
unknown origins and nature of dark matter and dark energy remain mysteries
which generate a steady flow of models and theories. Nonetheless, current
observations should be taken seriously, and viable cosmological models should
therefore presumably only deviate from $\Lambda$CDM cosmology marginally.
Moreover, although successful, the $\Lambda$CDM model still needs to be
observationally tested, which suggests that it should be continuously
deformed.
%or, preferably, a theory should admit a continuous set of models that each
%admit an open set of solutions that represents a continuous deformation of
%the $\Lambda$CDM models, thereby providing a natural testing ground for these
%models.
In addition, for a variety of reasons, it is of interest to consider a
dynamical dark energy content described by a field theoretical model instead
of a cosmological constant $\Lambda$ or a dark energy fluid. This leads to
the question: What phenomenological conditions must be imposed on an
effective classical field description of dark energy to continuously deform
$\Lambda$CDM cosmology for an open set of observationally viable solutions?
%What phenomenological conditions must be imposed on an effective classical
%field description of dark energy in order to obtain a continuous set of
%models where each model results in an open set of solutions that continuously
%deform $\Lambda$CDM cosmology?

Although observational compatibility arguably comes first, there also exist
other issues that are of interest such as various fine-tuning problems and if
there exist observationally viable models that emerge from, or at least have
ties to, a fundamental theory. The rather modest purpose of this paper,
however, is to
\begin{itemize}
\item[(i)] describe general features and determine asymptotic behavior of
    flat Friedmann-Lema\^{i}tre-Robertson-Walker (FLRW) models with a
    perfect fluid and a scalar field $\phi$, minimally coupled to
    gravity, with a positive monotonic potential $V(\phi)$,
\item[(ii)] shed some light on the above ``observationally viable field
    theoretical $\Lambda$CDM-deformation issue'' by considering some
    simple models of the above type with the equation of state for the
    perfect fluid specialized to dust, representing nonrelativistic
    matter (mainly baryons and nonbaryonic dark matter).
\end{itemize}
Since the literature about minimally coupled scalar fields and a perfect
fluid in a flat FLRW spacetime geometry is vast, we will start with only a
few general sample references: the influential
papers~\cite{peerat88,ratpee88,steetal99} and the recent review~\cite{tsu13},
and references therein.

It is of interest for discussions and future purposes to first consider a
perfect fluid with a general barotropic equation of state $p_m =
(\gamma_m-1)\rho_m$, where $p_m$ and $\rho_m \geq 0$ are the pressure and
energy density, respectively, and $\gamma_m=\gamma_m(\rho_m)$, together with
a scalar field $\phi$, minimally coupled to gravity, with a self-interaction
potential $V(\phi)$. Thus, the Einstein and matter field
equations for the present flat FLRW models can be written as (see,
e.g.,~\cite{waiell97,col03,wei08})
\begin{subequations}\label{Hphieq}
\begin{align}
3H^2 &= \frac12 \dot{\phi}^2 + V(\phi) + \rho_m,\label{Gauss1}\\
\dot{H} &= -\frac{1}{2}\left(\dot{\phi}^2 + \gamma_m\rho_m\right),\label{Ray1}\\
0 &= \ddot{\phi} + 3H\dot{\phi} + V_{\phi},\label{KG}\\
\dot{\rho}_m &= - 3H\gamma_m\rho_m ,\label{rhoeq}
\end{align}
\end{subequations}
where $V_\phi = dV/d\phi$ and where overdots denote derivatives with respect
to cosmic time $t$. Units are such that $8\pi G = 1 = c$, where $G$ is the
gravitational constant and $c$ the speed of light in vacuum. It is assumed
throughout the paper that the Universe is expanding, i.e., $H>0$.

Equation~\eqref{KG} can be heuristically interpreted as an equation for a
particle of unit mass with a one-dimensional coordinate $\phi$, moving in a
potential $V(\phi)$ with a friction force $-3H\dot{\phi}$.
Equation~\eqref{Gauss1} shows that $H$ can be expressed in terms of $\phi$,
$\dot{\phi}$, and $\rho_m$. Introducing $\dot{\phi}$ as an independent
variable therefore leads to a three-dimensional dynamical system for
$\phi,\dot{\phi},\rho_m$. Alternatively one can solve Eq.~\eqref{rhoeq}
to obtain $\rho_m=\rho_m(a)$ and consider the variables $\phi$, $\dot{\phi},
a$ as the dependent variables, where $a$ obeys the equation $\dot{a} = aH$,
where $H$ can be expressed in terms of $\phi,\dot{\phi},a$ by means
of~\eqref{Gauss1}.

It is of some interest to introduce effective equation of state parameters,
defined by
\begin{equation}
\gamma_\phi = \frac{\rho_\phi + p_\phi}{\rho_\phi} =
\frac{\dot{\phi}^2}{\frac12\dot{\phi}^2 + V(\phi)},\qquad
\gamma_{\mathrm{tot}} = \frac{\rho_{\mathrm{tot}} + p_{\mathrm{tot}}}{\rho_{\mathrm{tot}}},
\end{equation}
or, alternatively, $w_* = p_*/\rho_* = \gamma_* - 1$ ($*=\phi,\,
\mathrm{tot}$), where
\begin{subequations}
\begin{alignat}{2}
\rho_\phi &= \frac12\dot{\phi}^2 + V(\phi),&\qquad p_\phi &=
\frac12\dot{\phi}^2 - V(\phi),\\
\rho_{\mathrm{tot}} &= \rho_\phi + \rho_m, &\qquad p_{\mathrm{tot}} &= p_\phi
+ p_m.
\end{alignat}
\end{subequations}

Adding a cosmological constant to a perfect fluid with a barotropic equation
of state can be regarded as a problem with a single fluid with a changed
barotropic equation of state.
%, e.g., in the case of $\Lambda$CDM cosmology
%$\gamma_{\mathrm{tot}}$ is given by $\gamma_{\mathrm{tot}} = \rho_m/(\rho_m +
%\Lambda)$, which yields $\gamma_{\mathrm{tot}} =
%\Omega_{m0}\exp(-3\tau)/(\Omega_{m0}\exp(-3\tau) + \Omega_{\Lambda0})$.
Replacing a cosmological constant in a general barotropic fluid description
with a scalar field therefore adds the dynamical degrees of freedom, $\phi$
and $\dot{\phi}$, to $\rho_m$ (or $a$), which implies a complication, since
this leads to a larger set of solutions with different histories. In the
context of $\Lambda$CDM deformations this problem is somewhat alleviated by
the following assumption: To obtain an evolutionary history such that primordial 
nucleosynthesis gives observationally compatible light
nuclear abundances, we will assume that there exists a mechanism such as
inflation that sets initial conditions for the present models at high
redshifts so that the matter content greatly dominates over the scalar field
(dark energy) content in the early Universe.
%We stress that our main goal is to consider as simple classes of models as
%possible in order to clearly display general features that lead to open sets
%of $\Lambda$CDM deformed families of solutions.

For simplicity we will from now on consider a linear equation of state with a
constant $\gamma_m$ restricted to the range $0<\gamma_m<2$, which avoids
bifurcations at $\gamma_m=0$, corresponding to a cosmological constant, and
$\gamma_m=2$, which yields a stiff perfect fluid with the speed of sound
equal to the speed of light; $\gamma_m=1$ corresponds to a fluid without
pressure, which will be the focus when describing and discussing $\Lambda$CDM
deformations, while $\gamma_m=4/3$ corresponds to a radiation fluid. It
follows from Eq.~\eqref{rhoeq} that
\begin{equation}\label{rhosol}
\rho_m=\rho_0(a/a_0)^{-3\gamma_m}.
\end{equation}

Furthermore, most of the scalar field potentials that have been 
considered as possible alternatives to a cosmological constant as dark energy
have strictly monotonic scalar field potentials, as exemplified by, e.g., the popular 
inverse power-law potential.  
%Furthermore, since we are primarily interested in
%continuous scalar field deformations of $\Lambda$CDM cosmology this requires
%that the potential $V(\phi)$ is such that it does not give rise to solutions
%with different qualitative character, which occurs if the potential has
%extrema and/or inflection points. 
Apart from the model with $V=\Lambda = \mathrm{const.}$, we 
will therefore assume that $V$ is defined, differentiable,
positive, and strictly monotone for $\phi \in (\phi_-,\phi_+)$. Without loss
of generality, we assume that $V$ is monotonically decreasing. We also assume
that the potential is such that $\phi_+=\infty$. If $\phi_-$ is finite or
$-\infty$ depends on how fast $V(\phi)$ increases when $\phi\rightarrow
\phi_-$, e.g., for an inverse power-law potential, $V \propto
\phi^{-\alpha}$, $\phi_-=0$, while an exponential potential, $V \propto
\exp(-\lambda \phi)$, leads to $\phi_-=-\infty$.

A constant or monotonically strictly decreasing potential makes it convenient
to use a particular type of dynamical systems formulation that brings
\begin{equation}
\lambda(\phi) = - \frac{V_\phi}{V}
\end{equation}
into focus, where $\lambda$ is zero for a constant potential and positive for
$\phi \in (\phi_-,\infty)$ when $V(\phi)$ is strictly monotonically
decreasing. Within this context it is natural to regard $\lambda$ as more
``fundamental'' than $V$ itself, where $V$ is obtained from $\lambda(\phi)$ via
\begin{equation}\label{Vfromlambda}
V = V_0\exp\left(-\int_{\phi_0}^\phi \lambda(\tilde{\phi}) d\tilde{\phi}\right).
\end{equation}
In the special case where $\lambda$ and $\lim_{\phi \rightarrow
\phi_-}\lambda$ are bounded it follows that the potential can be bounded by
an exponential and therefore $\phi_- = -\infty$. If $\lambda$ is unbounded,
as in the case of an inverse power-law potential, $\phi_-$ is finite.

%Here, however, we will only attempt to shed some light on the above `field
%theoretical $\Lambda$CDM-deformation issue' with some simple quintessence
%examples by means of a dynamical systems approach, which also will situate
%some previous work in a hopefully useful new context.

Finally we note that the system~\eqref{Hphieq} has unbounded variables and 
right-hand sides that blow up. It is therefore not suitable for 
a global or asymptotic analysis of
the solution space and its properties. Below we will therefore make a change of
variables in order to obtain a dynamical system on a compact state space that
allows these issues to be addressed, as well as giving global illustrative
pictures of the entire solution spaces of the models under
consideration.\footnote{For a few examples of work dealing with dynamical
systems methods in cosmology,
see~\cite{waiell97,col03,giamir10,ugg13,leofad14,tam14,alhugg15,alhetal15},
and references therein.} 

The outline of the paper is as follows. In the next section we (i) formulate
the dynamical systems approach that is used throughout the paper to deal with
a fluid and a minimally coupled scalar field; (ii) describe the state space
features; (iii) point out some global properties and determine asymptotic
behavior for models with a perfect fluid and a scalar field such that
$0<\lambda(\phi) <\infty$, $\phi \in (\phi_-,\infty)$, and
\begin{equation}
\lim_{\phi \rightarrow \infty} \lambda = \lambda_+,\qquad \lim_{\phi \rightarrow \phi_-}\lambda = \lambda_-,
\end{equation}
where $\lambda_+$ is finite and $\lambda_-$ is finite or $\infty$.

In Sec.~\ref{sec:frozenlambda} we (i) situate $\Lambda$CDM cosmology,
which occurs for $\lambda=0$, in the three-dimensional state space, which
leads to the concept of an attracting separatrix surface rather than
individual ``attractor solutions''; (ii) consider the $\lambda=\mathrm{const.}$
(i.e., an exponential potential) and dust models in the present context and
give a specific example that shows that some of these models are completely
solvable; (iii) discuss the symmetries of the above models and how the
associated structures might be somewhat misleading for the general picture.
%This is partly due to that the symmetries make it possible to obtain
%projected two-dimensional representations of the dynamics of these models,
%which is usually what is done, while it is necessary to consider them in the
%generic three-dimensional state space in order to situate them in a generic
%context.

In Sec.~\ref{sec:dynlambda} we turn from the previous ``frozen $\lambda$''
models to some ``dynamical $\lambda$'' models. We consider (i) the inverse
power-law potentials, characterized by $\lambda= \alpha/\phi$, in a global
dynamical systems setting, tying local results to the global features
discussed in Sec.~\ref{sec:dynsys}; in particular we point out that
observational viability requires fine-tuning of $\alpha$ to small values,
since $\lambda \rightarrow \infty$ when $\phi \rightarrow 0$ produces a
``memory'' that affects the entire evolution of, e.g., ``tracker'' (attractor)
solutions. For this reason we therefore (ii) consider the arguably simplest
possible ``$\lambda$-regularization'' of the inverse power-law potential,
namely, $\lambda = \alpha/\sqrt{C^2 + \phi^2}$, which, in contrast to the
inverse power-law case, gives rise to continuous deformations of the
attracting $\Lambda$CDM separatrix surface, and which for $C \gtrsim \alpha$ 
yields much less stringent observational compatibility
conditions on $\alpha$. Finally, the section is concluded with (iii) a
general discussion about observational viability conditions that $\lambda$
should satisfy for \emph{any} potential that might arise from some fundamental
theory.

Throughout we use dynamical systems pictures which clearly display general
key features, and, in order to compare with $\Lambda$CDM cosmology, these
pictures are complemented by diagrams that indicate the evolutionary history
of physically important quantities. Finally, we outline how it is possible to
treat two fluids, e.g. dust and radiation, and a scalar field in
the Appendix.

%%%%%%%%%%%%%%%%%%%%%%%%%%%%%%%%%%%%%%%%%%%%%%%%%%%%%%%%%%%%%%%%%%
\section{Dynamical systems approach}\label{sec:dynsys}
%%%%%%%%%%%%%%%%%%%%%%%%%%%%%%%%%%%%%%%%%%%%%%%%%%%%%%%%%%%%%%%%%%

%-----------------------------------------------------------------
\subsection{Dynamical systems formulation}
%-----------------------------------------------------------------

The presently used dynamical systems formulation is based on the following
dependent variables,
\begin{equation}\label{vardef1}
(x,\Omega_V,\Omega_m,Z) =
\left(\frac{\dot{\phi}}{\sqrt{6}H}, \frac{V}{3H^2}, \frac{\rho_m}{3H^2},Z(\phi)\right).
\end{equation}
These variables are by no means suitable for all scalar field potentials,
e.g., for potentials with a zero minimum such as monomial potentials it is
advisable to replace the scalar field variable $Z$ with an $H$-based variable
that takes into account a varying averaged oscillatory time scale at late
times, as done in~\cite{alhetal15}. Nevertheless, the above variables are
useful for positive monotonic potentials. %positive,
%monotone, differentiable potential $V(\phi)$ on an interval $\phi \in
%(\phi_-,\phi_+)$, where $\phi_+$ ($\phi_-$) may or may not be equal to
%$\infty$ ($-\infty$), and where $\lim_{\phi\rightarrow \phi_+}V(\phi)$ has a
%non-negative limit, which we henceforth assume.
An optimal choice of the variable $Z(\phi)$ depends on the potential that is
studied. However, to obtain a suitable global dynamical systems formulation
$Z$ should always be chosen to be a globally invertible monotone function in
$\phi$, defined on a \emph{bounded} interval $Z\in (Z_-,Z_+)$, $Z_\pm =
Z(\phi_\pm)$; without loss of generality, we choose $Z$ to be monotonically
increasing in $\phi$ so that $dZ/d\phi>0$ for $Z\in (Z_-,Z_+)$.

Apart from the dependent variables we also need to choose a new time
variable. In order to obtain a regular dynamical system this choice depends
on if $\lambda$ is bounded or not. In the case $\lambda$ is bounded, as
exemplified by, e.g., exponentially bounded potentials, it is convenient to use
${\tau}$, defined by $dt=H^{-1}d{\tau}$, where ${\tau}= \ln(a/a_0)$ sometimes
is referred to as $N$, the number of $e$-folds from a reference time $t_0$
[$a_0 = a(t_0)$], which for the present time leads to $\tau = -\ln(1+z)$,
where $z$ is the redshift. Thus, the field equations can be
written as the following coupled system:
\begin{subequations}\label{dynsys0}
\begin{align}
x^\prime &= -(2-q)x + \sqrt{\frac{3}{2}}\lambda(Z)\,\Omega_V,\label{xeq}\\
\Omega_m^\prime &= 3\left[2x^2 - \gamma_m(1 - \Omega_m)\right]\Omega_m,\label{Omeq}\\
Z^\prime &= \sqrt{6}\frac{dZ}{d\phi}\,x,\label{Zeq0}
\end{align}
\end{subequations}
where ${}^\prime$ denotes derivatives with respect to $\tau$. With some slight
abuse of notation we have written $\lambda(Z) = \lambda(\phi(Z)) =
-V_\phi/V$. Furthermore, the Gauss constraint~\eqref{Gauss1} is used to
globally express the Hubble-normalized scalar field potential energy,
$\Omega_V$, in terms of the state space variables:
\begin{equation}\label{Gauss2}
\Omega_V = 1 - x^2 - \Omega_m,
\end{equation}
while the deceleration parameter $q$, defined via ${H}^\prime = - (1+q)H$, is
given by
\begin{equation}\label{q}
q = - 1 + 3x^2 + \frac32\gamma_m\Omega_m.
\end{equation}

The above system is suitable when $\lambda$ is bounded, but not when
$\lim_{\phi \rightarrow \phi_-}\lambda = \infty$, as exemplified by, e.g.,
inverse power-law potentials. In order to obtain a regular dynamical system
when $\lim_{\phi \rightarrow \phi_-}\lambda = \infty$ we therefore choose a
new time variable $\bar{\tau}$, defined by $d\tau = g(Z)d\bar{\tau}$, where
$g(Z)$ is a suitable positive bounded function of $Z$ such that
$\lim_{Z\rightarrow Z_-}g(Z)=0$,
\begin{equation}
\lim_{Z\rightarrow Z_-}g_\lambda(Z)  = g_{\lambda_-} =\mathrm{const}. < \infty, \quad \text{where} \quad
g_\lambda(Z) := g(Z)\lambda(Z),
\end{equation}
and $\lim_{Z\rightarrow Z_+}g(Z)=1$, where, e.g., $g(Z) =1/(1 + \lambda(Z))$ is
a suitable choice for inverse power-law potentials. This results in the
system
\begin{subequations}\label{dynsysZ}
\begin{align}
\frac{dx}{d\bar{\tau}} &= -g(Z)(2-q)x + \sqrt{\frac{3}{2}}g_\lambda(Z)\Omega_V,\label{xeqg}\\
\frac{d\Omega_m}{d\bar{\tau}} &= 3g(Z)\left[2x^2 - \gamma_m(1 - \Omega_m)\right]\Omega_m,\label{Omeqg}\\
\frac{dZ}{d\bar{\tau}} &= \sqrt{6}g(Z)\frac{dZ}{d\phi}\,x.\label{Zeq0g}
\end{align}
\end{subequations}
Note that~\eqref{dynsys0} is obtained by setting $g(Z)=1$, hence
$\bar{\tau}=\tau$, and so the above system formally covers both cases.

%-----------------------------------------------------------------
\subsection{State space structures}
%-----------------------------------------------------------------

The above assumptions lead to the following domain for the state vector
$(x,\Omega_m,Z)$ for the \emph{interior state space} ${\bf S}$ for models
with a perfect fluid and a scalar field:
\begin{equation}
\Omega_m>0, \qquad \Omega_V = 1- x^2 - \Omega_m > 0, \qquad Z_-<Z<Z_+.
\end{equation}
Furthermore, it follows from~\eqref{dynsys0}, and the following auxiliary
equation for $\Omega_V$,
\begin{equation}\label{OmV}
\Omega_V^\prime = 2\left(1 + q - \sqrt{\frac32}\lambda(Z)\,
x\right)\Omega_V,
\end{equation}
that it is possible to extend the state space and include the following
boundaries:
\begin{itemize}
\item The boundary $\Omega_m=0$ (i.e. $\rho_m=0$) yields the (pure)
    \emph{scalar field boundary subset} [see Eq.~\eqref{Omeq}], and thus
    \begin{equation}\Omega_\phi = x^2 + \Omega_V =1,
    \end{equation}
    with an interior state space ${\bf S}_\phi$ given by $\Omega_V = 1 -
    x^2 > 0$, $Z_-<Z<Z_+$.
\item The boundary $\Omega_V = 1 - x^2 - \Omega_m=0$ corresponds to a
    model with a perfect fluid and a massless scalar field; we refer to
    this subset as the \emph{massless scalar field boundary subset}. Note
    that this boundary also contains the pure massless scalar field
    boundary subset $\Omega_m=0$, $x=\pm 1$, and the \emph{perfect fluid
    boundary subset} $\Omega_m=1$ ($\Omega_\phi=0$, i.e.,
    $\Omega_V=0=x$), which is described by a line of fixed points,
    $\mathrm{FL}_Z$, with constant $Z$.
\item In addition we assume that the potential and $Z$ are such that the state space ${\bf S}$
can be extended to not only include the boundaries $\Omega_m=0$,
$\Omega_V=0$, but also the boundaries $Z=Z_-$ and $Z=Z_+$, which furthermore
are assumed to constitute invariant boundary subsets.
\end{itemize}
%

%With the exception of the inverse power-law potential, we assume that
%$\lambda$ is \emph{bounded} when $\phi\in (\phi_-,\phi_+)$, $\phi \rightarrow
%\phi_\pm$. In addition
%We , so that 
Thus the \emph{extended compact state spaces} $\bar{\bf S}$ and $\bar{\bf S}_\phi$ are characterized by
\begin{subequations}
\begin{align}
\Omega_m &\geq 0, &\quad \Omega_V &= 1 - x^2 - \Omega_m \geq 0, &\quad &Z_-\leq Z \leq Z_+ ,\\
\Omega_V &= 1- x^2 \geq 0, &\quad & Z_-\leq Z \leq Z_+, & &
\end{align}
\end{subequations}
respectively, and the dynamical systems on $\bar{\bf S}$ and $\bar{\bf
S}_\phi$ are of differentiability class $C^1$ or higher. We therefore assume
that the potential is such that it is possible to find a scalar field
variable $Z$ and, if needed, a function $g(Z)$ so that
\begin{itemize}
\item[(i)] $g_\lambda(Z)$ \emph{and} $g(Z)dZ/d\phi$ are bounded and
    differentiable on the extended interval $Z\in [Z_-,Z_+]$,
\item[(ii]) $g(Z)dZ/d\phi|_{Z=Z_\pm}=0$, i.e., $Z=Z_-$ and $Z=Z_+$ are
    invariant boundary subsets.
\end{itemize}

Although the above restrictive assumptions require that the potential
$V(\phi)$ is of differentiability class $C^2$, they cover a vast class of
potentials $V(\phi)$. Moreover, the extended state space treatment of the
admissible class of potentials can be used to provide bounds on asymptotic
behavior of potentials that do not admit $Z$-extended state spaces, and thus
they provide a natural starting point for very general situations. It should
also be noted that for many qualitative dynamical aspects, as will be seen,
there is no need to state any more information about $Z(\phi)$ or $V(\phi)$,
although quantitative treatments of course need specific functions and thus
we will give specific examples when dealing with continuous $\Lambda$CDM
deformations.

To physically compare solutions that are obtained in the dynamical systems
approach with $\Lambda$CDM cosmology, we will complement the state space
description with diagrams involving physically important quantities.
%To physically understand the dynamics given by the state space formulation we
%complement it by considering physically interesting quantities.
%Once one knows the dynamics it has to be physically interpreted. This can be
%done by plotting physically interesting quantities.
Apart from the physically important state space variable $\Omega_m$ we have
given the deceleration parameter $q$ in terms of state space variables in
Eq.~\eqref{q}, and below we will also express $H$ in state space variables.
In addition it is of interest to give the jerk parameter
$j$~\cite{blaetal04,rapetal07}, since even though the observational
feasibility of $j$ is questionable, it is still useful in order to
theoretically compare models with $\Lambda$CDM cosmology for which $j=1$; for
constant $\gamma_m$ the jerk parameter is given by
\begin{equation}
j = 1 + 9x^2 - \frac92\gamma_m(1-\gamma_m)\Omega_m - 3\sqrt{6}\lambda(Z)\,\Omega_V\,x.
\end{equation}

Next we turn to structures that are helpful when it comes to determining global
and asymptotic state space features.

%-----------------------------------------------------------------
\subsection{Nonlocal features}
%-----------------------------------------------------------------

%-----------------------------------------------------------------
\subsubsection*{The monotonicity of $H$}
%-----------------------------------------------------------------

The Hubble parameter $H$, or more conveniently $H^2$, can be expressed in the
state space variables in ${\bf S}$ and ${\bf S}_\phi$ and is given by
\begin{equation}\label{H2}
H^2 = \frac{V(\phi(Z))}{3\Omega_V} = \frac{V(\phi(Z))}{3(1 - x^2 - \Omega_m)},
\end{equation}
as follows from the definition of $\Omega_V$. This has far-reaching
consequences for the global dynamics due to
\begin{equation}
H^\prime = - (1+q)H \quad \Rightarrow  \quad (H^2)^\prime = - 2(1+q)H^2,
\end{equation}
since Eqs.~\eqref{Gauss2} and~\eqref{q} lead to
\begin{equation}\label{eqqineq}
1+q=3x^2 + \frac32\gamma_m\Omega_m, \qquad
2-q=3\Omega_V + \frac32(2-\gamma_m)\Omega_m,
\end{equation}
which together with $0<\gamma_m<2$ yield
\begin{equation}
-1\leq q\leq 2.
\end{equation}
As will be seen, the above inequalities for $q$, induced by energy and causality conditions,
play a central role for the dynamics, which strongly suggests that the field
equations should be expressed in terms of $q$, as done in~\eqref{dynsys0}.

The above relations imply that $q=2$ only occurs on the pure massless scalar
field subsets $x=\pm 1$, while $q=-1$ requires $\Omega_m=x=0$, $\Omega_V=1$.
Since $1+q>0$ on ${\bf S}$, $H^2$ is a monotonically decreasing function on
${\bf S}$. This is also true for ${\bf S}_\phi$, as can be seen as follows.
On ${\bf S}_\phi$ we have the inequalities $-1\leq q \leq 2$. As a
consequence $H^2$ is a monotonically decreasing function also on this subset,
except when $1+q=0$. However, when $1+q=0$ is not an invariant subset, then
$1+q=0$ just corresponds to an inflection point in the graph of $H^2$, and
$H^2$ is thereby still monotonically decreasing.\footnote{For an elaboration
of this in the case of monomial potentials, see~\cite{alhetal15}.} It is only
when $1+q=0$ is an invariant subset that $H^2$ stops decreasing and for which
$1+q=0$ can be an asymptotic state. For this to happen requires that $x=0$,
since $1+q=3x^2$ on ${\bf S}_\phi$, also means that $x=0$ must be
an invariant subset. Since $x^\prime = \sqrt{\frac{3}{2}}\lambda(Z)$ when
$\Omega_m=0=x$, $\Omega_V=1$, this leads to the condition that
$\lambda(Z)=0$, which by our assumptions about monotonically decreasing
potentials can only happen on the $Z_\pm$ boundaries. Hence $H^2$ is also
monotonically decreasing on ${\bf S}_\phi$. This implies that there can be no
fixed points or recurring orbits (solution trajectories) in ${\bf S}$ and
${\bf S}_\phi$, and due to~\eqref{H2} all solution trajectories in ${\bf S}$
and ${\bf S}_\phi$ originate and end at $Z_\pm$ and $\Omega_V=0$, which shows
that it is essential to extend the state space to include these boundaries 
in order to fully describe the dynamics of the present models .

%-----------------------------------------------------------------
\subsubsection*{The $Z_\pm$ boundaries}
%-----------------------------------------------------------------

Let us begin by considering the $Z_\pm$ boundaries for the case where
$\lim_{Z\rightarrow Z_\pm}\lambda = \lambda_\pm$ is finite. Dropping the
subscript $\pm$ on $\lambda_\pm$ leads to the system
\begin{subequations}\label{dynsysred}
\begin{align}
x^\prime &= -(2-q)x + \sqrt{\frac{3}{2}}\lambda\,\Omega_V,\label{xeqp}\\
\Omega_m^\prime &= 3\left[2x^2 - \gamma_m(1 - \Omega_m)\right]\Omega_m,\label{Omeqp}
\end{align}
\end{subequations}
on $Z_\pm$ where we recall that $\Omega_V = 1 - x^2 - \Omega_m$ and  $q = - 1
+ 3x^2 + \frac32\gamma_m\Omega_m$. As will be elaborated on in the next
section, these equations describe the reduced equations of an exponential
potential $V \propto \exp(-\lambda\phi)$ when $\lambda \neq 0$ and those of a
constant potential when $\lambda=0$. The above system has an extended state
space $\bar{\bf S}_\mathrm{red}$ given by
\begin{equation}
\Omega_m\geq 0, \qquad \Omega_V = 1 - x^2 - \Omega_m \geq 0,
\end{equation}
and admits a number of fixed points given in Table~\ref{tab:fixedpoint}.
\begin{table}[ht!]
\begin{center}
\begin{tabular}{|c|c|c|c|}\hline
% &  Fixed point values & &  \\ \hline
$\mathrm{FL}$ & $x = 0$, $\Omega_m = 1$ & $q = (3\gamma_m-2)/2$ & - \\ \hline
$\mathrm{EM}$ & $x = \sqrt{3/2}\gamma_m\lambda^{-1}$, $\Omega_m = 1 - 3\gamma_m\lambda^{-2}$; $\lambda >\sqrt{3\gamma_m}$ & $q =(3\gamma_m-2)/2$
& $\gamma_\phi=\gamma_m$ \\ \hline
$\mathrm{M}^\pm$ & $x = \pm 1$, $\Omega_m =0$ & $q = 2$ & $\gamma_\phi=2$ \\ \hline
$\mathrm{PL}$ & $x = \lambda/\sqrt{6}$, $\Omega_m = 0$; $\lambda <\sqrt{6}$ & $q = -1 + \lambda^2/2$ & $\gamma_\phi = \lambda^2/3$ \\ \hline
$\mathrm{dS}$ & $x = \Omega_m = 0$; $\lambda=0$ & $q = -1$ & $\gamma_\phi =0$ \\ \hline
\end{tabular}
\end{center}
\caption{Table depicting the type of fixed points that can occur for a FLRW
model with a monotonically decreasing scalar field potential with a finite
$\lambda$ and a perfect fluid with a linear equation of state parameter
$0<\gamma_m<2$.} \label{tab:fixedpoint}
\end{table}

We have here introduced a notation for the fixed points where the kernel
$\mathrm{M}$ stands for a massless scalar field state; $\mathrm{FL}$ stands
for a Friedmann-Lema\^{i}tre perfect fluid state, while $\mathrm{dS}$ stands
for a de Sitter state (for interpretation of various types of de Sitter
states in scalar field cosmology, see~\cite{alhetal15}); the kernel
$\mathrm{PL}$ stands for power law (inflation, when $\lambda <\sqrt{2}$, since
this leads to an accelerating state with $q<0$), while $\mathrm{EM}$ stands
for a scaling solution with an exponential potential and matter in the form of a
perfect fluid with a linear equation of state. In addition a superscript
describes the value of $x$, while in the full three-dimensional treatment we
also add a subscript that describes the value for $Z$.

\begin{figure}[ht!]
\begin{center}
\subfigure[Solution space for $\lambda=0$ and $\gamma_m=1$.]{\label{fig:SS_CC2D}
\includegraphics[width=0.45\textwidth, trim = 0cm 2cm 0cm 0cm]{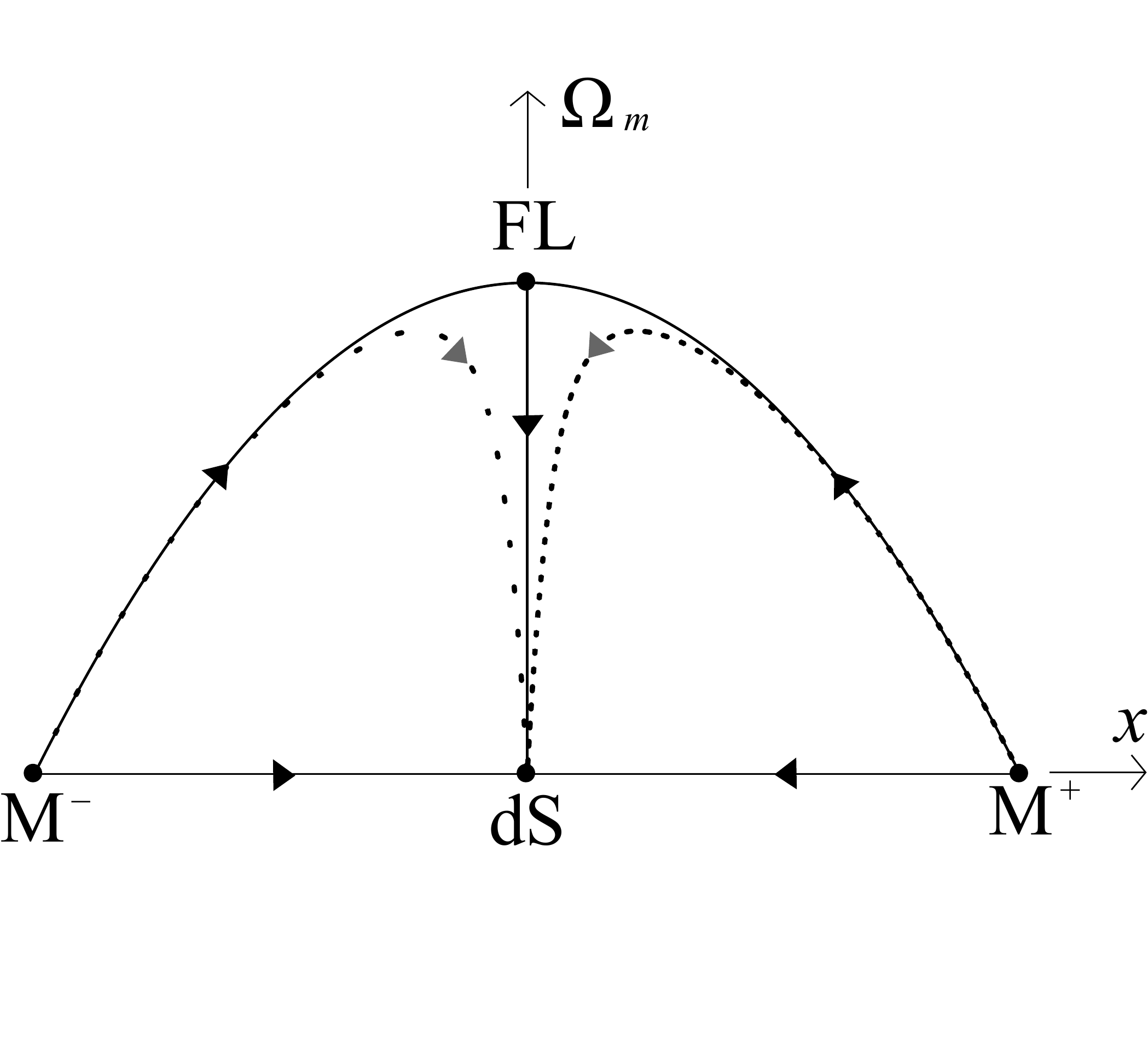}} \quad
\subfigure[Solution space for ${\lambda}=\sqrt{\frac{3}{2}}$ and $\gamma_m=1$.]{\label{fig:SSCC2D}
\includegraphics[width=0.45\textwidth, trim = 0cm 2cm 0cm 0cm]{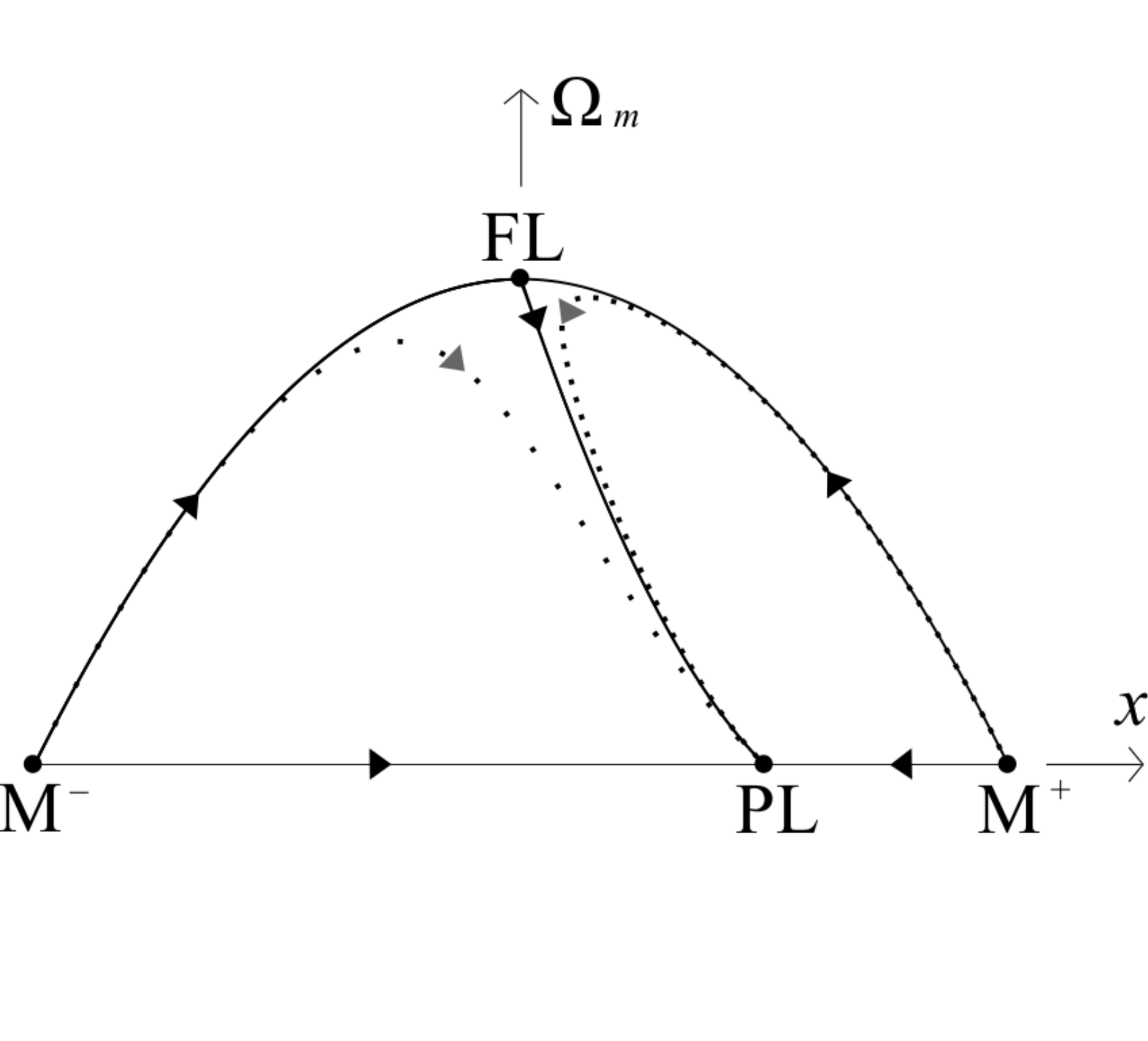}}\\
\subfigure[Solution space for ${\lambda}=2$ and $\gamma_m=1$.]{\label{fig:SSCCnum}
\includegraphics[width=0.45\textwidth, trim = 0cm 2cm 0cm 0cm]{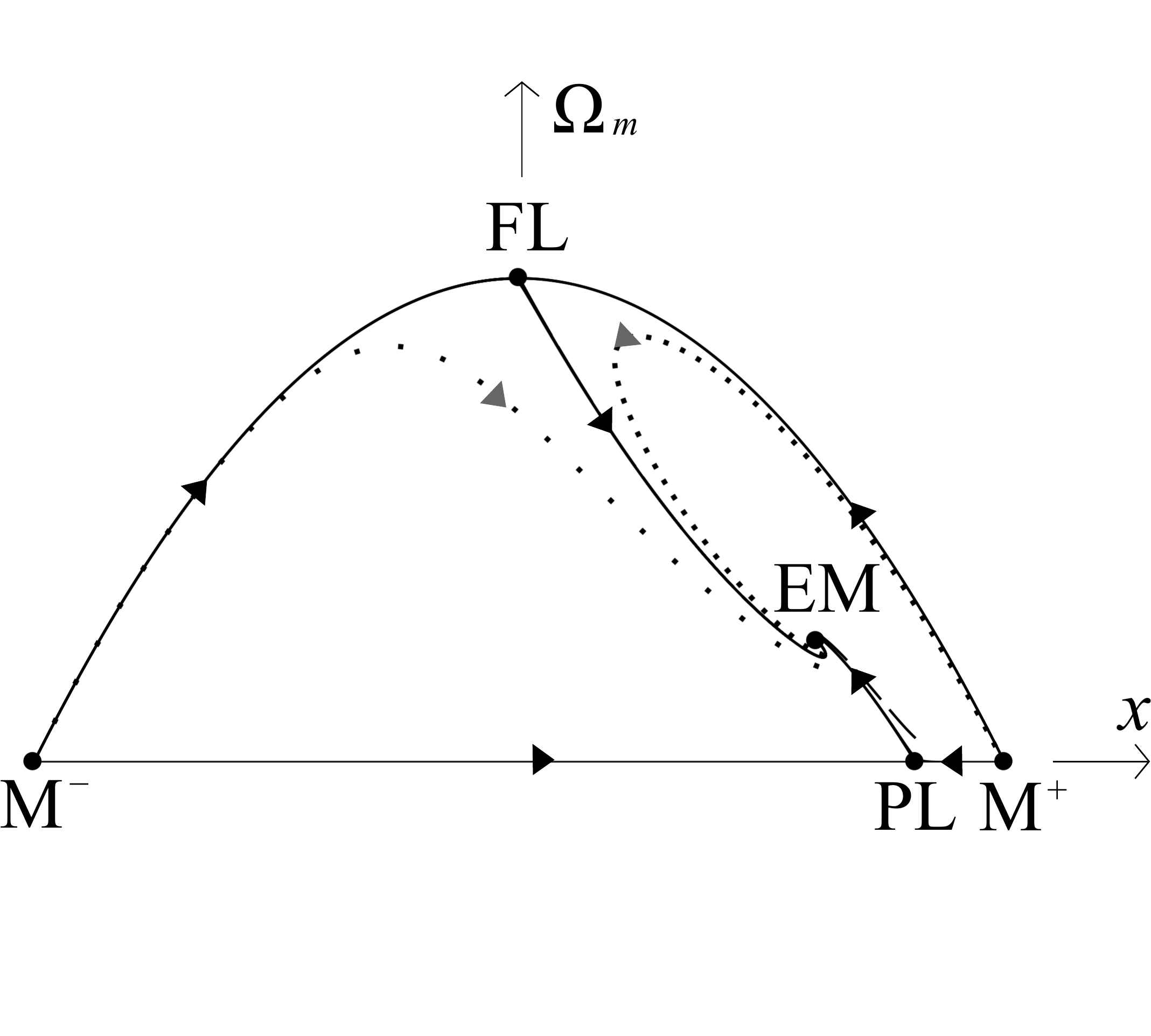}} \qquad
\subfigure[Solution space for ${\lambda}=3$ and $\gamma_m=1$.]{\label{fig:SSCCLamda2}
\includegraphics[width=0.45\textwidth, trim = 0cm 2cm 0cm 0cm]{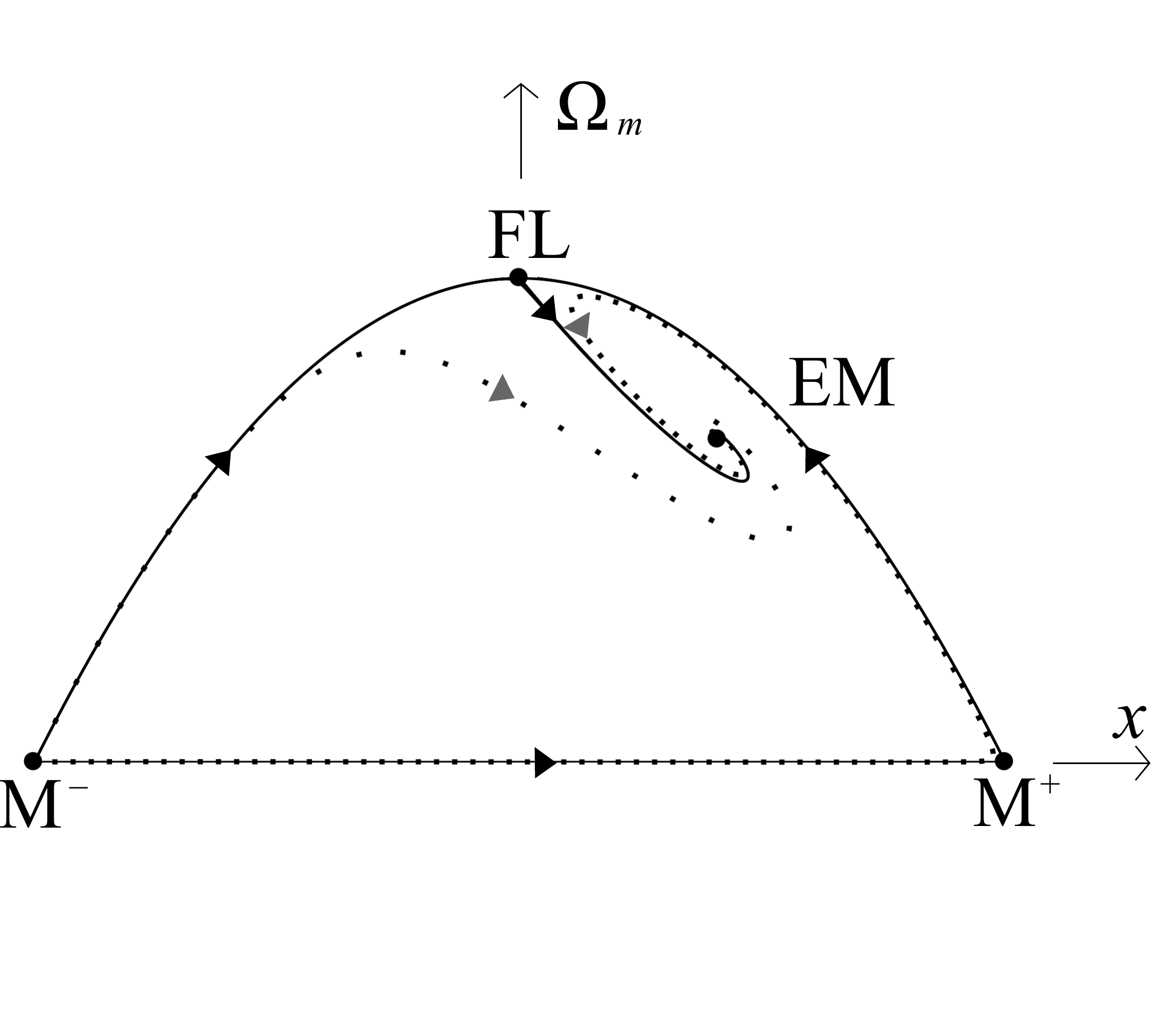}}
\vspace{-0.5cm}
\end{center}
\caption{Representative solution space structures for $\bar{\bf
S}_\mathrm{red}$.}
\label{fig:redrep}
\end{figure}

A one-parameter set of orbits in $\bar{\bf S}_\mathrm{red}$ 
originates from each of the sources $\mathrm{M}^+$ and $\mathrm{M}^-$ when
${\lambda}<\sqrt{6}$, but when ${\lambda}>\sqrt{6}$ there are no solutions
that originate from $\mathrm{M}^+$ into ${\bf S}_\mathrm{red}$,
while one solution originates from $\mathrm{FL}$. Toward the future
$\mathrm{PL}$ is a sink when ${\lambda} < \sqrt{3\gamma_m}$
%(and the future attractor on $\bar{\bf S}_\mathrm{red}$
while $\mathrm{EM}$ is future stable when ${\lambda} > \sqrt{3\gamma_m}$.
%(and the future attractor on $\bar{\bf S}_\mathrm{red}$).
Thus bifurcations take place when ${\lambda}=\sqrt{6}$, which is when
$\mathrm{PL}$ leaves the physical state space via $\mathrm{M}^+$, and when
${\lambda}=\sqrt{3\gamma_m}$, which is when $\mathrm{EM}$ enters the state
space via $\mathrm{PL}$. In addition there is a bifurcation when $\lambda=0$,
which is when $\mathrm{PL}$ becomes the future sink $\mathrm{dS}$. The
solution space structures for representative examples of these different
cases are given in Fig.~\ref{fig:redrep}.

The case when $\lambda \rightarrow \infty$ when $\phi \rightarrow \phi_-$
yields the following system on $Z_-$:
\begin{subequations}
\begin{align}
\frac{dx}{d\bar{\tau}} &= \sqrt{\frac{3}{2}}g_{\lambda_-}\,\Omega_V,\label{xm}\\
\frac{d\Omega_m}{d\bar{\tau}} &= 0,\label{Omm}
\end{align}
\end{subequations}
where we recall that $g_{\lambda_-} = \lim_{Z\rightarrow Z_-}
(g(Z)\lambda(Z))>0$. It follows that $\Omega_m = \mathrm{const}.$ while the
subset $\Omega_V=1-x^2-\Omega_m=0$ forms a line of fixed points
$\mathrm{M}^x$, and that $x$ is monotonically increasing in ${\bf
S}_\mathrm{red}$. Thus $\mathrm{M}^x$ with $x<0$ is the source while
$\mathrm{M}^x$ with $x>0$ is the sink of all orbits in ${\bf
S}_\mathrm{red}$; finally, note that $\mathrm{M}^x$ with $x=0$ is the fixed
point $\mathrm{FL}$.

%-----------------------------------------------------------------
\subsubsection*{The $\Omega_V=0$ boundary}
%-----------------------------------------------------------------

We first note that the equation for $x$ on $\Omega_V=0$ is invariant under
the transformation $x\rightarrow - x$, which enables one to use
$\Omega_\mathrm{stiff}=x^2$ as a variable instead of $x$, and as a result
the dynamics of the massless scalar field explicitly takes the same form as
that of a stiff fluid, $p=\rho$. This leads to the equations
\begin{subequations}\label{dynsysOmV}
\begin{align}
\Omega_m^\prime &= 3(2-\gamma_m)(1 - \Omega_m)\Omega_m,\label{OmeqV}\\
Z^\prime &= \sqrt{6}\,\frac{dZ}{d\phi}\,x = \pm\sqrt{6}\,\frac{dZ}{d\phi}\,\sqrt{1-\Omega_m},\label{ZeqV}
\end{align}
\end{subequations}
where we have used the Gauss constraint to write $\Omega_\mathrm{stiff}= 1 -
\Omega_m$ and $x=\pm \sqrt{\Omega_\mathrm{stiff}}$. Since the equation for
$\Omega_m$ decouples from that of $Z$, it follows that all orbits (solution
trajectories) in the two-dimensional state space look the same as the orbits
of the one-dimensional problem associated with $\Omega_m$, when projected
onto $\Omega_m$. As follows from~\eqref{OmeqV}, $\Omega_m$ monotonically
increases from $0$ to $1$. Thus the past asymptotic state of all orbits on
$\Omega_V=0$ resides on the pure massless scalar field subset $\Omega_m=0$,
$x^2=1$ and thereby $q=2$; thus either $x=+1$, which from~\eqref{ZeqV} leads to
$Z=Z_-$ and thus a fixed point $\mathrm{M}^+_{Z_-}$, since $dZ/d\phi>0$, or
$x=-1$, which yields $Z=Z_+$ and a fixed point $\mathrm{M}^-_{Z_+}$. It also
follows that the future asymptotic state of all orbits on $\Omega_V=0$
resides on $\Omega_m=1$, $x=0$, $Z=\mathrm{const}.$, which thereby form a
line of fixed points $\mathrm{FL}_Z$. To determine what happens with $Z$
toward the future on $\Omega_V=0$ we note that it is possible to solve the
equation for $\Omega_m$ explicitly, but the solution can also be obtained by
using the fact that $\rho_\mathrm{stiff} \propto \exp(-6\tau)$, $\rho_m \propto
\exp(-3\tau)$, $3H^2 = \rho_\mathrm{stiff} + \rho_m$, which gives
\begin{subequations}
\begin{align}
\frac{H}{H_0} &= \left(\Omega_{\mathrm{stiff}0}\exp(-6\tau)+ \Omega_{m0}\exp(-3\gamma_m\tau)\right)^{1/2},\\
x &= \pm \Omega_{\mathrm{stiff}0}^{1/2}\,\exp(-3\tau)/(H/H_0), \\
\Omega_m &= \Omega_{m0}\exp(-3\gamma_m\tau)/(H/H_0)^2.
\end{align}
\end{subequations}
To obtain the solution for $Z$ requires a specification of $Z$. However, it
is easier to obtain $Z$ by integrating the equation for $\phi$,
\begin{equation}
\phi^\prime = \sqrt{6}x = \pm(6\Omega_{\mathrm{stiff}0})^{1/2}\,\exp(-3\tau)/(H/H_0),
\end{equation}
and then inserting the result in the chosen definition of $Z$. As follows
from the above equation, $\phi \rightarrow \mathrm{const.}$ when $\tau
\rightarrow +\infty$, and therefore also $Z\rightarrow \mathrm{const.}$ in
this limit; i.e., each point on the line $\mathrm{FL}_Z$ attracts one
solution with positive and one with negative initial $x$ on the $\Omega_V=0$
boundary. For a representative depiction of the $\Omega_V=0$ boundary, see
Fig.~\ref{fig:V0Dust}.

%-----------------------------------------------------------------
\subsubsection*{Role of the $\Omega_V=0$ boundary}
%-----------------------------------------------------------------

To study the neighborhood of $\Omega_V=0$ we note that
\begin{equation}
\Omega_V^{-1}\Omega_V^\prime|_{\Omega_V=0} = 2\left(1 + q - \sqrt{\frac32}\lambda(Z)\, x\right),
\end{equation}
and hence
\begin{equation}\label{MFL}
\Omega_V^{-1}\Omega_V^\prime|_{\mathrm{M}^\pm_{Z_\mp}} = \sqrt{6}\left(\sqrt{6} \mp \lambda(Z_\mp)\right), \qquad
\Omega_V^{-1}\Omega_V^\prime|_{\mathrm{FL}_Z} = 3\gamma_m.
\end{equation}
As a consequence it follows that $\mathrm{M}^-_{Z_+}$ is a source for orbits
in ${\bf S}$ and ${\bf S}_\phi$ as is $\mathrm{M}^+_{Z_-}$ when $\lambda(Z_-)
< \sqrt{6}$, but if $\lambda(Z_-) > \sqrt{6}$, then no orbits enter ${\bf S}$
or ${\bf S}_\phi$ from $\mathrm{M}^+_{Z_-}$;\footnote{Solutions originating
from $\mathrm{M}^-_{Z_+}$ corresponds to ``scalar field particles'' that come
from $\infty$ moving toward decreasing $\phi$, while if $\lambda(Z_-) <
\sqrt{6}$ the potential is shallow enough so that scalar field particles also 
can come from $-\infty$ moving toward increasing $\phi$. Solutions originating
from $\mathrm{FL}_Z$ correspond to scalar field particles that initially lie
still at some initial value $\phi$ and then roll down the potential
$V(\phi)$.} a one-parameter set of orbits, one from each point on
$\mathrm{FL}_Z$, enters ${\bf S}$, where $\mathrm{FL}_Z$ acts as a
``transversal saddle line,'' forming a two-dimensional invariant subset in
${\bf S}$. By demanding initial data for which $\Omega_m$ is close to 1,
this implies the important feature that for \emph{all} the models we consider
such solutions can be approximated by this (at least initially)
\emph{attracting invariant} surface \emph{subset} of codimension 1 with
respect to the state space dimension. We will later contrast this feature
with discussions about \emph{single} ``attracting'' solutions that appear for
various models in the literature.

%-----------------------------------------------------------------
\subsubsection*{Monotonicity of negative $x$}
%-----------------------------------------------------------------

Further restrictions on the global dynamics of the present models come from
eqs.~\eqref{xeq} and~\eqref{Zeq0}. It follows from~\eqref{xeq} that $x$ is
monotonically increasing when $-1<x<0$, since then $2-q>0$. This in turn
implies that if an orbit resides or partly resides in the region $-1<x<0$,
then the orbit originates from $x=-1$, which due to~\eqref{Zeq0} and
$dZ/d\phi>0$ implies that the past asymptotic state must be
$\mathrm{M}^-_{Z_+}$. Furthermore, the monotonicity of $x$ when $-1<x<0$
implies that all orbits in ${\bf S}$ and ${\bf S}_\phi$ must end at $x\geq
0$.

The equation for $x$ can be written on the following form:
\begin{equation}
x^\prime = -\frac32\left(2-\gamma_m\right)x\,\Omega_m - 3\left(x - \frac{\lambda(Z)}{\sqrt{6}}\right)\Omega_V,
\end{equation}
which shows that $x$ is monotonically decreasing if $1>x >
\lambda(Z)/\sqrt{6}$. In combination with the above result for $x$ this
implies that all orbits in ${\bf S}$ and ${\bf S}_\phi$ must end at $0 \leq x
\leq \lambda(Z)/\sqrt{6}$. However, in order for an orbit to end at $x=0$
requires that $x=0$ be an invariant subset. Since
\begin{equation}
x^\prime|_{x=0} = \sqrt{\frac32}\lambda(Z)(1-\Omega_m),
\end{equation}
and since $\Omega_m=1$ is the $\mathrm{FL}_Z$ subset, it follows that this
can only be the case if $\lambda(Z)=0$, which can only happen on the $Z_\pm$
boundaries for a monotonically strictly decreasing potential. The above
equation also shows that $x$ is monotonically increasing when $x=0$ when
$Z_-<Z<Z_+$. Furthermore, since $dZ/d\phi>0$ it follows from~\eqref{Zeq0}
that $Z$ is monotonically increasing (decreasing) in $\tau$, or $\bar{\tau}$,
when $x>0$ ($x<0$).

If $x$ is future asymptotically positive it follows that $Z\rightarrow Z_+$.
However, to exclude or show that any of the orbits from $\mathrm{M}^-_{Z_+}$
end up at $\mathrm{dS}_{Z_-}$ if $\lambda_- =0$ requires more information
about the potential. If $\lim_{\phi \rightarrow \phi_-} V(\phi) =
\mathrm{const}.$, then there are solutions that end at $Z_-$, but if $V
\rightarrow \infty$ when $\phi \rightarrow \phi_-$, there are not. This can
be understood by considering~\eqref{KG}. A solution with negative $x$ can be
viewed as a particle moving in a direction where the potential is monotonically 
increasing. At the same time it is losing energy due to the friction force 
$-3H\dot{\phi}$ until $\dot{\phi}=0$,
%($V\neq 0$ implies $H\neq 0$
%due to the Gauss constraint~\eqref{Gauss1},),
but then $\ddot{\phi} = -V_\phi$, where $V_\phi$ is strictly negative.
%, also
%in the limit $\phi_-$ when $V \rightarrow \infty$, and 
As a consequence the particle changes direction and $\dot{\phi}$ and thereby also $x$ become
positive, and therefore $Z\rightarrow Z_+$ toward the future.

Thus all orbits in ${\bf S}$ and ${\bf S}_\phi$ end at the invariant boundary
subset $Z_+$ when the potential is monotonically decreasing and $V
\rightarrow \infty$ when $\phi \rightarrow \phi_-$, which we for simplicity
from now on assume (apart from the case of a constant potential, which is
treated separately, all our explicit examples obey these conditions).

%-----------------------------------------------------------------
\subsubsection*{The role of the $Z_-$ subset}
%-----------------------------------------------------------------

The four dynamically qualitatively different cases on $Z_-$ when $0\leq
\lambda_- < \infty$ lead to different interior dynamics:
\begin{itemize}
\item[(i)] When $\lambda_-<\sqrt{3\gamma_m}$, a two-parameter
    (one-parameter) set of orbits originates from the source
    $\mathrm{M}_{Z_-}^+$ into ${\bf S}$ (${\bf S}_\phi$), while a single
    orbit originates from $\mathrm{PL}_{Z_-}$ into ${\bf S}_\phi$.
\item[(ii)] When $\sqrt{3\gamma_m} < \lambda_- < \sqrt{6}$, again a
    two-parameter (one-parameter) set of orbits originates from the
    source $\mathrm{M}_{Z_-}^+$ into ${\bf S}$ (${\bf S}_\phi$), but in
    this case there is also a one-parameter set (a single orbit) that
    enters ${\bf S}$ (${\bf S}_\phi$) from $\mathrm{PL}_{Z_-}$ as well as
    a single orbit from $\mathrm{EM}_{Z_-}$.
\item[(iii)] When $\lambda_->\sqrt{6}$, a single orbit enters ${\bf S}$
    from $\mathrm{EM}_{Z_-}$.
\item[(iv)] When $\lambda_-=0$, a two-parameter (one-parameter) set of orbits originates
from the source $\mathrm{M}_{Z_-}^+$ into ${\bf S}$ (${\bf S}_\phi$), while a
single orbit originates from $\mathrm{dS}_{Z_-}$ into ${\bf S}_\phi$.
\end{itemize}

When $\lambda_- \rightarrow \infty$ at $Z_-$, there is just one orbit that
enters ${\bf S}$ from $\mathrm{FL}_{Z_-}$, the so-called tracker or attractor
solution. The exclusion of orbits into ${\bf S}$ and ${\bf S}_\phi$ from the
fixed points $\mathrm{M}^x_{Z_-}$ with $x\neq 0$ can be understood as
follows: There are no solutions coming from $\mathrm{M}^x_{Z_-}$ when $x<0$
since $Z$ is decreasing toward the future when $x<0$. When $x>0$, it follows
that $\Omega_V$ is decreasing in the vicinity of $\mathrm{M}^x_{Z_-}$ and
hence there are no solutions coming from this part of $\mathrm{M}^x_{Z_-}$
into ${\bf S}$ or ${\bf S}_\phi$ either. Hence $\mathrm{M}^x_{Z_-}$ with
$x=0$, i.e., $\mathrm{FL}_{Z_-}$ is the only fixed point on
$\mathrm{M}^x_{Z_-}$ which can give rise to a solution into ${\bf S}$, and it
does give rise to the tracker solution, which can be established by center
manifold theory or by consideration of nearby solutions and continuity, since
this fixed point acts as a kind of saddle in the full state space.

%-----------------------------------------------------------------
\subsubsection*{The role of the $Z_+$ subset}
%-----------------------------------------------------------------

The only solutions that originate from the $Z_+$ subset into ${\bf S}$ (${\bf
S}_\phi$) are those from the source $\mathrm{M}_{Z_+}^-$, from which a
two-parameter (one-parameter) set of orbits originates into ${\bf S}$ (${\bf
S}_\phi$).

For the present monotonically decreasing potentials, all orbits in ${\bf S}$
and ${\bf S}_\phi$ end at $Z_+$ as follows when $0 \leq\lambda_+<\infty$:
\begin{itemize}
\item[(i)] When $\lambda_+<\sqrt{3\gamma_m}$, a two-parameter
    (one-parameter) set of orbits ends at the global sink
    $\mathrm{PL}_{Z_+}$ from ${\bf S}$ (${\bf S}_\phi$).
\item[(ii)] When $\sqrt{3\gamma_m} < \lambda_+ < \sqrt{6}$,  a
    two-parameter set of orbits in ${\bf S}$ ends at the sink
    $\mathrm{EM}_{Z_+}$ while a one-parameter set of orbits in ${\bf
    S}_\phi$ end at $\mathrm{PL}_{Z_+}$.
\item[(iii)] When $\lambda_+>\sqrt{6}$, a two-parameter set of orbits in
    ${\bf S}$ ends at the sink $\mathrm{EM}_{Z_+}$ while a one-parameter
    set of orbits in ${\bf S}_\phi$ end at $\mathrm{M}_{Z_+}^+$.
\item[(iv)] When $\lambda_+=0$, a two-parameter (one-parameter) set of orbits 
    ends at the global sink $\mathrm{dS}_{Z_+}$ from ${\bf S}$ (${\bf S}_\phi$).
\end{itemize}
%

%-----------------------------------------------------------------
\subsubsection*{An additional useful quantity}
%-----------------------------------------------------------------

Finally, consider
\begin{equation}
\xi = \frac{\Omega_m}{\Omega_V},
\end{equation}
which obeys
\begin{equation}
\xi^\prime = -\sqrt{6}\left(\sqrt{\frac32}\,\gamma_m -
\lambda(Z)\,x\right)\xi,
\end{equation}
as follows from~\eqref{Omeq} and~\eqref{OmV}. When $\sqrt{\frac32}\,\gamma_m
- \lambda(Z)\,x>0$, %, which e.g. is the case when $x$ is negative or zero or
%when $\lambda(Z) < \sqrt{\frac32}\gamma_m$,
$\xi$ is monotonically decreasing. For globally bounded $\lambda$ such that
$\lambda(Z) < \sqrt{\frac32}\gamma_m$ it follows that $\xi$ is strictly
monotonically decreasing and hence $\xi \rightarrow 0$ and thereby
$\Omega_m\rightarrow 0$ toward the future; i.e., the future dynamics resides
on the scalar field subset. Toward the past $\xi \rightarrow \infty$ and
hence $\Omega_V\rightarrow 0$ and as a consequence it follows that the past
dynamics asymptotically resides on the $\Omega_V=0$ subset, which of course
is already implied by our previous discussion in a somewhat more complicated
manner, which, however, includes more details and general situations.

We will now turn to explicit examples of potentials and a fluid with a dust
equation of state, $\gamma_m=1$. Throughout we will choose scalar field
variables so that $Z_-=0$ and $Z_+ = 1$, so that $\bar{\bf S}$ is given by
\begin{equation}
\Omega_m\geq 0, \qquad \Omega_V = 1 -x^2 - \Omega_m \geq 0, \qquad 0\leq Z \leq 1.
\end{equation}
It should be pointed out that this type of choice of scalar field variable is
by no means an optimal choice for all scalar field potentials that can be dealt with
by means of a scalar field variable $Z(\phi)$, as illustrated by the treatment of the
generalized Chaplygin gas in~\cite{ugg13}, where $-1\leq Z \leq 1$, but
choices such that $0\leq Z \leq 1$ are ``good enough'' for our present
purposes. Since $Z_-=0$ and $Z_+=1$ we will introduce $0$ and $1$ as
subscripts instead of $Z_\pm$ to distinguish fixed points on the two
boundaries (in addition we have the subscript $Z$ for the line of fixed
points $\mathrm{FL}_Z$ and, for a constant potential, the line of fixed
points $\mathrm{dS}_Z$).

%%%%%%%%%%%%%%%%%%%%%%%%%%%%%%%%%%%%%%%%%%%%%%%%%%%%%%%%%%%%%%%%%%
\section{Frozen $\lambda$CDM dynamics}\label{sec:frozenlambda}
%%%%%%%%%%%%%%%%%%%%%%%%%%%%%%%%%%%%%%%%%%%%%%%%%%%%%%%%%%%%%%%%%%

Before considering the constant and exponential potentials, let us first
recall some facts about $\Lambda$CDM cosmology. In their simplest form, the
$\Lambda$CDM models are solutions of Einstein's equations with (i) a
spatially isotropic and homogeneous flat FLRW geometry and (ii) a perfect fluid
that has negligible pressure, i.e., dust, representing nonrelativistic
matter, and a cosmological constant $\Lambda$. It is useful to describe
solutions by the following quantities:
\begin{subequations}
\begin{align}
H &= \dot{a}/a,\\
q &= -H^{-2}\left(\frac{\ddot{a}}{a}\right) = -\left(1 + \frac{H^\prime}{H}\right),\\
j &= H^{-3}\left(\frac{\dddot{a}}{a}\right) = -q^\prime + q + 2q^2,\\
\Omega_m &= \frac{\rho_m}{3H^2},\\
\Omega_\Lambda &= \frac{\Lambda}{3H^2}.
\end{align}
\end{subequations}
Here the cosmographic (or cosmokinetic) quantities $a(t), H(t), q(t), j(t)$
are the cosmological scale factor and the time-dependent Hubble,
%(i.e., $H$ is the time
%dependent Hubble parameter; also, recall that $H = \theta/3$, where $\theta$
%is the expansion of the symmetry surfaces and thereby also the fluid),
deceleration, and jerk parameters, respectively~\cite{blaetal04,rapetal07},
which take the following form for the $\Lambda$CDM models:
\begin{subequations}
\begin{align}
\frac{H}{H_0} &= \sqrt{\Omega_{m0}\exp(-3\tau) + \Omega_{\Lambda0}},\\
q &= -1 + \frac32\Omega_m = -1 + \frac32\left(\frac{\Omega_{m0}\exp(-3\tau)}{\Omega_{m0}\exp(-3\tau) + \Omega_{\Lambda0}}\right),\label{qLCDM}\\
j &= 1,\\
\Omega_m &= \frac{\Omega_{m0}\exp(-3\tau)}{\Omega_{m0}\exp(-3\tau) + \Omega_{\Lambda0}}, \\
\Omega_\Lambda &= \frac{\Omega_{\Lambda0}}{\Omega_{m0}\exp(-3\tau) + \Omega_{\Lambda0}},
\end{align}
\end{subequations}
where $\Omega_m + \Omega_\Lambda = 1$ and where $H_0$, $\Omega_{m0}$, and
$\Omega_{\Lambda0}$ are the values of $H$, $\Omega_{m}$, and
$\Omega_{\Lambda}$ at the reference time $t_0$ and hence
$\tau=0$.\footnote{Due to noisy data, the current observational feasibility
of the jerk parameter $j$ is questionable. Nevertheless, the fact that $j=1$
for the $\Lambda$CDM models makes it useful as a theoretical indicator of
deviations from $\Lambda$CDM cosmology, and it is for this reason we give
$j$. Moreover, as will be seen, the jerk parameter seems to be the parameter
that is most sensitive when it comes to deviations from $\Lambda$CDM
cosmology; it is unfortunate that it is also the most difficult quantity to
observationally measure of the presently discussed ones.} Throughout we will
use the following present values:
\begin{equation}
\Omega_{m0} = 0.3, \qquad \Omega_{\Lambda0} = 0.7.
\end{equation}
It follows from~\eqref{qLCDM} that $q_0=-0.55$.

\begin{figure}[ht!]
\begin{center}
\subfigure[$H(z)$-diagram for the $\Lambda$CDM models.]{\label{fig:hLCDM}
\includegraphics[width=0.48\textwidth, trim = 0cm 0cm 0cm 0cm]{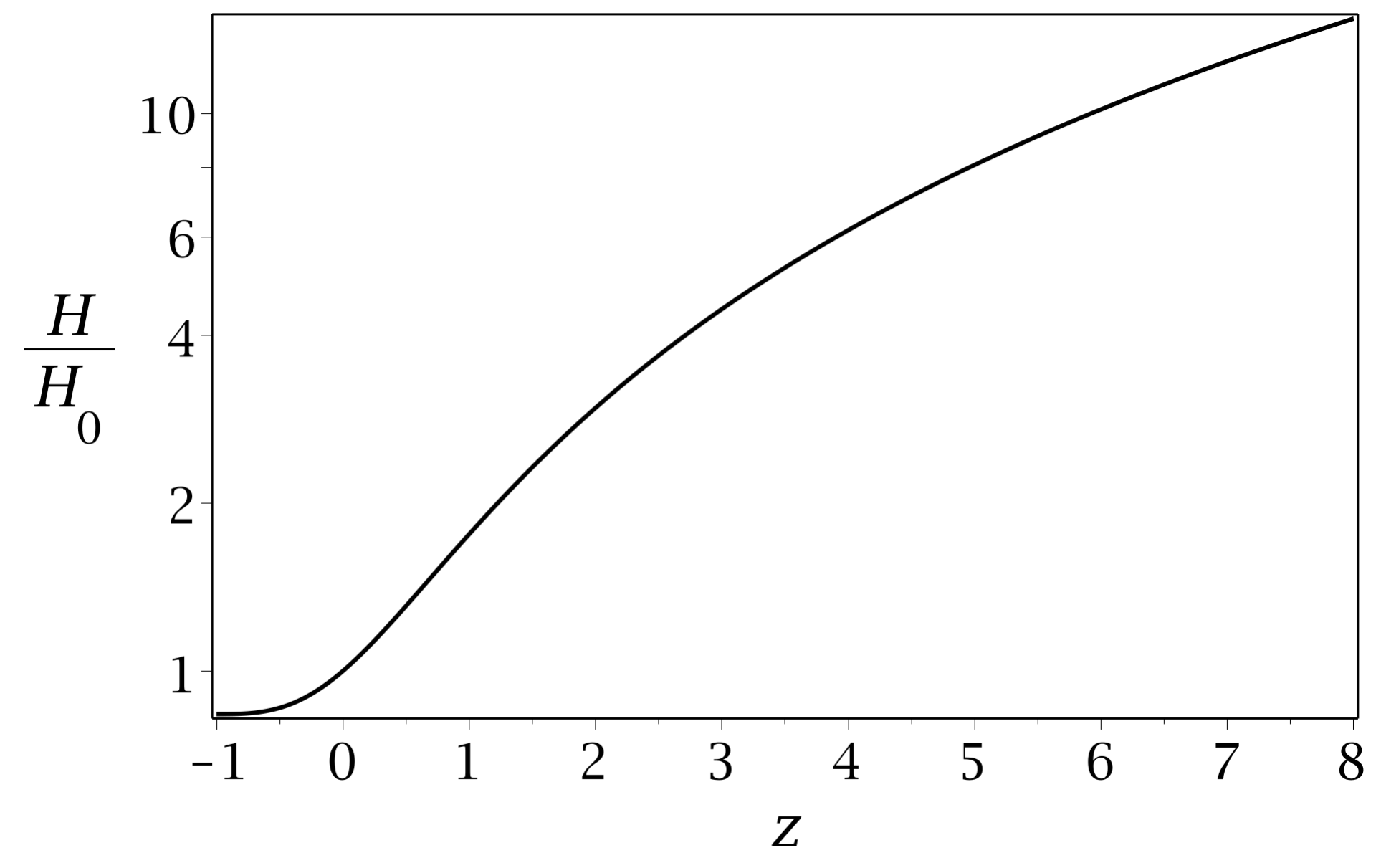}} \,
\subfigure[$q(z)$-diagram for the $\Lambda$CDM models.]{\label{fig:qLCDM}
\includegraphics[width=0.48\textwidth, trim = 0cm 0cm 0cm 0cm]{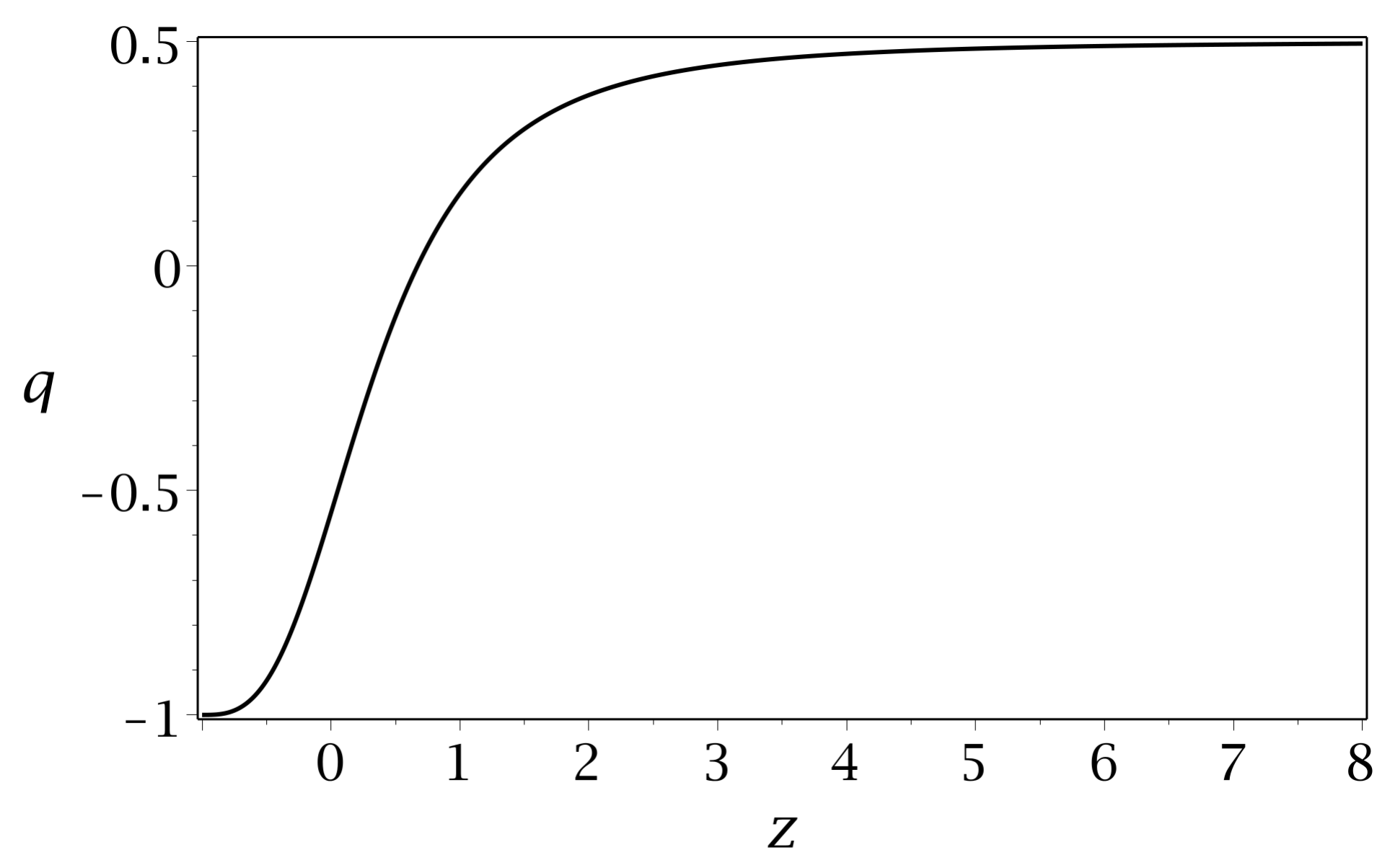}}
\subfigure[$\Omega_m(z)$-diagram for the $\Lambda$CDM models.]{\label{fig:OmegaLCDM}
\includegraphics[width=0.48\textwidth, trim = 0cm 0cm 0cm 0cm]{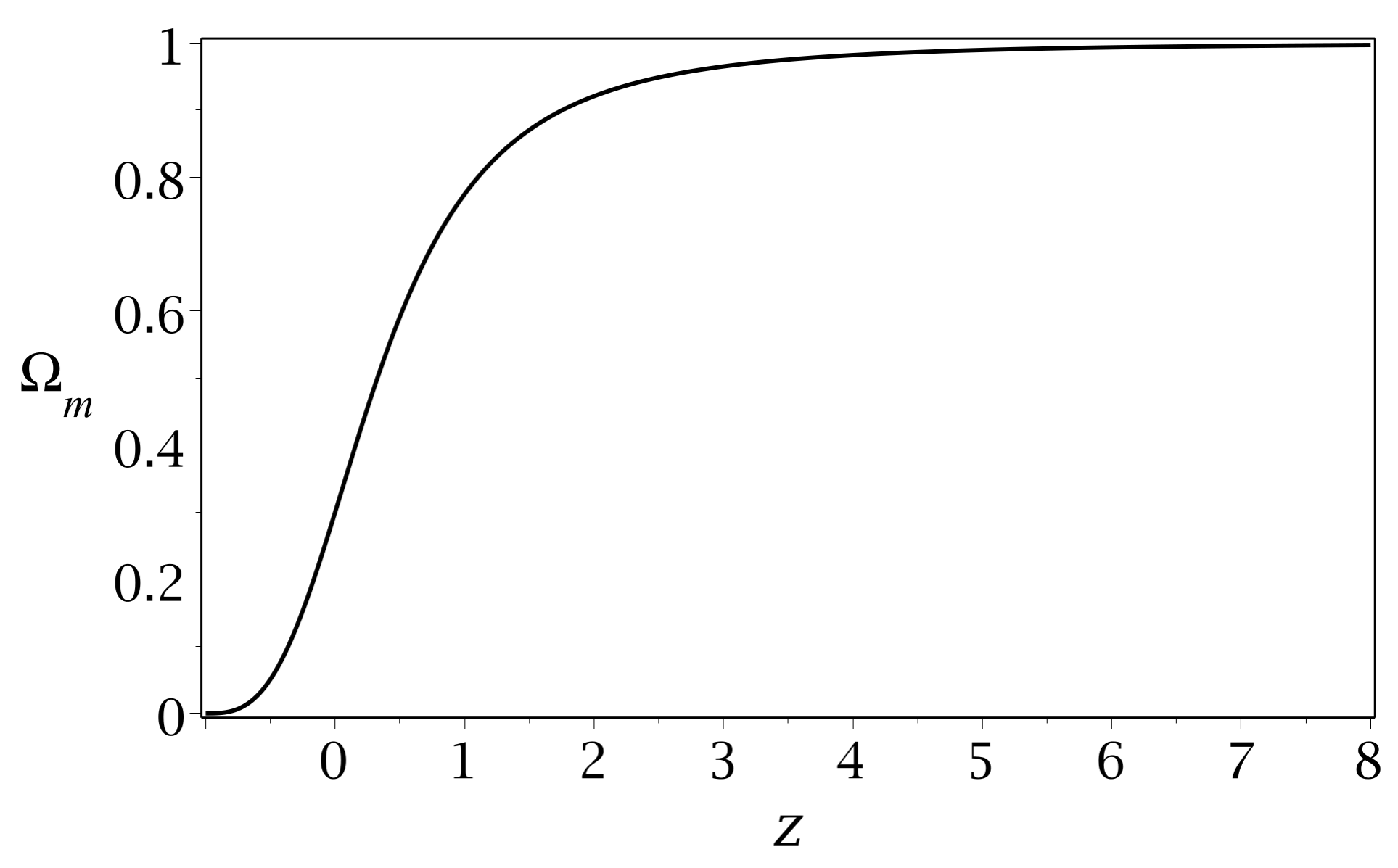}}\vspace{-0.5cm}
\end{center}
\caption{Depiction of the evolutionary history of $\Lambda$CDM cosmology for
$\Omega_{m0} = 0.3$, $\Omega_{\Lambda0} = 0.7$ in terms of the redshift $z$.}
\label{fig:LCDMgraphs}
\end{figure}

Figure~\ref{fig:LCDMgraphs} depicts the evolution of $H/H_0$, $q$, and
$\Omega_m$ in terms of the redshift $z$ for the $\Lambda$CDM models with the
above present values.
As can be seen, $\Omega_m$ monotonically decreases from 1 (corresponding to
the big bang limit $\tau \rightarrow -\infty$) to zero (corresponding to the
infinite future limit  $\tau \rightarrow + \infty$, described by a de Sitter
state with $\Omega_\Lambda=1$). Alternatively this also follows by
considering the differential equation for $\Omega_m$, which is given by (see,
e.g., p. 62 in~\cite{waiell97})
\begin{equation}\label{OmprimeLCDM}
\Omega_m^\prime = - 3(1-\Omega_m)\Omega_m.
\end{equation}
%

%-----------------------------------------------------------------
\subsection{$\Lambda$CDM dynamics as $\lambda=0$-CDM dynamics}
%-----------------------------------------------------------------

Setting $p_m=0$ and $V=\Lambda$ yields a model that can be interpreted as that
of dust, a massless scalar field, and a cosmological constant $\Lambda$. The
associated model does not solve any of the issues one might find unattractive
with the $\Lambda$CDM model, but it does show some aspects that are important
in a broader context. Since $\lambda=0$, one obtains a reduced coupled
two-dimensional system for $x$ and $\Omega_m$, while the decoupled equation
for $Z$ can be integrated once the evolution for $x$ and $\Omega_m$ is found.
However, instead of just considering the essential reduced extended
two-dimensional state space $\bar{\bf S}_\mathrm{red}$, we need to consider
the full three-dimensional extended state space $\bar{\bf S}$ to illustrate
how this model is connected with more general models for which $\lambda =
\lambda(Z)$. Note, however, that the decoupling of the equation for $Z$ from
the coupled system of $x$ and $\Omega_m$ results in that all solutions with the
same initial data of $x$ and $\Omega_m$ yield the same curves when projected
onto the $x-\Omega_m$-plane, irrespective of the initial value of $Z$. To
proceed requires that $Z$ be specified. There are many possible choices that
lead to an analytic dynamical system on a compact state space that cover the
domain $\phi_\pm = \pm \infty$, but here, due to future purposes, we choose
\begin{equation}\label{Zdef}
Z=\frac{1}{1+\exp(-{\lambda}\phi)},
\end{equation}
which is monotonically increasing in $\phi$ and where ${\lambda}$ is an
arbitrary positive constant; the scalar field variable $Z$ thereby has an
extended range $Z\in [0,1]$. This leads to the following dynamical system for
the present models:
\begin{subequations}\label{dynsys}
\begin{align}
x^\prime &= -(2-q)x,\label{xeq1}\\
\Omega_m^\prime &= 3\left[2x^2 - (1 - \Omega_m)\right]\Omega_m,\\
Z^\prime &= \sqrt{6}{\lambda} Z(1-Z)x,\label{Zeq1}
\end{align}
\end{subequations}
where $2-q=3(1-x^2) - \frac32\Omega_m = 3(1-x^2-\Omega_m) + \frac32\Omega_m$.
The reduced coupled system for $x$ and $\Omega_m$ is invariant under the
transformation $x \rightarrow -x$, which makes it possible to identify this
problem with that of a stiff fluid, characterized by $\Omega_\mathrm{stiff} =
x^2$, a cosmological constant, and dust. In a similar manner as for the
$\Omega_V=0$ subset, this leads to
\begin{subequations}
\begin{align}
\frac{H}{H_0} &= \left(\Omega_{\mathrm{stiff}0}\exp(-6\tau) + \Omega_{\Lambda 0} + \Omega_{m0}\exp(-3\tau)\right)^{1/2},\\
x &= \pm \sqrt{\Omega_{\mathrm{stiff}0}}\exp(-3\tau)/(H/H_0), \\
\Omega_m &= \Omega_{m0}\exp(-3\tau)/(H/H_0)^2.
%, \\
%\frac{x^2}{\Omega_m} &\propto \exp(-3\tau).
\end{align}
\end{subequations}
As for the $\Omega_V=0$ subset it is easiest to obtain $Z$ by integrating the
equation for $\phi$, given by
\begin{equation}
\phi^\prime = \sqrt{6}x = \pm(6\Omega_{\mathrm{stiff}0})^{1/2}\,\exp(-3\tau)/(H/H_0),
\end{equation}
and then inserting the solution into~\eqref{Zdef}.

We have previously dealt with the global dynamics of monotonically decreasing
potentials. We here give a complete description for the constant potential. 
The past and future states of all orbits follow from the exact solution, or from 
the monotone function $H^2$ in combination with the dynamical structure at the
boundaries and the local properties of the fixed points. The result is that a
two-parameter set of orbits in ${\bf S}$ and a one-parameter set in ${\bf
S}_\phi$ originate from each of the hyperbolic sources $\mathrm{M}_0^+$ and
$\mathrm{M}_1^-$ while a one-parameter set of solutions originates from the
transversally hyperbolic line of fixed points $\mathrm{FL}_{Z}$ into ${\bf
S}$; orbits in ${\bf S}$ and ${\bf S}_\phi$ end at the transversally
hyperbolic line of fixed points $\mathrm{dS}_{Z}$, which thereby constitute
the future attractor. The orbits that come from $\mathrm{M}_0^+$
($\mathrm{M}_1^-$) correspond to solutions with increasing (decreasing)
$\phi$ from $\phi \rightarrow - \infty$ ($\phi \rightarrow +\infty$),
starting from a massless state [heuristically this can be understood from
Eq.~\eqref{KG} where the friction force $-3H\dot{\phi}$ generates energy
toward the past, making the energy content of $\Lambda$ (and $\rho$) 
negligible compared to the kinetic energy of the scalar field], while
solutions that originate from $\mathrm{FL}_{Z}$ reside on the invariant
subset $x=0$.

Since the $x=0$ subset corresponds to solutions with different constant
$\phi$, and thereby also a constant $\rho_\phi = \Lambda$, these models are
identical to the $\Lambda$CDM models. We therefore denote the $x=0$ invariant
subset surface as the $\Lambda$CDM surface or subset. Furthermore, this
surface divides the state space into two disconnected domains with $x>0$ and
$x<0$, respectively, and the $\Lambda$CDM subset thereby acts as a separatrix
surface. Note that it is a necessary condition that the potential is a
constant for $x=0$ to be an invariant subset in ${\bf S}$ and thus no other
potentials yield a $\Lambda$CDM model.

\begin{figure}[ht!]
\begin{center}
\subfigure[The massless scalar field boundary $\Omega_V=0$.]{\label{fig:V0Dust}
\includegraphics[width=0.44 \textwidth, trim = 0cm 2.5cm 0cm 0cm]{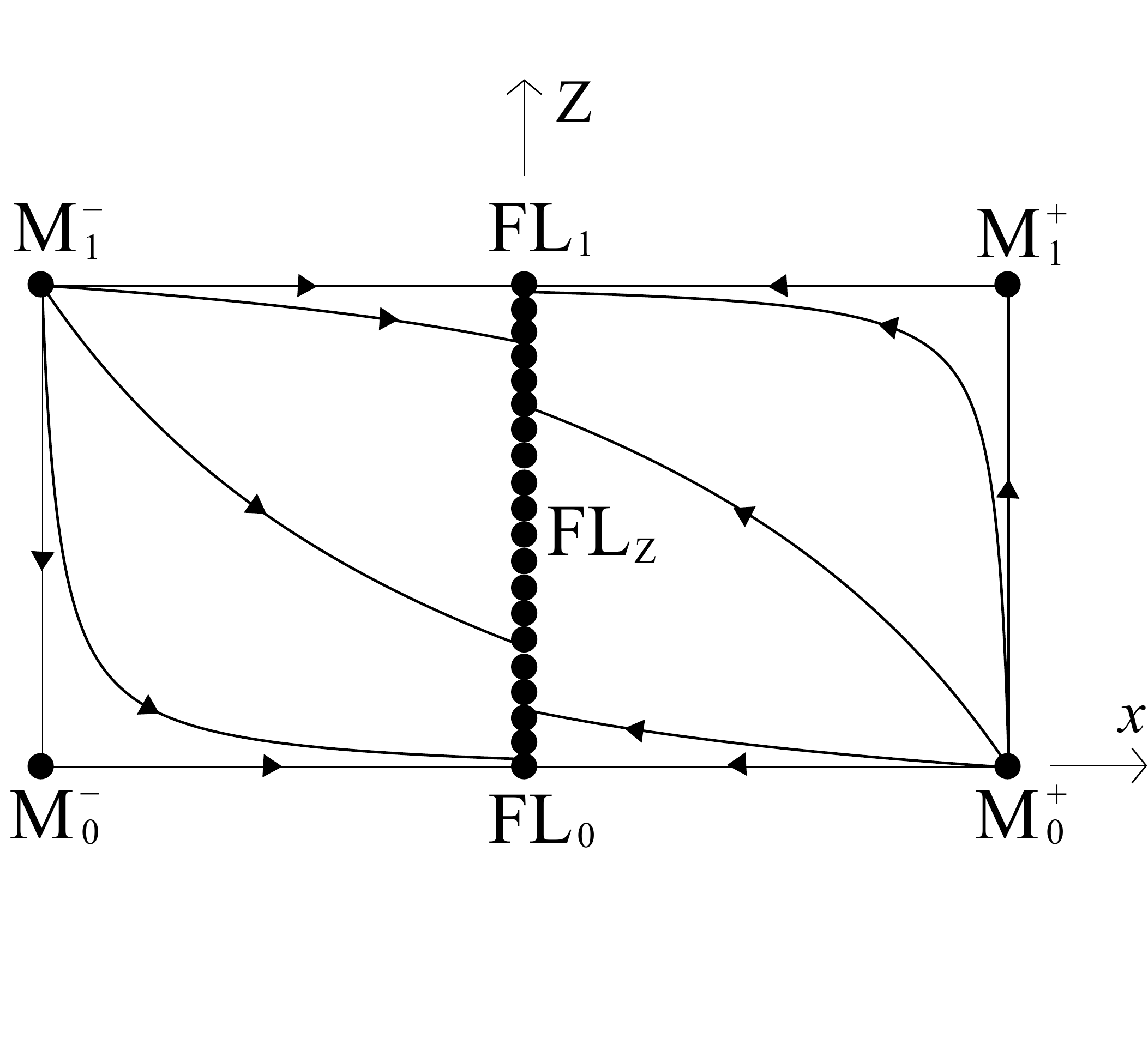}}\qquad
\subfigure[The scalar field boundary $\Omega_m=0$.]{\label{fig:M0CC}
\includegraphics[width=0.44\textwidth, trim = 0cm 2.5cm 0cm 0cm]{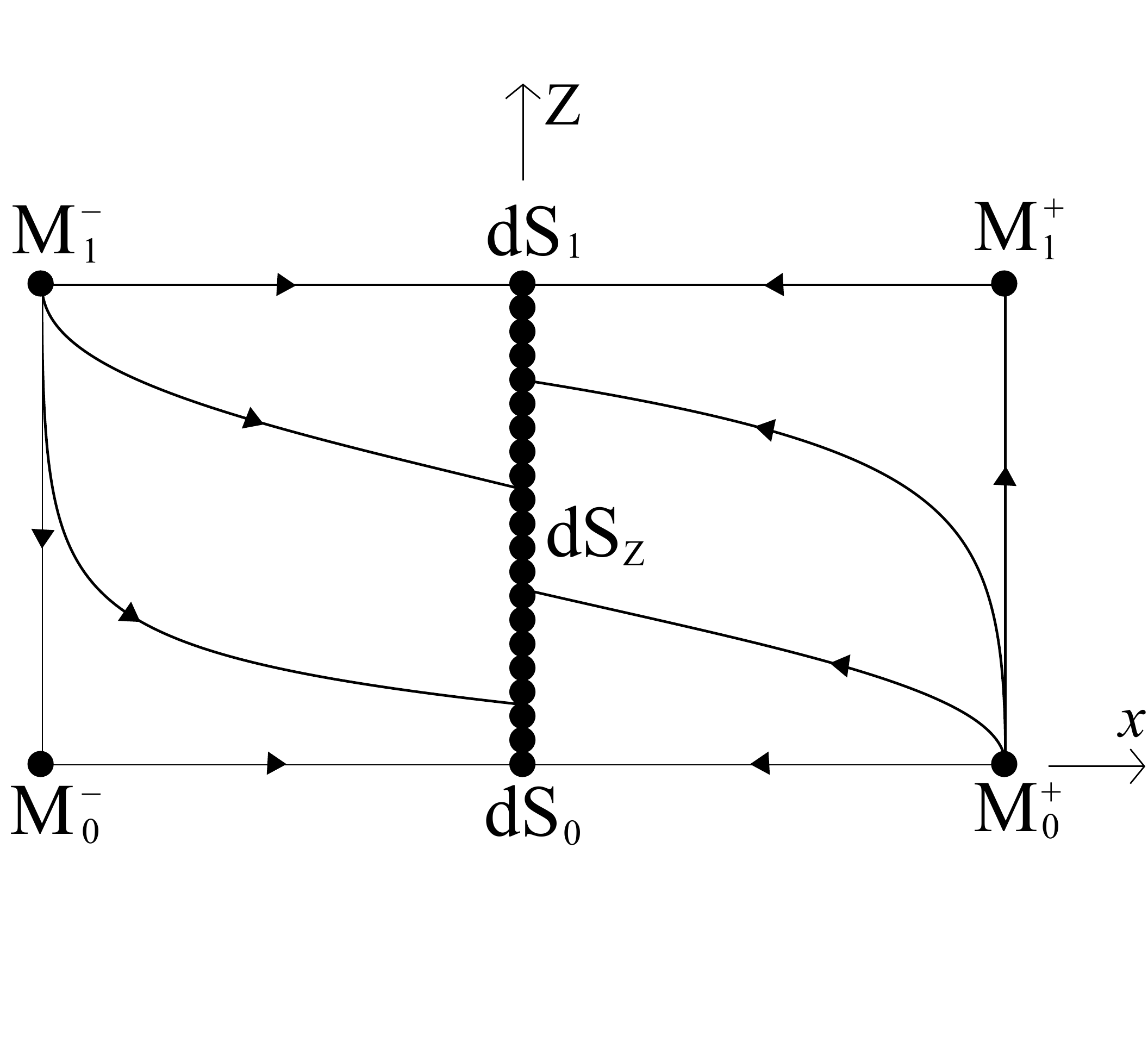}}
\subfigure[The $(x,\Omega_{m},Z)$ state space for a constant potential and dust.]{\label{fig:SSCC}
\includegraphics[width=0.5\textwidth, trim = 0cm 2.5cm 2cm 0cm]{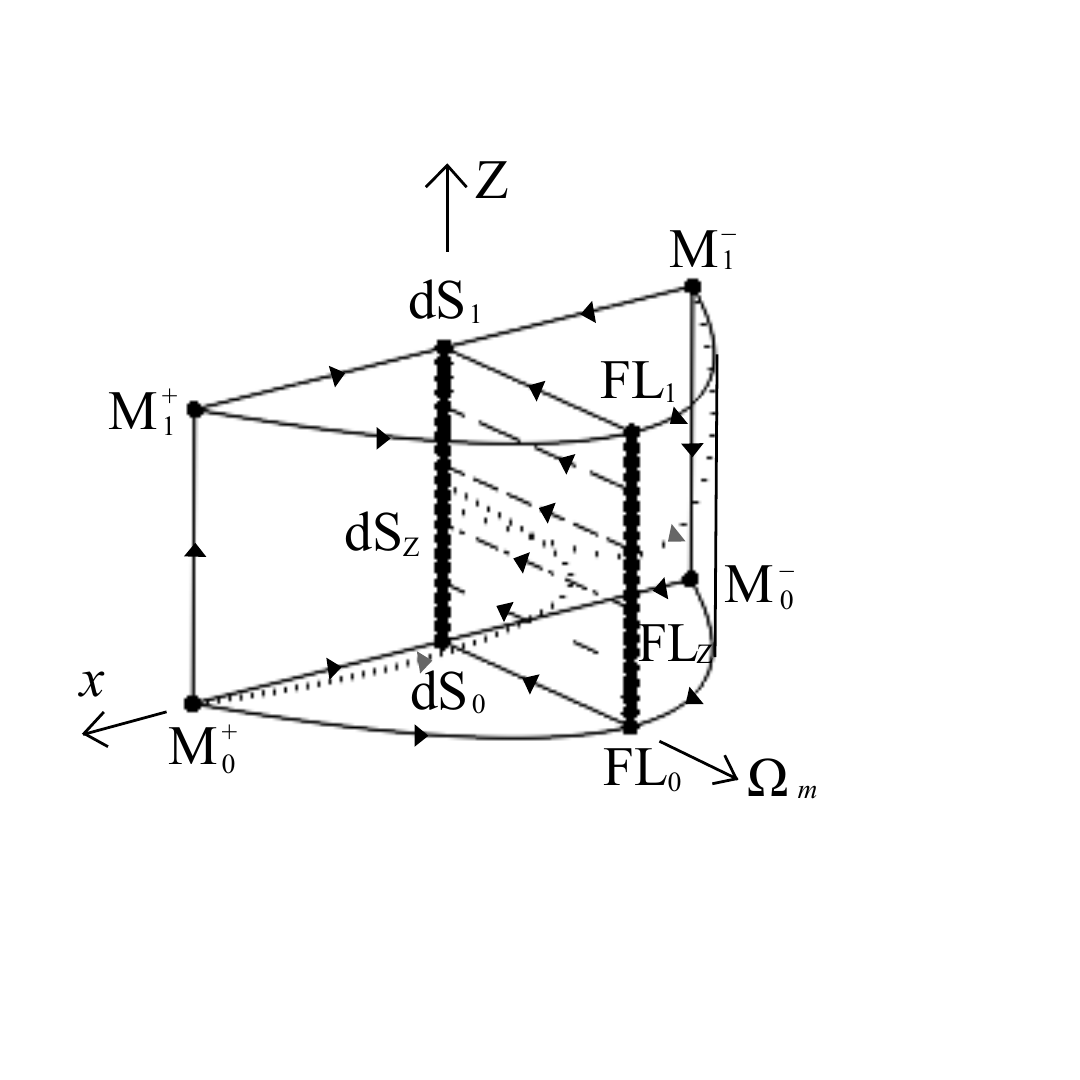}} \vspace{0.0cm}
%\subfigure[Projected state space picture onto the $(x,\Omega_{m})$-plane for a constant potential and dust.]{\label{fig:SSCCTop}
%\includegraphics[width=0.45\textwidth, trim = 0cm 3cm 0cm 0cm]{SS_CC_TopView.pdf}} \vspace{-0.5cm}
\end{center}
\caption{Depiction of the solution space of a fluid without pressure and a scalar field
with a constant potential $V(\phi)=\Lambda>0$, where 
$Z=(1+\exp(-\sqrt{3/2}\phi))^{-1}$, illustrating the $x=0$ $\Lambda$CDM
separatrix surface subset and its attracting nature.}
\label{fig:CC}
\end{figure}

Due to the regularity of the dynamical system and that each fixed point on
$\mathrm{FL}_{Z}$ acts as a transversal saddle point (exemplifying the
previous general discussion concerning the $\Omega_V=0$ subset and the
vicinity of $\Omega_V=0$ and $\mathrm{FL}_Z$), and since $\mathrm{dS}_{Z}$ is
the future attractor, it follows that the $\Lambda$CDM separatrix surface
$x=0$ is
%attracting nearby solutions,
%indeed, it follows from the dynamics that it is a temporally globally
%attracting surface, i.e., all solutions that are near the surface
%\emph{anywhere} are decreasing their distance to the surface; we therefore
%refer to the $\Lambda$CDM separatrix surface $x=0$ as
an ``attracting'' surface or invariant subset (indeed, $x^2$ is monotonically
decreasing). As a consequence, demanding that $\Omega_m$ is close to 1
initially leads to solutions with almost $\Lambda$CDM behavior. The solution
space for this set of models is depicted in Fig.~\ref{fig:CC} where we have
chosen ${\lambda} = \sqrt{3/2}$.
Note that the structure on the $\Omega_V=0$ subset is qualitatively the same
for all the potentials we consider, as discussed in the previous section.

The present models in effect do not really have a \emph{dynamical} dark
energy since the dark energy content ``freezes'' toward an ``unnatural'' value
$\Lambda>0$ toward the future for models with $x\neq 0$, and they therefore
do not solve any coincidence or energy scale problems one might have with the
$\Lambda$CDM models. Nevertheless, since models with $\Omega_m$ close to 1
as an initial condition basically have the same evolution as the $\Lambda$CDM
models, these models with their \emph{attracting} $\Lambda$CDM
\emph{separatrix surface subset}, in an evolutionary history sense, are the
best a scalar field and dust model can accomplish when it comes to mimicking
$\Lambda$CDM cosmology, and they therefore, from a purely time development
perspective, set the standards for any truly dynamical dark energy model.
Next we turn to the well-known case of an exponential potential (see,
e.g.,~\cite{copetal98,Nunes2000}), but in the present three-dimensional context.

%-----------------------------------------------------------------
\subsection{Constant $\lambda$ $\lambda$CDM dynamics}
%-----------------------------------------------------------------

An exponential potential
\begin{equation}
V=V_0\exp(-{\lambda}\phi), \qquad {\lambda}>0,
\end{equation}
and a dust matter equation of state results in the dynamical system
\begin{subequations}\label{dynsysexp}
\begin{align}
x^\prime &= -(2-q)x + \sqrt{\frac{3}{2}}{\lambda}(1 - x^2 - \Omega_m),\label{Sigexp}\\
\Omega_m^\prime &= 3\left[2x^2 - (1 - \Omega_m)\right]\Omega_m,\\
Z^\prime &= \sqrt{6}{\lambda} Z(1-Z)x,
\end{align}
\end{subequations}
where we have used the definition~\eqref{Zdef} for $Z$ (also, recall that
$2-q=3(1-x^2) - \frac32\Omega_m$). Once again there is a decoupling between
the scalar field variable $Z$ and a coupled system for $x$ and $\Omega_m$ on
a reduced state space ${\bf S}_\mathrm{red}$, parametrized by ${\lambda}$.
As for a constant potential, it therefore follows that solutions with the
same initial $x$ and $\Omega_m$, but with different initial $Z$, have the
same trajectories when projected onto the $x-\Omega_m$-plane.

The qualitative structure of the solution space of $\bar{\bf S}$ is entirely
determined by the general monotonic features and the structures on the
boundary subsets of ${\bf S}$ and their neighborhoods described earlier. Thus
there are two-parameter sets of orbits in ${\bf S}$ that originate from the
sources $\mathrm{M}_0^+$ and $\mathrm{M}_1^-$ when ${\lambda}<\sqrt{6}$, but
only from $\mathrm{M}_1^-$ when ${\lambda}>\sqrt{6}$, while a one-parameter
set of solutions originates from the transversally hyperbolic line of fixed
points $\mathrm{FL}_{Z}$. Toward the future $\mathrm{PL}_1$ is a sink when
${\lambda} < \sqrt{3}$ (and the future attractor on ${\bf S}$ and ${\bf
S}_\phi$) while $\mathrm{EM}_1$ is future stable when ${\lambda} > \sqrt{3}$
(and the future attractor on ${\bf S}$).
%The present case
%can therefore only mimic $\Lambda$CDM dynamics for quite small
%$\bar{\lambda}$.
As previously shown, it is a general feature that there is a one-parameter
set of solutions that originate from the line $\mathrm{FL}_Z$, forming an
attracting invariant subset, but it is only when ${\lambda}<\sqrt{3}$ that this
invariant subset forms an ``attracting \emph{separatrix} surface,'' which in
the limit ${\lambda}\rightarrow 0$ leads to the previous $\Lambda$CDM
surface.

%Since $x$ is monotonically increasing when $-1<x<0$, it follows that the
%separatrix surface is located at $x\geq 0$; note that the same holds true for
%the attracting separatrix surfaces discussed in the future examples.

It is of historical interest to point out that it was the case with
${\lambda} > \sqrt{6}$ with $\mathrm{EM}_1$ as the future attractor (although
in the two-dimensional projected context) that was the original role model
for concepts such as attractor (the fixed point $\mathrm{EM}_1$ \emph{is} the
formal future attractor) and tracker solutions in scalar field cosmology
[recall that $\gamma_\phi = \gamma_m$ for general $\gamma_m$ (see,
Table~\ref{tab:fixedpoint}); i.e., the effective scalar field equation of
state ``tracks'' the matter equation of state]. As we will see, this is rather
misleading. Moreover, from current observational constraints these models are
no longer viable cosmological models since $q=1/2$ for the future attractor
$\mathrm{EM}_1$. Since models with small ${\lambda}$ are arguably more
relevant for understanding more general and observationally competitive
scalar field models, we focus on this class of models next.
%, while models with sufficiently small $\lambda$ are,
%although, as will be argued below, they are unlikely candidates for solving
%any problems one might have with $\Lambda$CDM cosmology. Nevertheless, from a
%general dynamical systems perspective they are of interest, and we therefore
%focus on this class next.

\begin{figure}[ht!]
\begin{center}
\subfigure[The scalar field boundary $\Omega_m=0$; ${\lambda}=\sqrt{3/2}$]{\label{fig:M0EXP}
\includegraphics[width=0.44 \textwidth, trim = 0cm 2.5cm 0cm 0cm]{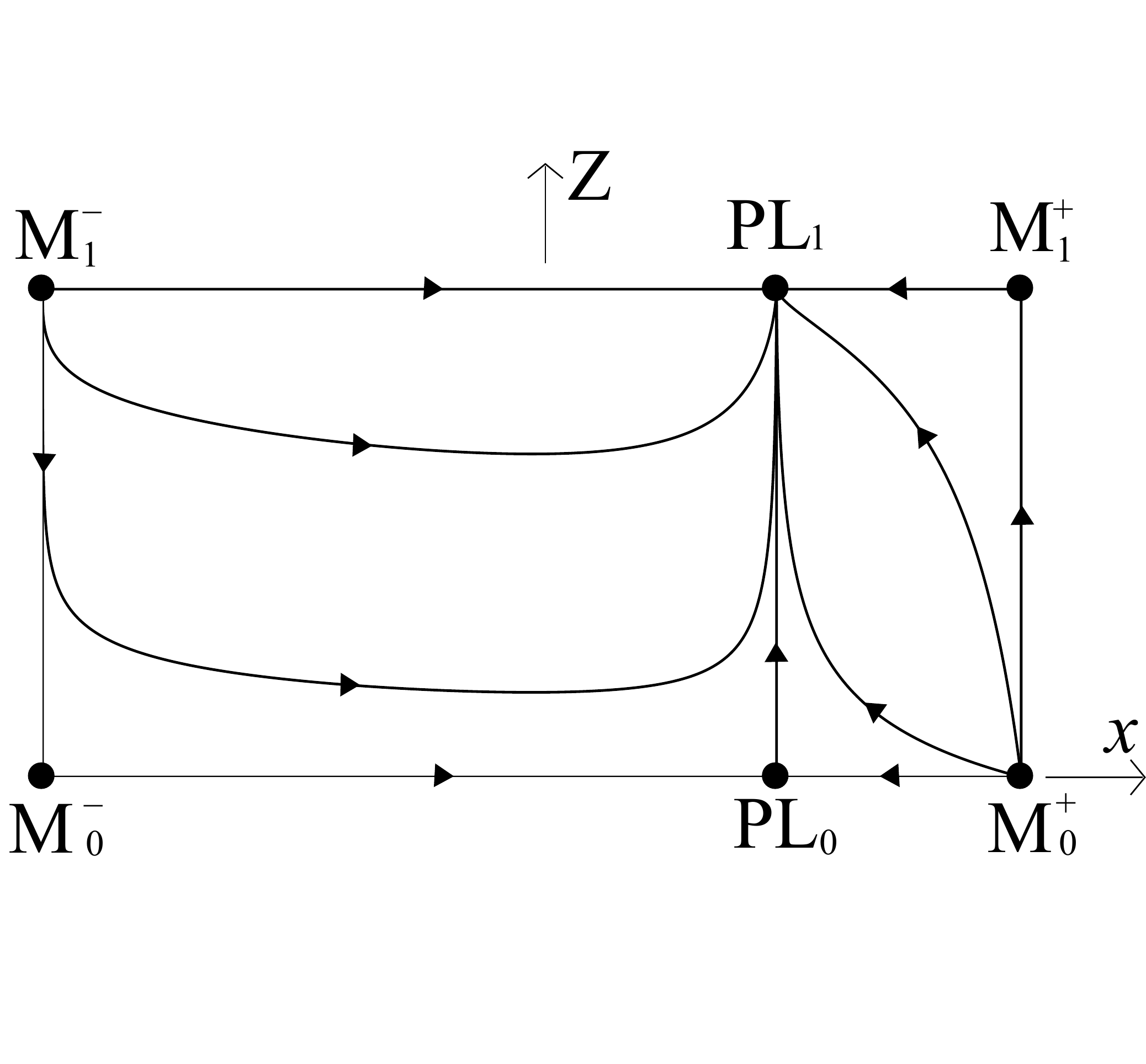}} \,
\subfigure[The $(x,\Omega_{m},Z)$ state space for dust and ${\lambda}=\sqrt{3/2}$.]{\label{fig:SSEXP}
\includegraphics[width=0.5\textwidth, trim = 0cm 2.5cm 2cm 0cm]{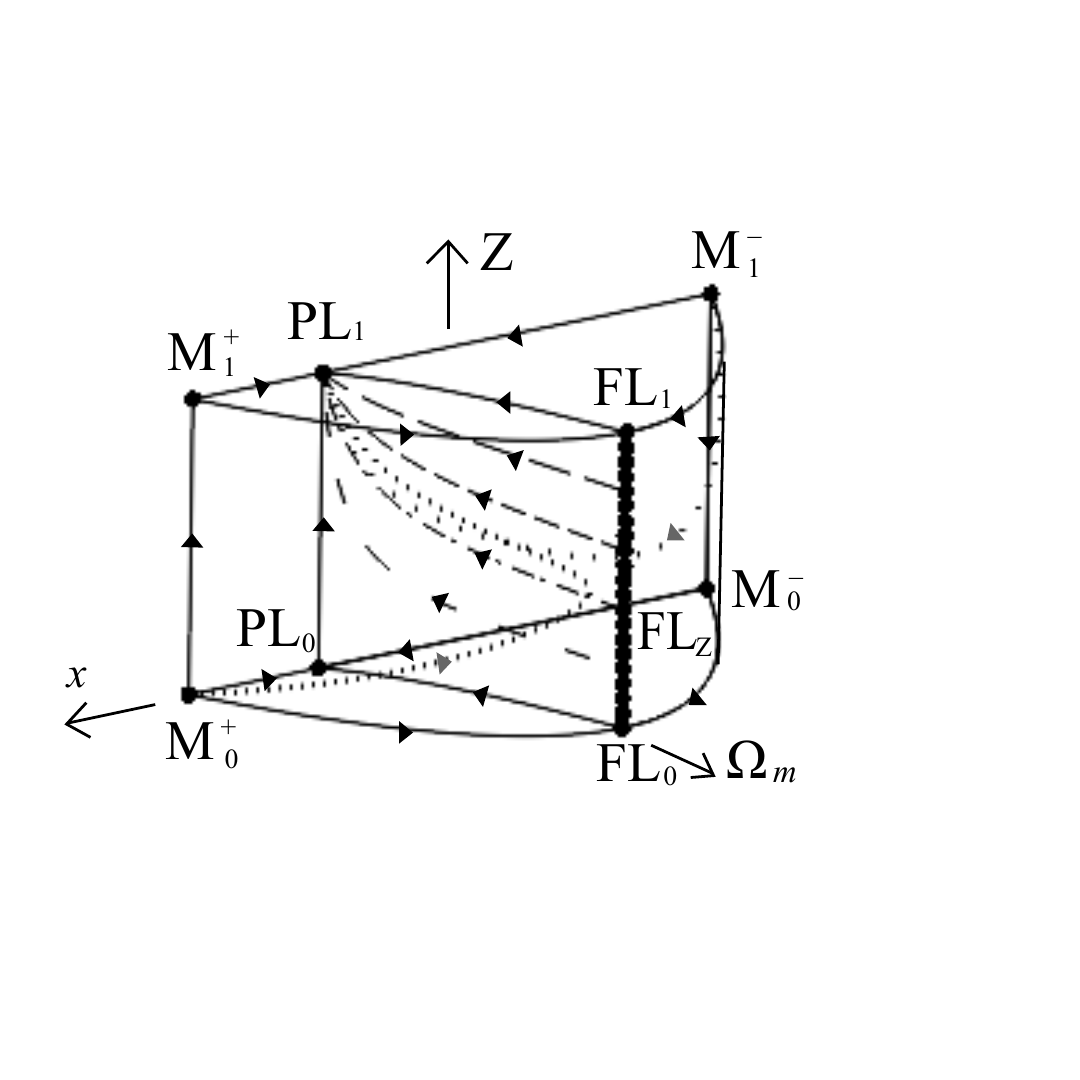}} \vspace{-0.5cm}
\end{center}
\caption{State space structures for dust and $V=V_0\exp(-{\lambda}\phi)$ with
${\lambda}=\sqrt{3/2}$.
}
\label{fig:EXP1}
\end{figure}

It is worthwhile to point out that since this does not seem to be generally
known, there are models with a perfect fluid and a scalar field with an
exponential potential that admits simple explicit solutions, as shown by
Uggla \emph{et al.} in~\cite{uggetal95} (where many other solvable scalar
field and modified gravity models can be found as well). One such example is
for dust when ${\lambda}=\sqrt{3/2}$. Using the methods
in~\cite{uggetal95}, we see that $x$ and $\Omega_m$ can be found in terms of
cosmological time $t$:
\begin{equation}\label{expsol}
x = \frac{2t^3-c}{4t^3+6t+c}, \qquad
\Omega_{m}=\frac{1-x^2}{1+t^2}=\frac{12t\left(t^3+3t+c\right)}{\left(4t^3+6t+c\right)^2},
\end{equation}
which thus explicitly gives the trajectories in the $x-\Omega_m$-plane in
parametrized form, where the constant $c$ characterizes the different
solutions, with $c=0$ yielding the separatrix subset, i.e., the heteroclinic
orbits from $\mathrm{FL}_Z$ to $\mathrm{PL}_1$ in ${\bf S}$ (a heteroclinic
orbit is a solution trajectory that connects two distinct fixed points).
%Although of course not necessary, we will specialize the potential several
%times so that this particular example occurs as a boundary subset when
%illustrating various properties of the solution space of different
%potentials.
The solution space structure is depicted in Fig.~\ref{fig:EXP1} [see also the
reduced state space in Fig.~\ref{fig:SSCC2D}].
%The
%interior solution space for models with ${\lambda}<\sqrt{3}$ for the
%representative value ${\lambda}=\sqrt{3/2}$ is depicted in
%Fig.~\ref{fig:SSEXP}, with focus on the `attractive separatrix surface.'
%In addition
%figures~\ref{fig:SSEXPTop1}, \ref{fig:SSEXPTop2}, \ref{fig:SSEXPTop3} show
%the projected dynamics onto the reduced $x-\Omega_m$ state space for
%${\lambda}=\sqrt{3/2}$, ${\lambda}=2$, and ${\lambda}=3$, respectively,
%representing the three classes of exponential potential models.
%
%\begin{figure}[ht!]
%\begin{center}
%\subfigure[$\gamma_{m}=1$ and $\lambda_{0}=\sqrt{\frac{3}{2}}$]{\label{fig:Z0Dust_1}
%\includegraphics[width=0.45\textwidth, trim = 0cm 3cm 0cm 0cm]{Z0_Boundary_Final.pdf}} \hspace{0.2cm}
%\subfigure[$\gamma_{m}=1$ and $\lambda_{0}=0.2$]{\label{fig:Z0Dust_2}
%\includegraphics[width=0.45\textwidth, trim = 0cm 3cm 0cm 0cm]{Z0_Boundary_v2_Final.pdf}}
%\end{center}
%\caption{Solutions on the invariant $Z=0$ boundary subset of $\bar{\bf S}$.
%In Fig. (a) the space-dot line corresponds to $a=10$, the dotted line to $a=1$, the dash-dotted line to $a=0$,
%the dashed line to $a=-1/2$ and the space-dotted line to $a=-2$. Fig. (b) represent numerical solutions.}
%\label{fig:exp}
%\end{figure}
%
%\begin{figure}[ht!]
%\begin{center}
%\subfigure[$\gamma_{m}=1$ and $\lambda_{0}=\sqrt{\frac{3}{2}}$]{\label{fig:V0Dust}
%\includegraphics[width=0.45\textwidth, trim = 0cm 3cm 0cm 0cm]{V0_Boundary_Final.pdf}} \qquad
%\end{center}
%\caption{Solutions on the invariant $\Omega_V=0$ boundary subset of $\bar{\bf S}$.
%}
%\label{fig:V0}
%\end{figure}
%
%
%
%\begin{figure}[ht!]
%\begin{center}
%\subfigure[The $(x,\Omega_{m},Z)$ state space for dust and ${\lambda}=\sqrt{\frac{3}{2}}$.]{\label{fig:SSEXP}
%\includegraphics[width=0.5\textwidth, trim = 0cm 2cm 0cm 0cm]{SS_EXP.pdf}} \quad
%\subfigure[Projection of the $(x,\Omega_{m},Z)$ state space onto $(x,\Omega_{m})$ for dust and ${\lambda}=\sqrt{\frac{3}{2}}$.]{\label{fig:SSEXPTop1}
%\includegraphics[width=0.4\textwidth, trim = 0cm 2cm 0cm 0cm]{SS_EXP_TopView.pdf}}\\
%\subfigure[Projection of the $(x,\Omega_{m},Z)$ state space onto $(x,\Omega_{m})$ for dust and ${\lambda}=2$.]{\label{fig:SSEXPTop2}
%\includegraphics[width=0.44\textwidth, trim = 0cm 2cm 0cm 0cm]{SS_EXP_TopView_Lambda2.pdf}} \qquad
%\subfigure[Projection of the $(x,\Omega_{m},Z)$ state space onto $(x,\Omega_{m})$ for dust and ${\lambda}=3$.]{\label{fig:SSEXPTop3}
%\includegraphics[width=0.4\textwidth, trim = 0cm 2cm 0cm 0cm]{SS_EXP_TopView_Lambda3.pdf}}
%\vspace{-0.5cm}
%\end{center}
%\caption{State space structure for $\bar{\bf S}$ for dust and
%$V=V_0\exp(-{\lambda}\phi)$.}
%\label{fig:EXP2}
%\end{figure}
%

\begin{figure}[ht!]
\begin{center}
\subfigure[$\Omega_{m}-q$ diagram]{\label{fig:Oqec}
\includegraphics[width=0.48\textwidth, trim = 0cm 0cm 0cm 0cm]{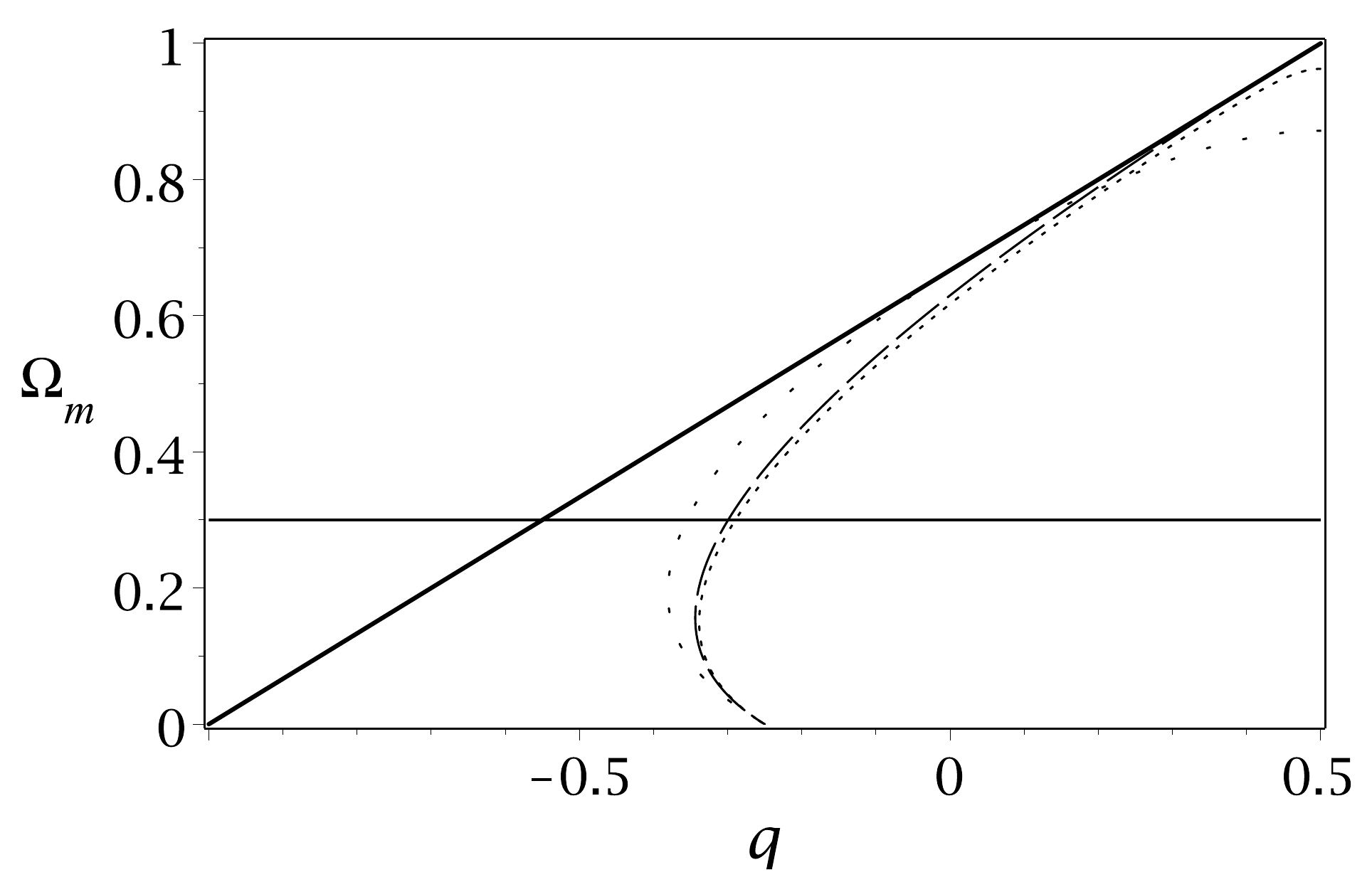}} 
\subfigure[$H-q$ diagram]{\label{fig:Hqec}
\includegraphics[width=0.48\textwidth, trim = 0cm 0cm 0cm 0cm]{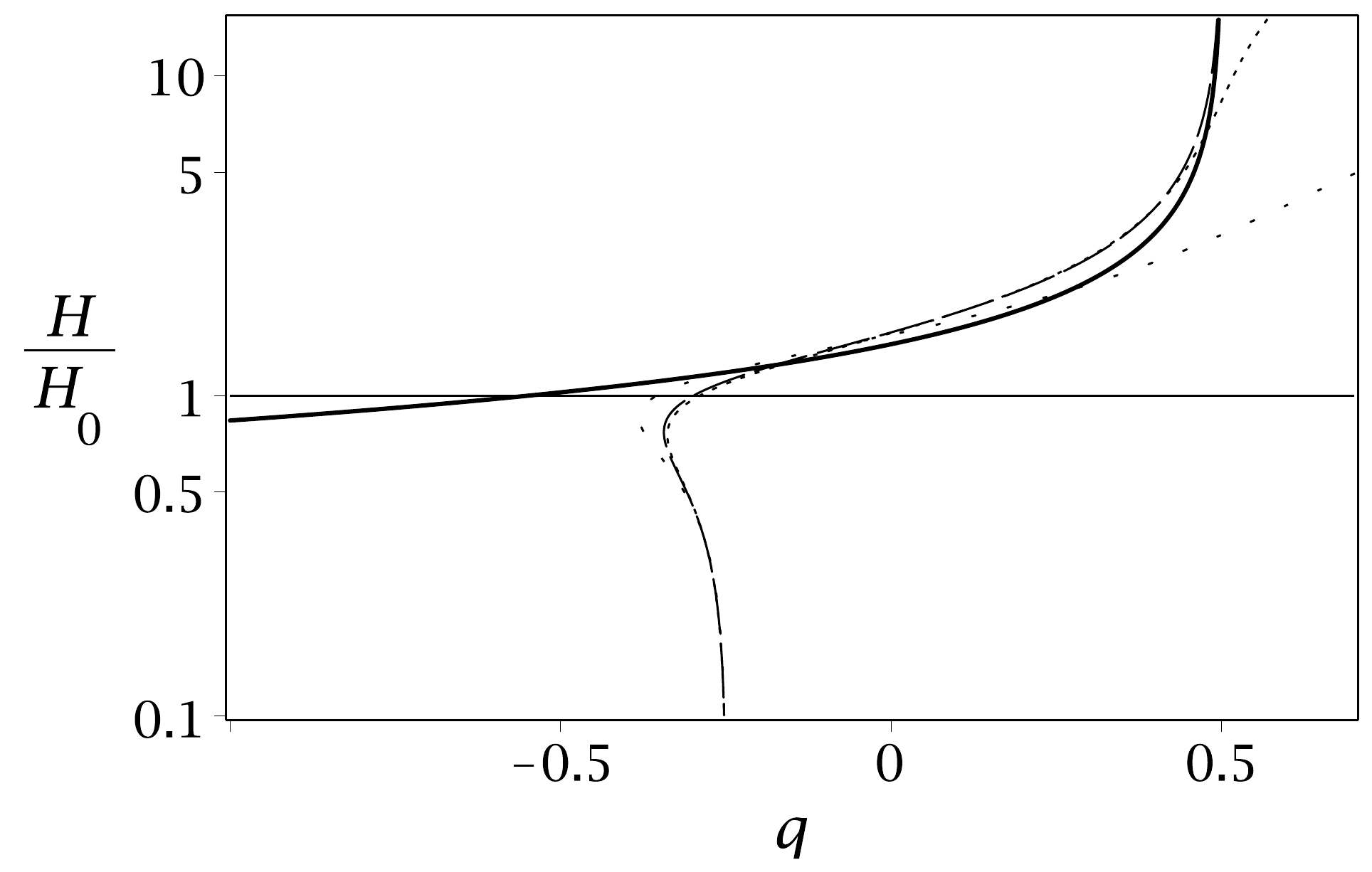}} \\
\subfigure[$H-j$ diagram]{\label{fig:Hjec}
\includegraphics[width=0.48\textwidth, trim = 0cm 0cm 0cm 0cm]{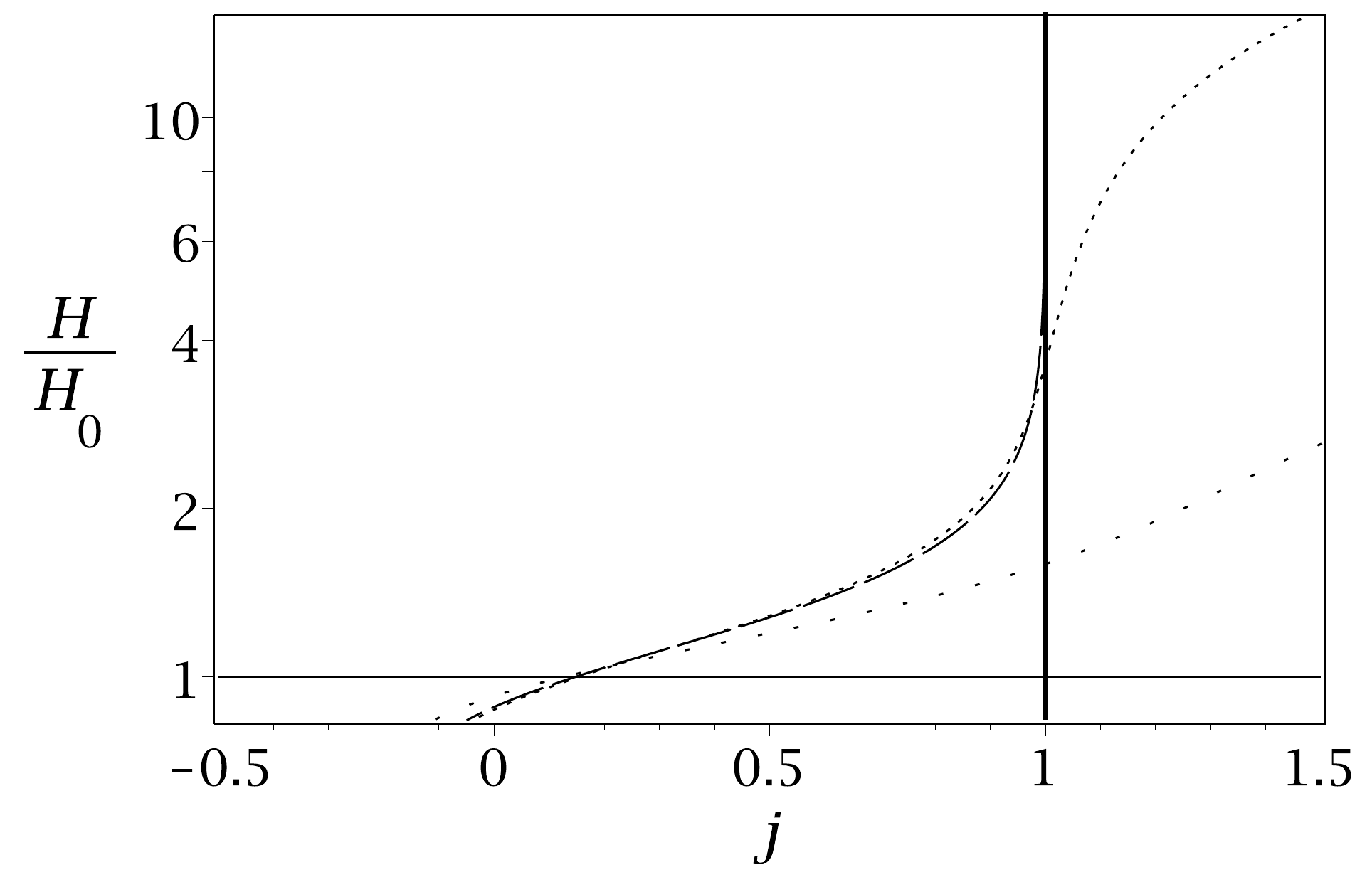}}\,\,
\subfigure[$q-j$ diagram]{\label{fig:qjec}
\includegraphics[width=0.48\textwidth, trim = 0cm 0cm 0cm 0cm]{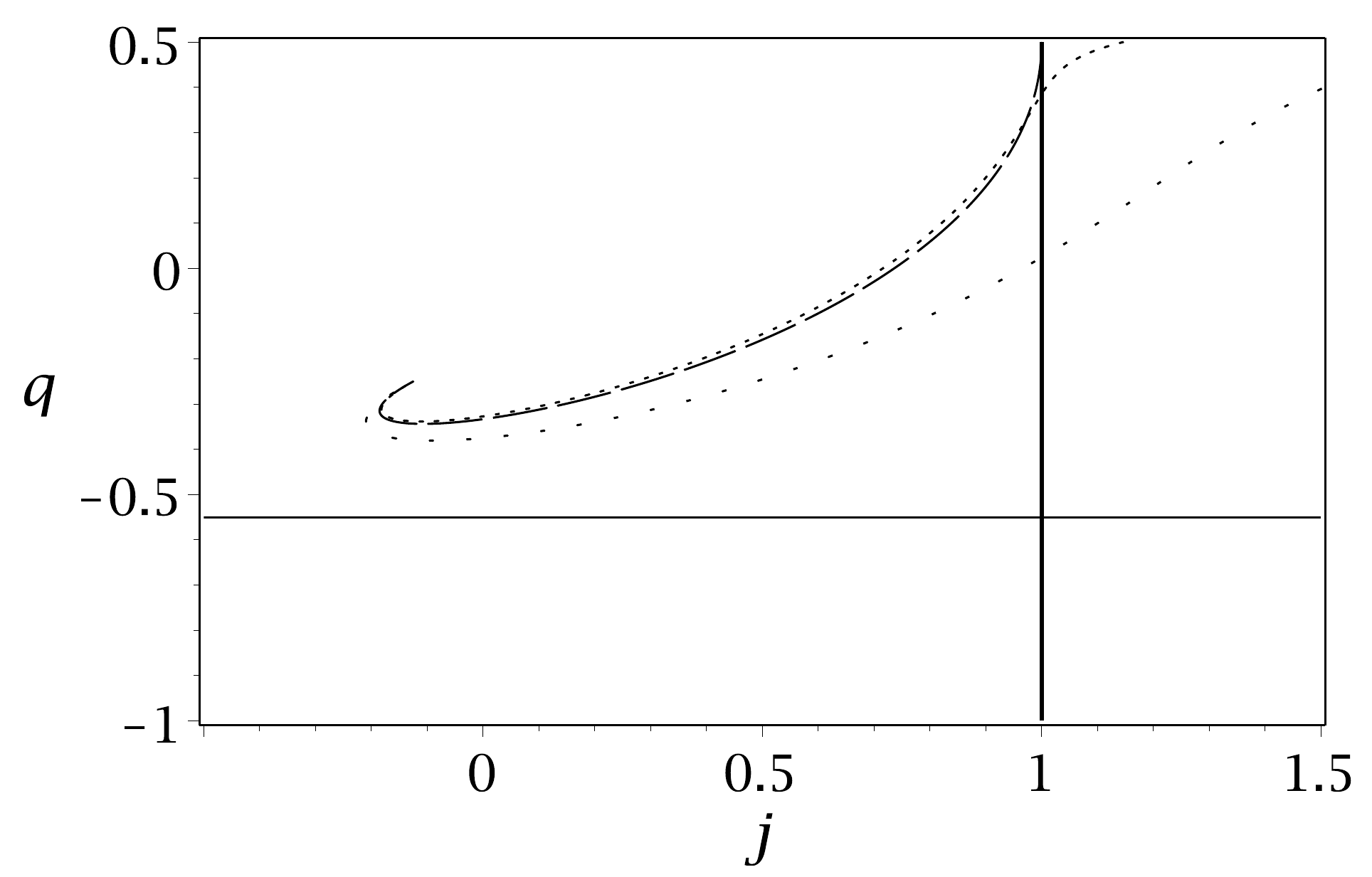}}
\end{center}
\vspace{-0.5cm}
\caption{Diagrams for the observable $\Omega_m$ and the cosmographic parameters $H$, $q$, $j$ (the Hubble,
deceleration, and jerk parameters) for dust and a potential $V = V_0 \exp(-\sqrt{3/2}\,\phi)$
for the solutions (with the same notation) in Fig.~\ref{fig:SSEXP}.
The thick line represents $\Lambda$CDM cosmology with $\Omega_{m0}=0.3$, $\Omega_{\Lambda 0}=0.7$.
The horizontal lines with $\Omega_{m}=0.3$, $H/H_0=1$, and $q=-0.55$ represent the present situation, 
i.e., $z=0$.}
\label{fig:EXP2c}
\end{figure}

Let us now consider the interior solutions in Fig.~\ref{fig:SSEXP} with
${\lambda}=\sqrt{3/2}$ in an $\Omega_m-q$ diagram and in several
\emph{cosmographic diagrams} (see Fig.~\ref{fig:EXP2c}), together with the
$\Lambda$CDM model with $\Omega_{m0}=0.3$ and $\Omega_{\Lambda0}=0.7$. To
have a feeling for the evolution in terms of redshift one can compare with
the graphs in Fig.~\ref{fig:LCDMgraphs}, which gives a good estimate for the
history also for the other solutions when expressed in the redshift $z$, as
long as they are close to the $\Lambda$CDM model.
%
%
%\begin{figure}[ht!]
%\begin{center}
%\includegraphics[width=0.4\textwidth]{h_q_j_EXP.pdf} \vspace{-1.5cm}
%\end{center}
%\caption{$H-q-j$; $\lambda_0=\sqrt{\frac{3}{2}}$.}
%\label{fig:EXPHqj}
%\end{figure}
%
As can be seen, these models are deviating rather significantly from the
$\Lambda$CDM model, and therefore a considerably smaller value of ${\lambda}$
is needed to obtain an evolution that is close to $\Lambda$CDM dynamics, even
when $\Omega_m$ is close to 1 initially. Furthermore, in a certain sense,
discussed next, the present class of models is as special as the scalar
field models with a cosmological constant, and they are therefore unlikely
candidates for solving any of the issues one might have with $\Lambda$CDM
cosmology.
%, even though models with very small
%$\bar{\lambda}$, and with $\Omega_m$ close to one initially, are going to
%have an evolutionary history very similar to that of the $\Lambda$CDM models.

%-----------------------------------------------------------------
\subsection{Symmetries of frozen $\lambda$CDM dynamics}
%-----------------------------------------------------------------

It is only the cases of a constant or an exponential potential that
effectively reduce the problem from three to two dimensions by decoupling the
equation for the scalar field variable. The underlying reason for this is
that these models admit scaling symmetries.

In the case of an exponential potential, a translation of the scalar field
leads to a scaling of the potential, which in combination with an appropriate
scaling of $t$ yields a scaling of $\rho_\phi$; furthermore, a scaling of the
spatial coordinates leads to a scaling of the scale factor $a$ and thereby
also of $\rho_m$ in the case of a linear equation of state with $\gamma_m\neq
0$, which results in a scaling symmetry of Einstein's field equations. This
in turn gives rise to a one-parameter set of equivalent solutions, being the
reason for the decoupling property, and it is also the underlying reason for
why there exist ``scaling solutions,'' which more appropriately should be
referred to as homothetic self-similar solutions since the spacetime geometry
of these solutions admits a homothetic Killing vector field. For models with
both a perfect fluid with a linear equation of state and a scalar field with
an exponential potential, a scaling solution can only exist if $\rho_m$ and
$\rho_\phi$ behave in the same manner, which requires $\gamma_m=\gamma_\phi$.
The existence of a fixed point $\mathrm{EM}$ on the reduced state space
$x-\Omega_m$ therefore necessarily depends on this feature, but it turns out
to not always be possible since $\mathrm{EM}$ only exists when ${\lambda} >
\sqrt{3\gamma_m}$; furthermore, the \emph{global} attractive property of
$\mathrm{EM}$, which does \emph{not} follow from a fixed point analysis, follows
from a monotone function that can be \emph{derived} from another kind of
symmetry of the field equations that is associated with coordinate scalings
(this is a general mechanism, discussed for anisotropic models and perfect
fluids in~\cite{heiugg10}). For a constant potential scalings of the
coordinates instead bring a solution with a given $V=\Lambda$ to a solution
with a different $\Lambda$, and because $\Lambda$ carries dimension
(length$^{-2}$), there are no solutions admitting a homothetic symmetry when
$\Lambda \neq 0$; i.e., this scaling symmetry is not quite the same as that
in the exponential case.

The existence of symmetries is what makes these models tractable, and therefore
also popular. However, the symmetries endow these models with special
properties, which suggests that perhaps it is not a good idea to consider
these models as role models, since other models do not admit similar
symmetries and properties, or at least that some careful considerations may
be needed. Moreover, originally, before one realized that the deceleration
parameter $q$ is evolving, it was thought that scale invariance and thereby a
constant $q$ was quite desirable, since this would solve fine-tuning problems
such as coincidence and energy scale problems. On the other hand, the
symmetries also make the models quite inflexible. Due to the changing
evolutionary history of $q$, where $q$ is currently negative, the scaling
symmetries must be broken, while still leading to models that in some way at
least alleviate various fine-tuning problems; for these reasons, and from a
phenomenological perspective, scale-invariant models are no longer the favorite
candidates as dark energy models.

%Next we consider the simplest possible generalization by breaking the scaling
%symmetry by having a potential that consists of an exponential and a constant
%term. Although these models are unlikely to serve as useful dynamical dark
%energy models, they do exhibit some features that are characteristic of more
%general classes of models, and it is for this reason we consider them next.

%%%%%%%%%%%%%%%%%%%%%%%%%%%%%%%%%%%%%%%%%%%%%%%%%%%%%%%%%%%%%%%%%%
\section{Dynamical $\lambda$CDM dynamics}\label{sec:dynlambda}
%%%%%%%%%%%%%%%%%%%%%%%%%%%%%%%%%%%%%%%%%%%%%%%%%%%%%%%%%%%%%%%%%%

%-----------------------------------------------------------------
\subsection{Inverse power-law potentials}
%-----------------------------------------------------------------

In the influential papers~\cite{peerat88,ratpee88} Peebles and Ratra argued
that dark energy can be phenomenologically modeled by a minimally coupled
scalar field with a potential with a shallow tail, which results in the dark field 
energy density decreasing more slowly than the matter energy density to
its ``natural'' value -- zero. In particular they considered an inverse
power-law potential,
\begin{equation}
V=\frac{V_0}{\phi^{\alpha}}, \qquad \phi >0, \quad \alpha > 0 \quad \Rightarrow \quad \lambda = \frac{\alpha}{\phi}.
\end{equation}
In addition they assumed a matter dominated flat FLRW universe after
inflation, which resulted in a set of approximate flat FLRW equations for
which they found a particular solution, which was shown to be linearly stable
within the approximate context. The associated solution of the exact FLRW
equations has subsequently been referred to as an attractor or tracker
solution~\cite{steetal99}, and has resulted in numerous papers.
In~\cite{tam14} a thorough local dynamical systems analysis of the present
models was performed, based on previous work,
e.g.,~\cite{ngetal01,ure12,gon14} (for additional references see,
e.g.,~\cite{tsu13,tam14}). Here we will use a slight variation of this
approach and use the following bounded scalar field variable:
\begin{equation}
Z=\frac{1}{1+\lambda} = \frac{\phi}{\phi + \alpha} \quad \Rightarrow \quad \lambda=\frac{1-Z}{Z},
\quad \phi = \alpha\left(\frac{Z}{1-Z}\right),
\end{equation}
where $Z$ thereby is monotonically increasing in $\phi$ (in~\cite{tam14}
$1-Z$ was used as the scalar field variable).

The present case has an unbounded $\lambda$, and therefore we use the
dynamical system~\eqref{dynsysZ}, where we choose $g(Z) = Z$, and thus
$g_{\lambda_-}=1$ and
\begin{equation}
\frac{d{\tau}}{d\bar{\tau}} = Z, \qquad \frac{dt}{d\bar{\tau}} = H^{-1}Z.
\end{equation}
and the following dynamical system for the state vector $(x,\Omega_m,Z)$:
\begin{subequations}
\begin{align}
\frac{dx}{d\bar{\tau}} &=-(2-q)xZ+\sqrt{\frac{3}{2}}\left(1-Z\right)\left(1-x^2-\Omega_m\right), \\
\frac{d\Omega_{m}}{d\bar{\tau}} &= 3\left[2x^2-(1-\Omega_{m})\right]\Omega_m Z, \\
\frac{dZ}{d\bar{\tau}} &=\frac{\sqrt{6}}{\alpha}\left(1-Z\right)^2 Z x,
 \end{align}
\end{subequations}
where $2-q=3(1-x^2) - \frac32 \Omega_m$.

\begin{figure}[ht!]
\begin{center}
\subfigure[The $(x,\Omega_{m},Z)$ state space.]{\label{fig:SSRP_alpha6}
\includegraphics[width=0.45\textwidth, trim = 1cm 2cm 2cm 0cm]{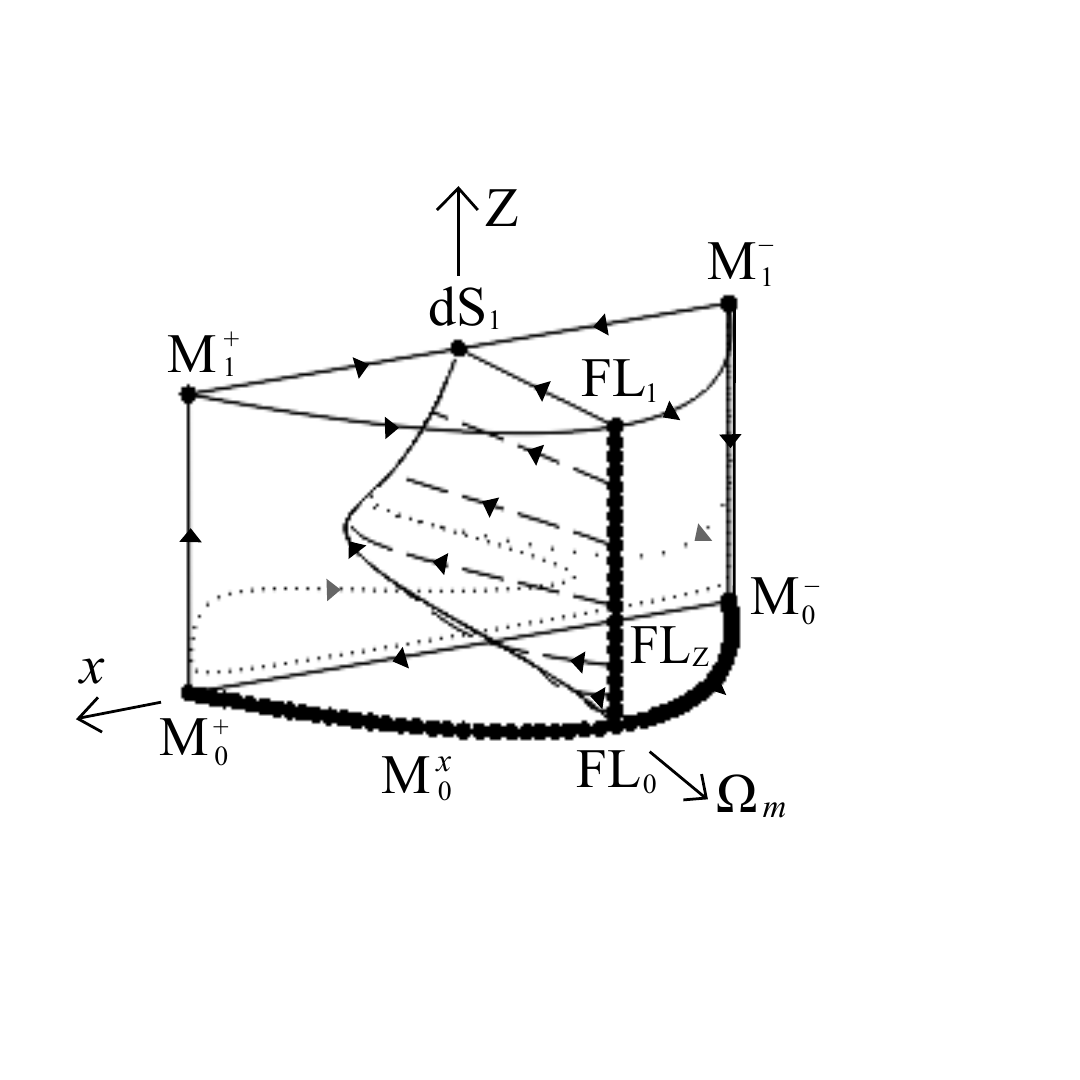}}
\subfigure[Projection of $(x,\Omega_{m},Z)$ state space onto  $(x,\Omega_{m})$.]{\label{fig:SSRPTop_alpha6}
\includegraphics[width=0.45\textwidth, trim = 0.5cm 2cm 1cm 0cm]{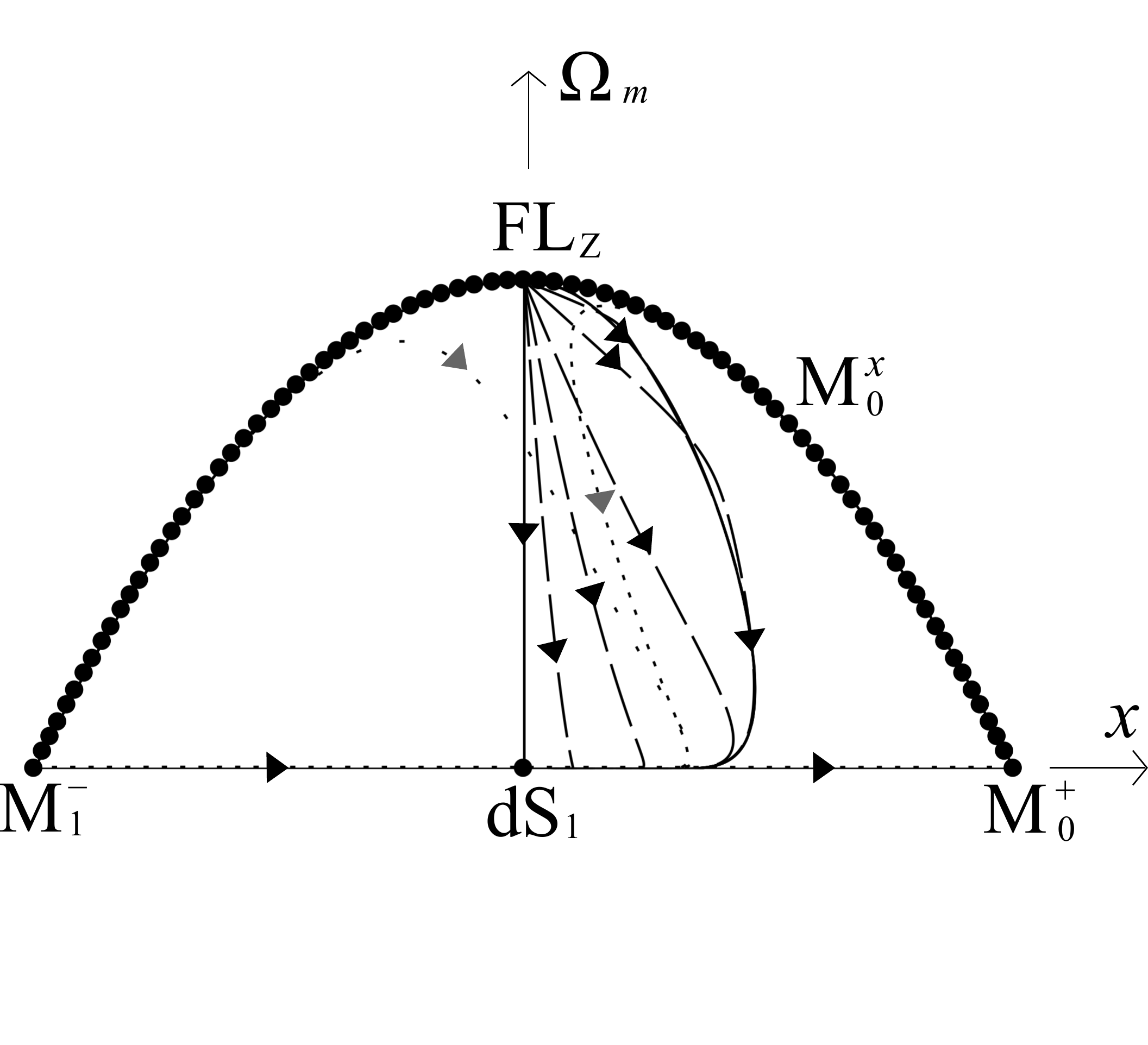}}
\end{center}
\vspace{-0.5cm}
\caption{Solution structure for the state space for models with dust and a scalar field
with inverse power-law potential $V=V_0\phi^{-\alpha}$ with $\alpha=6$.}
\label{fig:RP_alpha6}
\end{figure}

Due to the discussion about global dynamics in Sec.~\ref{sec:dynsys}, it
follows that the conclusions obtained from the local fixed point results
in~\cite{tam14} correspond to global asymptotic features. Hence a
two-parameter set of solutions originates from $\mathrm{M}^-_1$, and a
one-parameter set from $\mathrm{FL}_Z$, where the tracker solution originates
from $\mathrm{FL}_0$. Toward the future all orbits on ${\bf S}$ and ${\bf
S}_\phi$ end at the global future attractor $\mathrm{dS}_1$.\footnote{Since
$\lim_{\phi\rightarrow +\infty}V(\phi)= 0$ it follows that the future
asymptotic state $\mathrm{dS}_1$ now corresponds to the Minkowski spacetime
instead of the de Sitter spacetime. However, because the matter energy
density goes to zero faster than the scalar field energy density it still
follows that $q\rightarrow -1$; the asymptotic evolution thus resembles that
of the de Sitter spacetime, since asymptotically the spacetime approaches
that of a Minkowski spacetime in a (local) foliation where $q\rightarrow -1$.
A similar statement holds for the models in the next subsection.} The
solution structure for $\alpha=6$, and especially that of the subset that
originates from $\mathrm{FL}_Z$, is depicted in Fig.~\ref{fig:RP_alpha6}.

It is of some interest to complement the dynamical systems picture with a
heuristic description where we regard the scalar field as a particle moving
in a potential, while losing energy due to the friction force
$-3H\dot{\phi}$. Due to the existence of a potential wall that is so steep
that $\lambda \rightarrow \infty$ when $\phi\rightarrow 0$, all ``scalar field
particles'' (solutions) either come from $\phi \rightarrow -\infty$ and bounce
against the potential wall (the two-parameter set of solutions from
$\mathrm{M}^-_1$) or from being initially still at some $\phi$ (no initial
kinetic scalar field energy, i.e., $x=0$) and then rolling down the potential
in the positive $\phi$-direction; the tracker solution corresponds to the
solution that starts by being initially still with an infinite scalar field
energy associated with the limit $\phi \rightarrow 0$ (note that this result
also follows from the asymptotic explicit results concerning this solution
given originally in~\cite{peerat88,ratpee88}).

All solutions therefore eventually move in an increasingly shallow potential,
slowed by ``friction,'' and hence with decreasing scalar field energy,
although the matter energy decreases faster, thus leading to a future
asymptotic (quasi) de Sitter state at $\phi \rightarrow +\infty$ (i.e.,
$\mathrm{dS}_1$).

Loosely speaking, the tracker solution is connected with a bifurcation
associated with going from a potential with finite $\lambda$ to infinite
$\lambda$, transforming the attracting focus of the exponential case with
large $\lambda$ to a center. The orbits on $Z=0$ are characterized by
$\Omega_m = \mathrm{const}$. They therefore correspond to the periodic
orbits found in the monomial case at late times, given in different variables
in~\cite{alhetal15}, but in the present case they are interrupted by being
cut in half by the line of fixed points, $\mathrm{M}^x_0$, which is a formal
consequence of the new time variable. The physical reason for this feature is
that in contrast to the monomial case there are no scalar field oscillations
at late times for inverse power-law potentials.

As can be seen from Fig.~\ref{fig:RP_alpha6}, the tracker solution forms part
of the boundary of the $\mathrm{FL}_Z$ saddle subset, which therefore is not a
separatrix surface. Because of its initial linear stability property in the Peebles-Ratra 
formulation~\cite{peerat88,ratpee88}, which was seen as alleviating the
coincidence problem, nearby solutions are attracted to it.
However, there exists an open set of solutions that are further away from it
that are not, even if $\Omega_m$ is close to 1 initially (those with fairly
large initial $Z$). This is to be expected since the Peebles-Ratra
stability analysis is only local. It should also be pointed out that there is
no {\it a priori} reason for believing that the local attracting property
holds globally in time.\footnote{See, e.g.~\cite{ugg13} for an example of an
attracting solution that is locally but not globally attracting.} However,
all solutions near the tracker solution, indeed all solutions that are near
or on the subset that originates from $\mathrm{FL}_Z$, are eventually
shadowing the center manifold of the fixed point $\mathrm{dS}_1$ on ${\bf
S}_\phi$\footnote{Incidentally, it is the center manifold solution on ${\bf
S}_\phi$ that the slow-roll approximation, given in, e.g.,~\cite{wei08},
describes approximately; see~\cite{alhetal15,alhugg15} for a similar
situation at early times instead of late times.} and it is the center
manifold solution on ${\bf S}_\phi$ that is an ``attractor solution'' rather than the
tracker solution, since any solution close to this solution is attracted to
it because it is a center manifold, but note that it is $\mathrm{dS}_1$
that is the formal future attractor.

\begin{figure}[ht!]
\begin{center}
\subfigure[$\Omega_{m}-q$ diagram.]{\label{fig:q_M_RP_alpha6}
\includegraphics[width=0.45\textwidth, trim = 0cm 0cm 0cm 0cm]{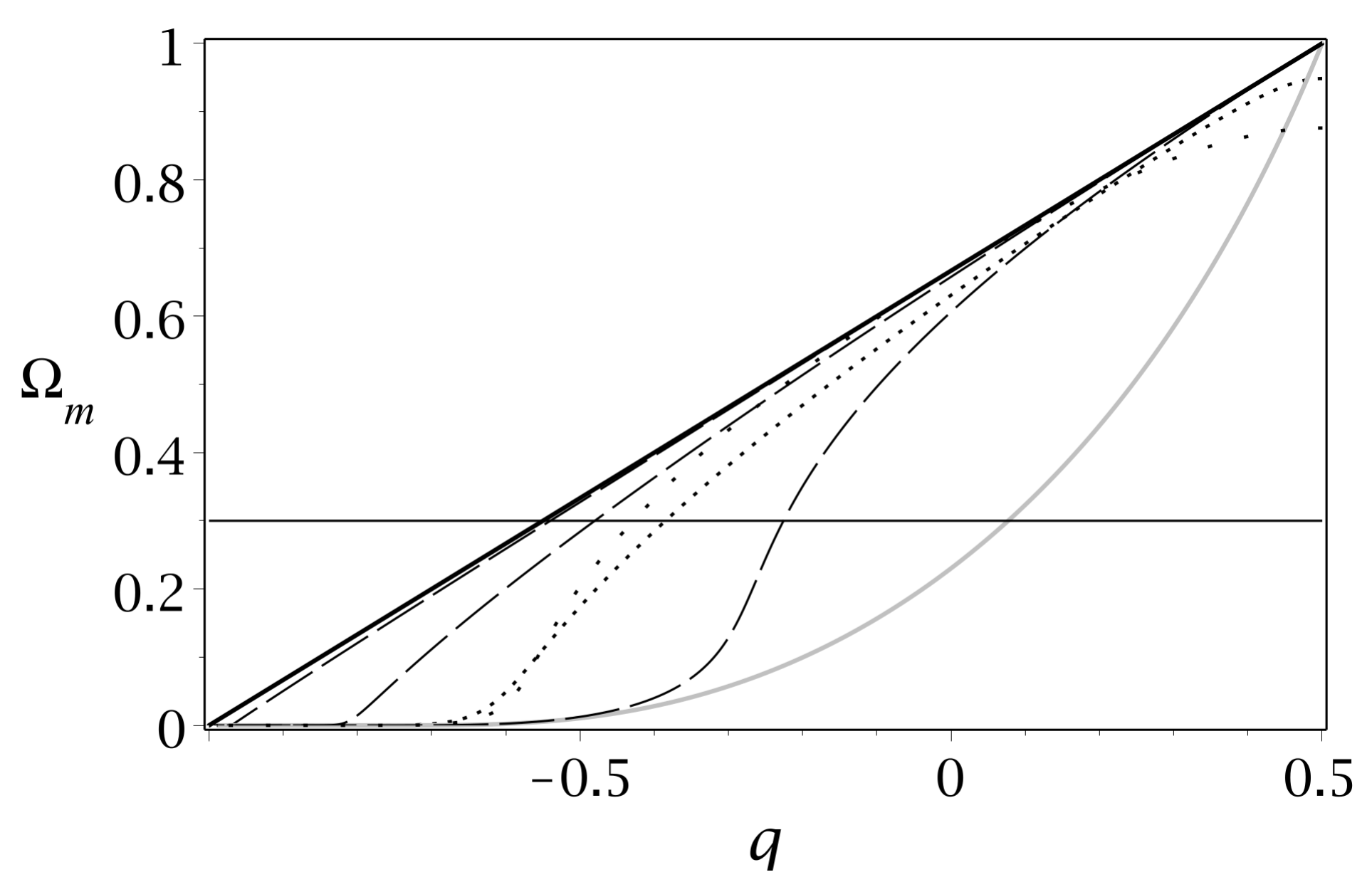}}
\subfigure[$h-q$ diagram.]{\label{fig:h_q_RP_alpha6}
\includegraphics[width=0.45\textwidth, trim = 0cm 0cm 0cm 0cm]{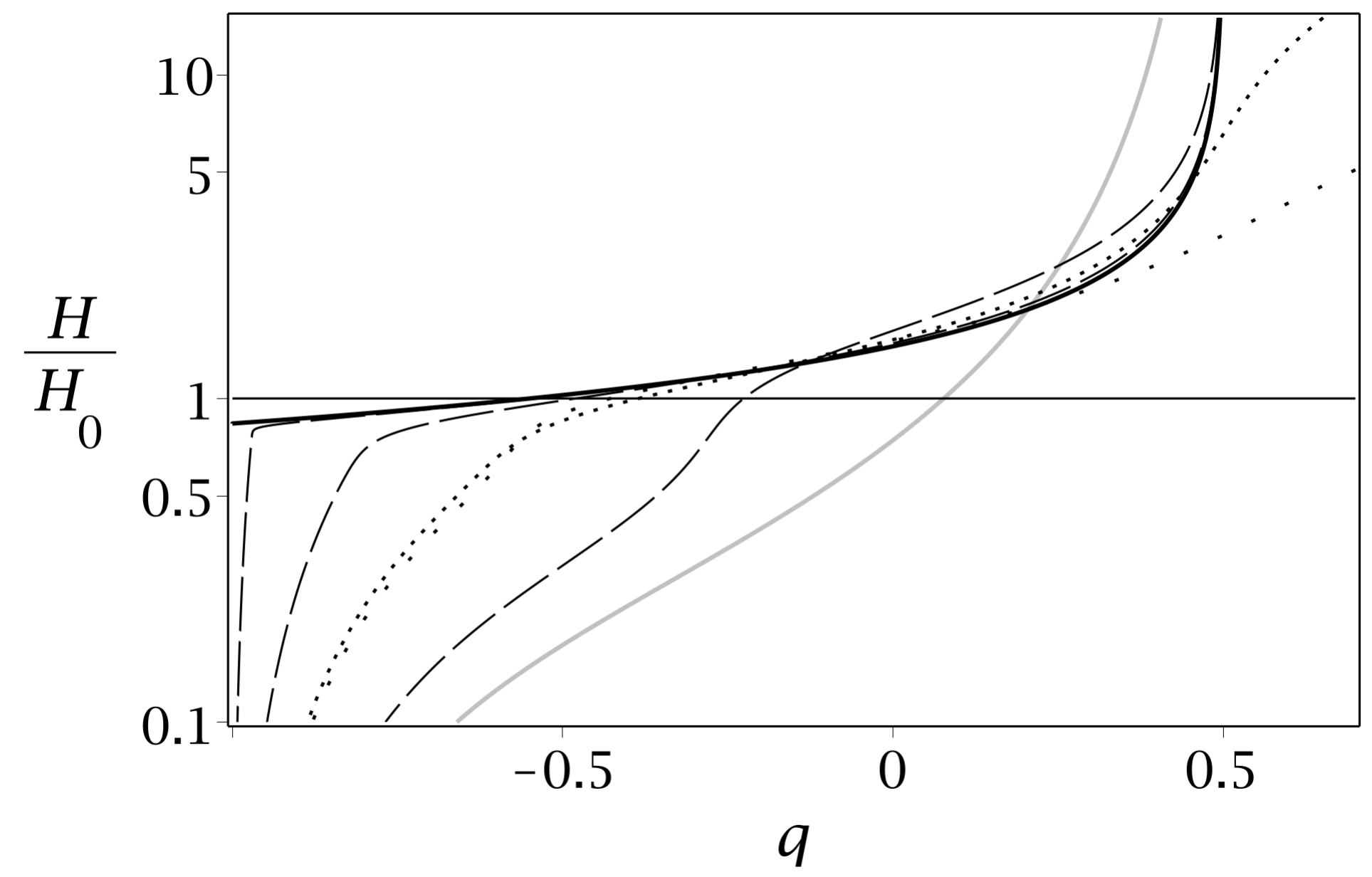}}
\subfigure[$h-j$ diagram.]{\label{fig:h_j_RP_alpha6}
\includegraphics[width=0.45\textwidth, trim = 0cm 0cm 0cm 0cm]{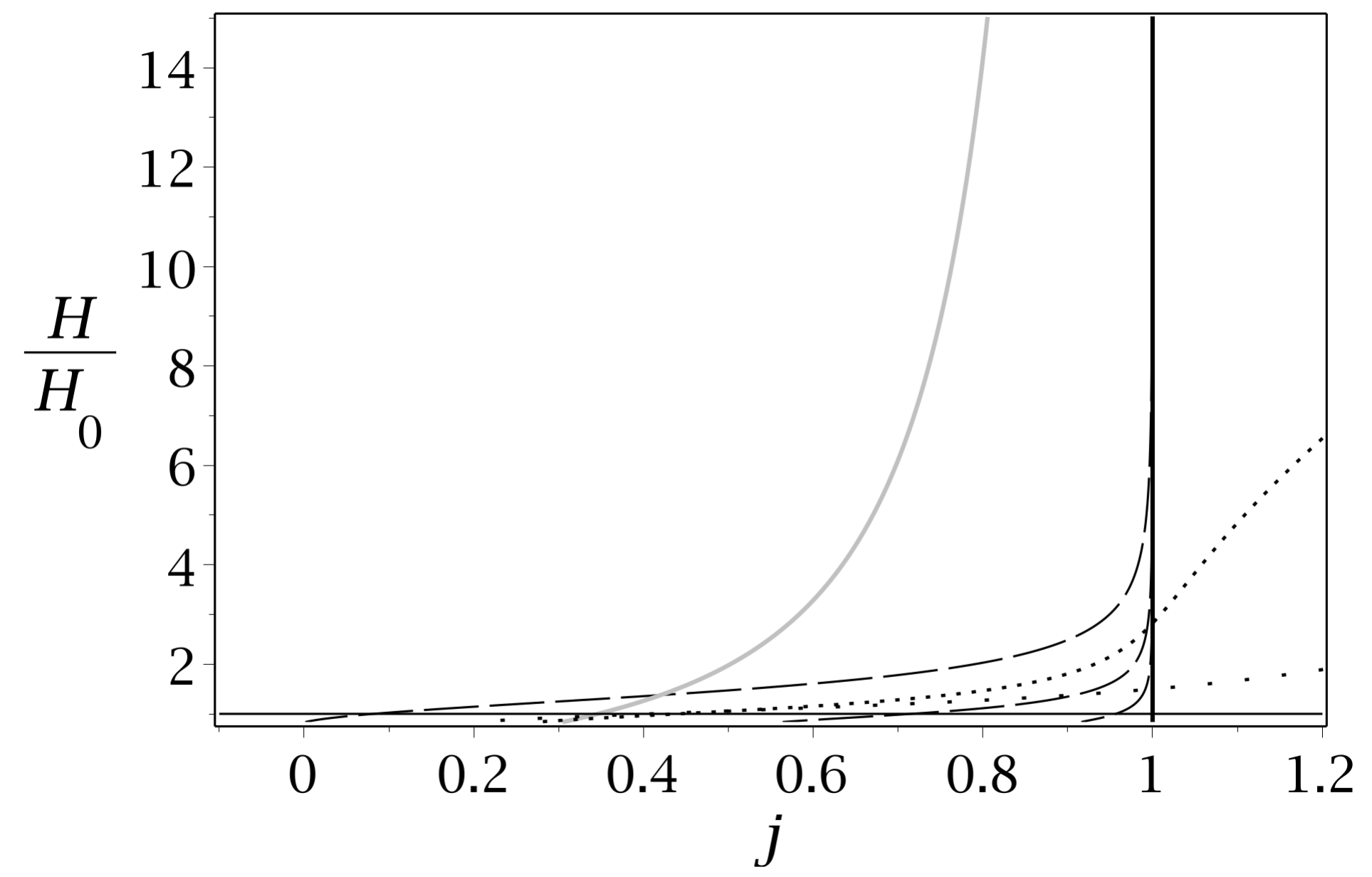}}
\subfigure[$q-j$ diagram.]{\label{fig:j_q_RP_alpha6}
\includegraphics[width=0.45\textwidth, trim = 0cm 0cm 0cm 0cm]{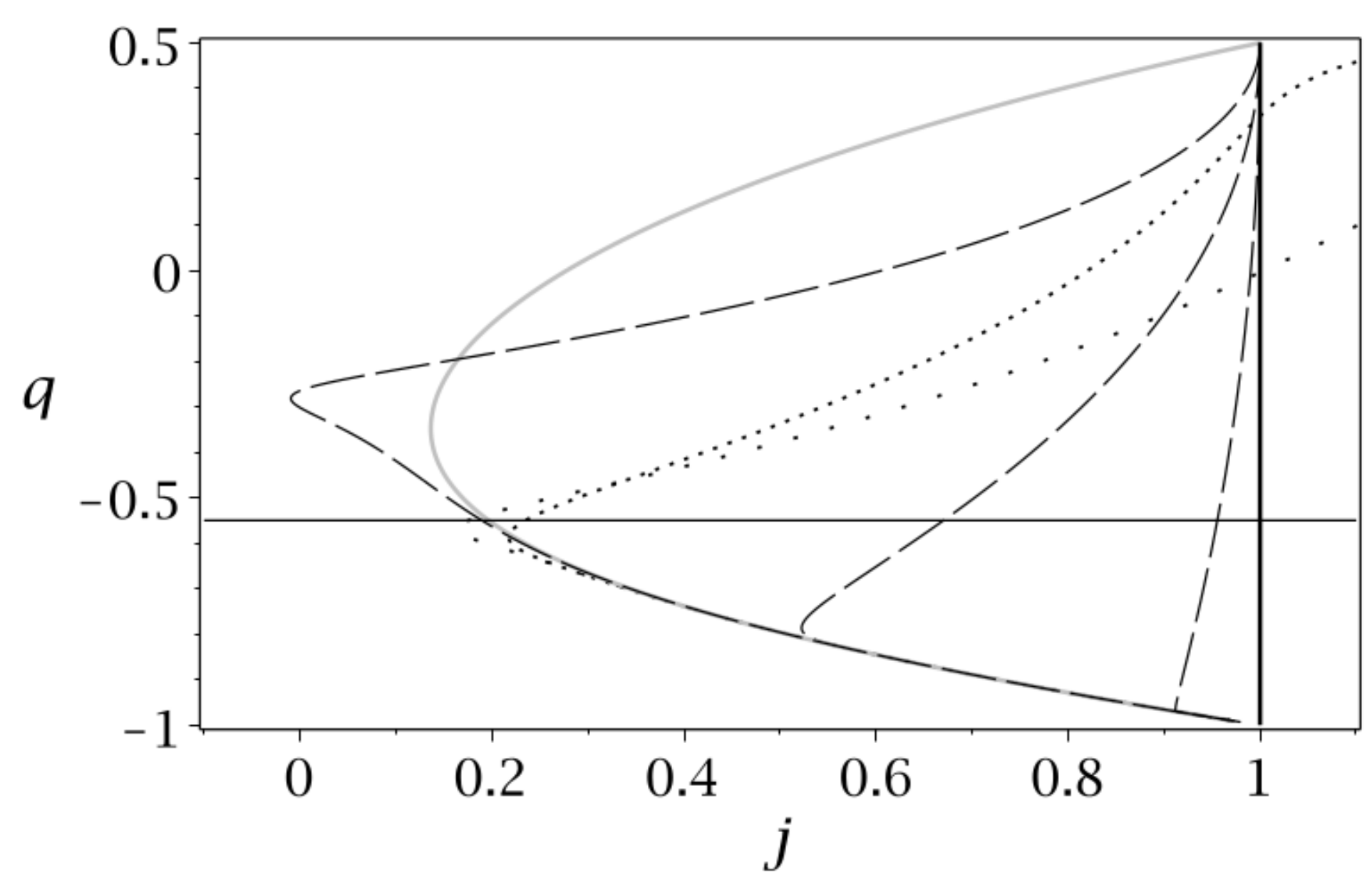}}
%\subfigure[$h-q-j$.]{\label{fig:h_q_j_RP_alpha6}
%\includegraphics[width=0.45\textwidth, trim = 0cm 0cm 0cm 0cm]{h_q_j_RP6.pdf}}
\end{center}
\caption{Observables for dust and an inverse power-law potential $V=V_0\phi^{-\alpha}$ with $\alpha=6$. 
The thick solid line represents $\Lambda$CDM cosmology; the horizontal lines represent now, i.e., zero redshift; 
and the other lines correspond to the solutions in Fig.~\ref{fig:RP_alpha6}, where the grey line corresponds 
to the tracker solution. 
}
\label{fig:Cosm_RP_alpha6}
\end{figure}

As seen from Fig.~\ref{fig:RP_alpha6}, the tracker solution and the solutions
that are near to it on the subset that originates from $\mathrm{FL}_Z$ are
those solutions on this subset that diverge from $\Lambda$CDM dynamics the
most (in this sense they are the observationally worst solutions in this
subset of solutions, which is also seen by plotting the solutions in diagrams
for the observables; see Fig.~\ref{fig:Cosm_RP_alpha6}) 
and quite substantially when $\alpha=6$. Let us
therefore consider the dynamics for models with lower $\alpha$.

\begin{figure}[ht!]
\begin{center}
\subfigure[$(x,\Omega_{m},Z)$ state space.]{\label{fig:SSRP_alpha1_2}
\includegraphics[width=0.45\textwidth, trim = 1cm 2cm 2cm 0cm]{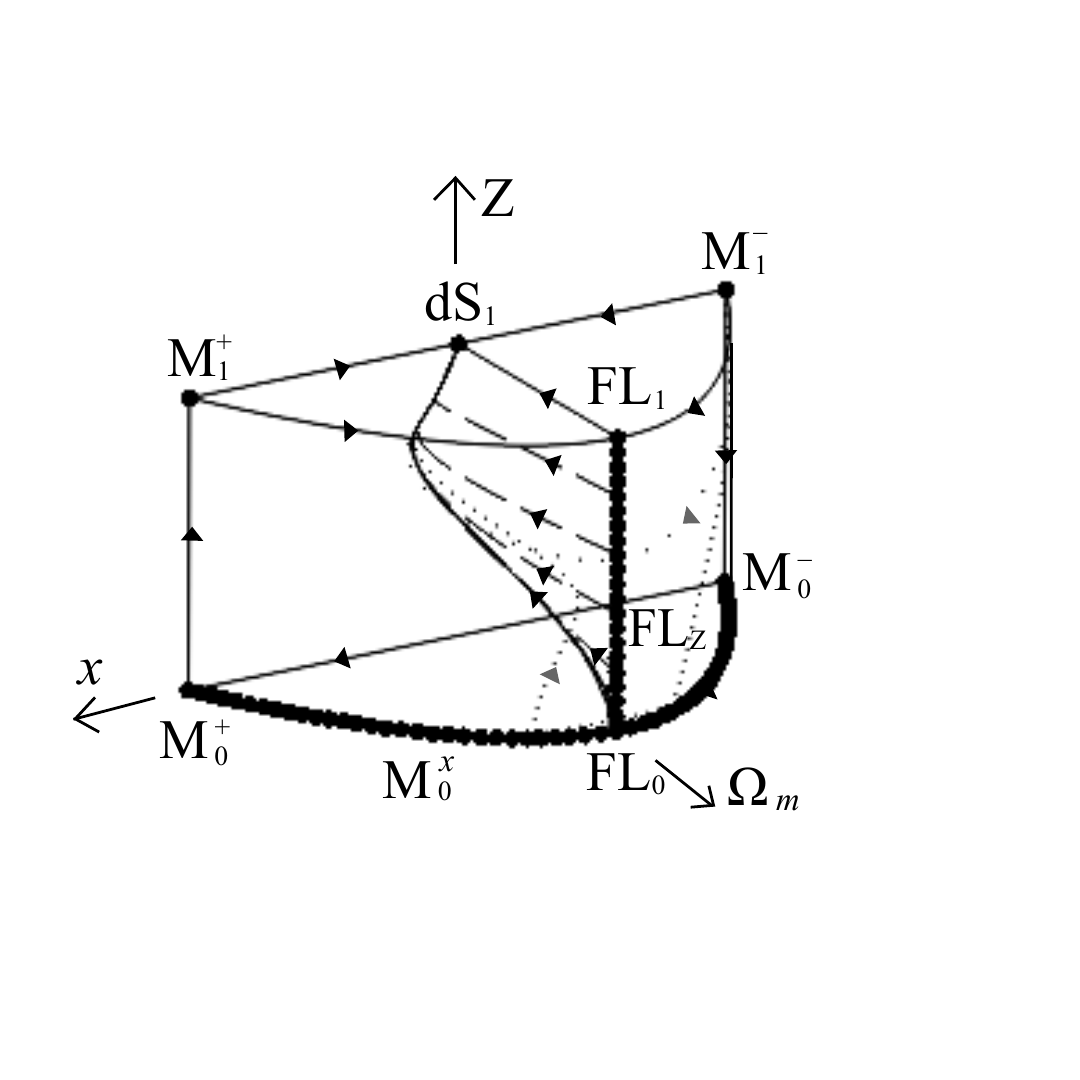}}
\subfigure[Projection  of $(x,\Omega_{m},Z)$ state space onto  $(x,\Omega_{m})$.]{\label{fig:SSRPTop_alpha1_2}
\includegraphics[width=0.45\textwidth, trim = 0.5cm 2cm 1cm 0cm]{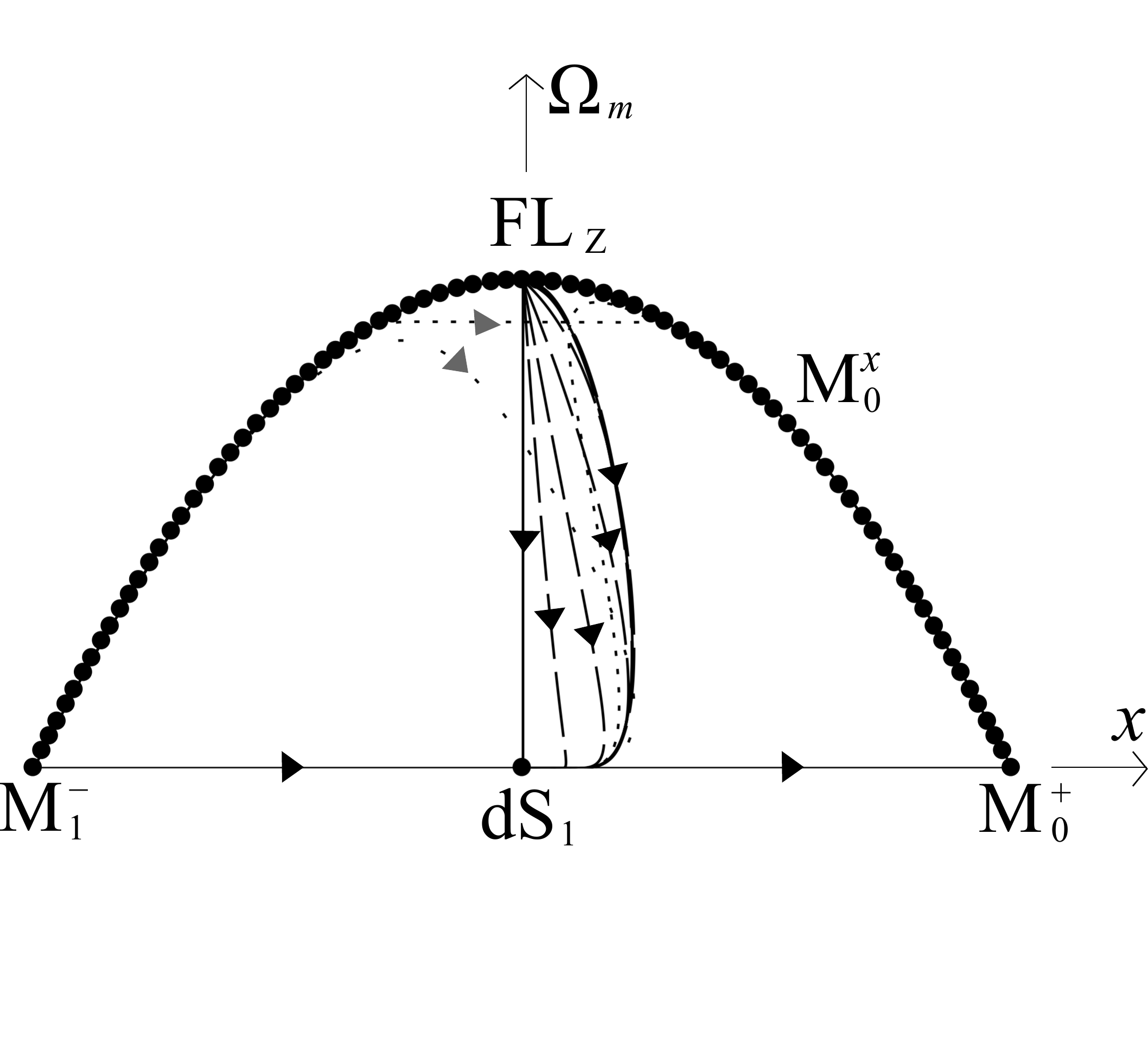}}
\end{center}\vspace{-0.5cm}
\caption{The solution space of dust and an inverse power-law potential $V=V_0\phi^{-\alpha}$ with $\alpha=1/2$.
}
\label{fig:RP_alpha1_2}
\end{figure}

\begin{figure}[ht!]
\begin{center}
\subfigure[$\Omega_{m}-q$ diagram]{\label{fig:q_M_RP_alpha1_2}
\includegraphics[width=0.48\textwidth, trim = 0cm 0cm 0cm 0cm]{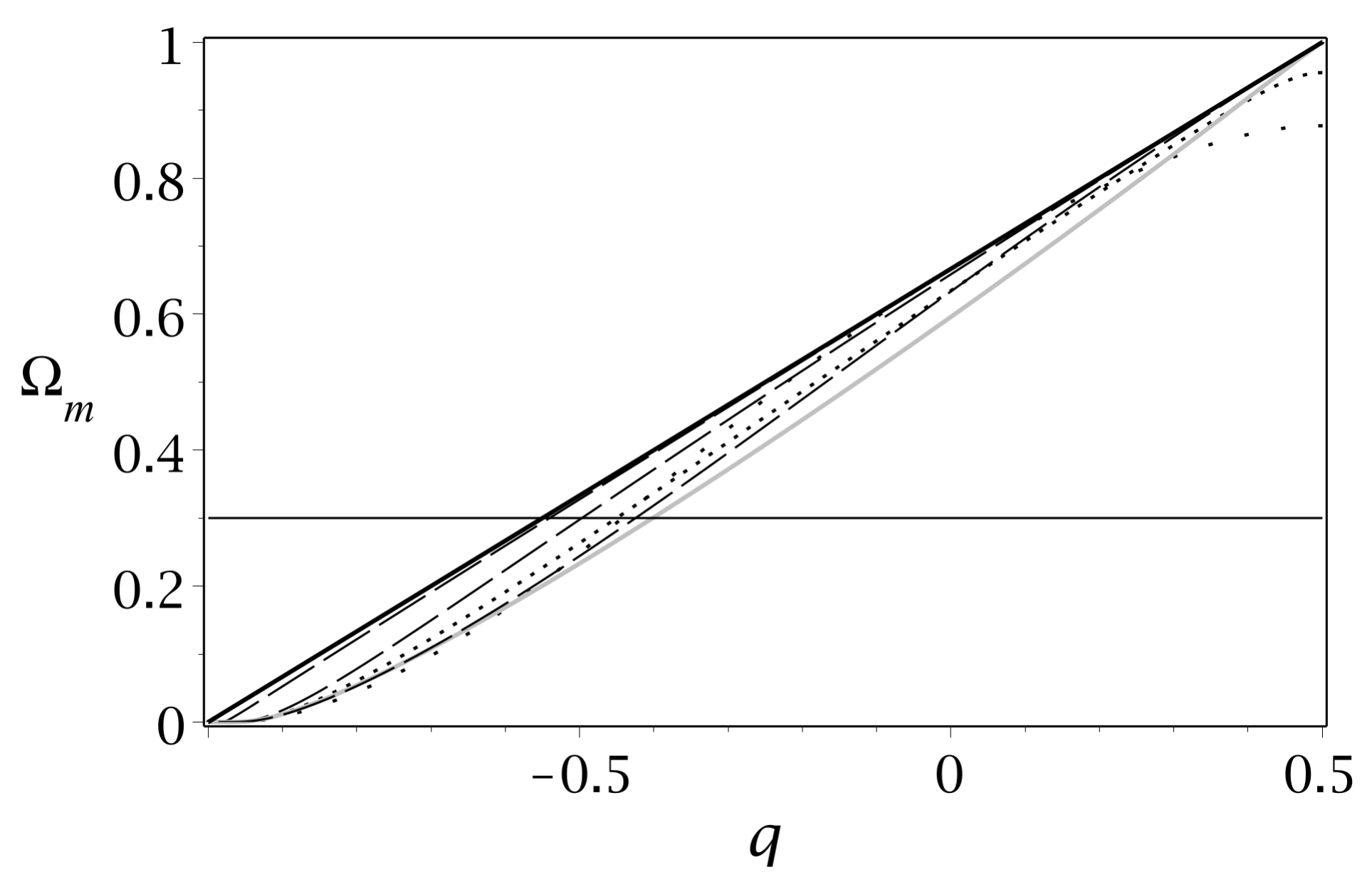}}
\subfigure[$h-q$ diagram]{\label{fig:h_q_RP_alpha1_2}
\includegraphics[width=0.48\textwidth, trim = 0cm 0cm 0cm 0cm]{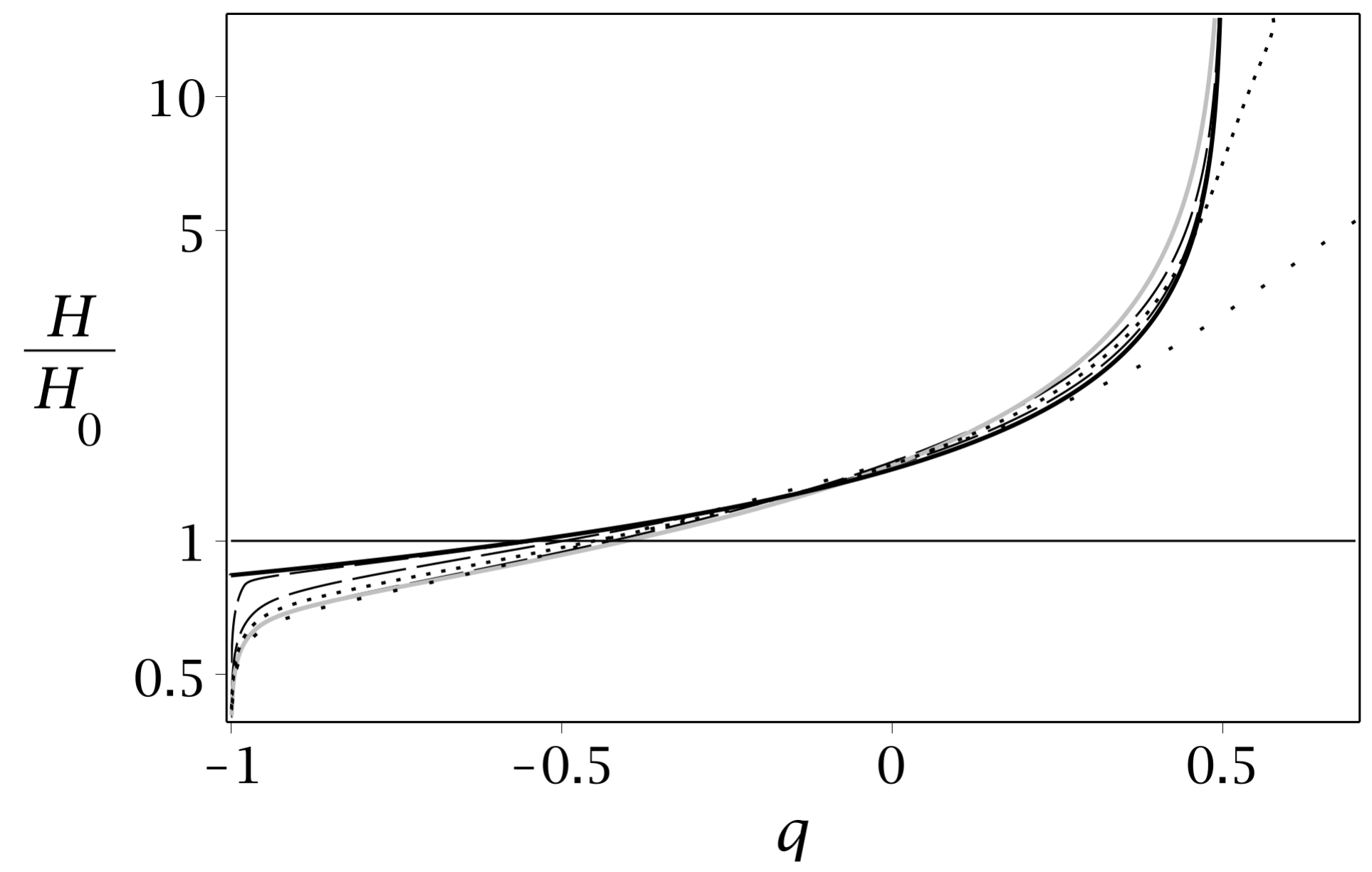}}
\subfigure[$h-j$ diagram]{\label{fig:h_j_RP_alpha1_2}
\includegraphics[width=0.48\textwidth, trim = 0cm 0cm 0cm 0cm]{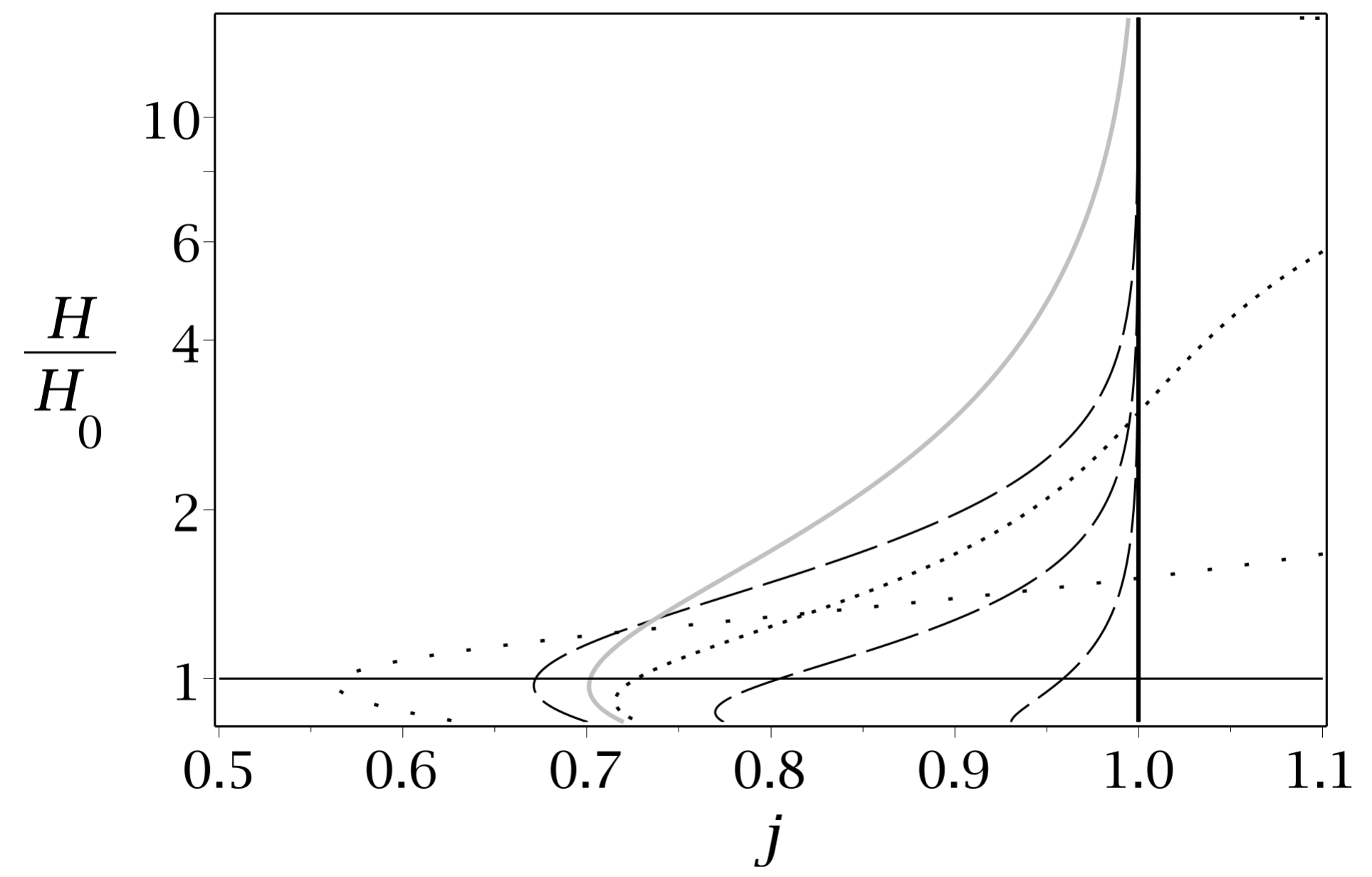}}\,\,
\subfigure[$q-j$ diagram]{\label{fig:j_q_RP_alpha1_2}
\includegraphics[width=0.48\textwidth, trim = 0cm 0cm 0cm 0cm]{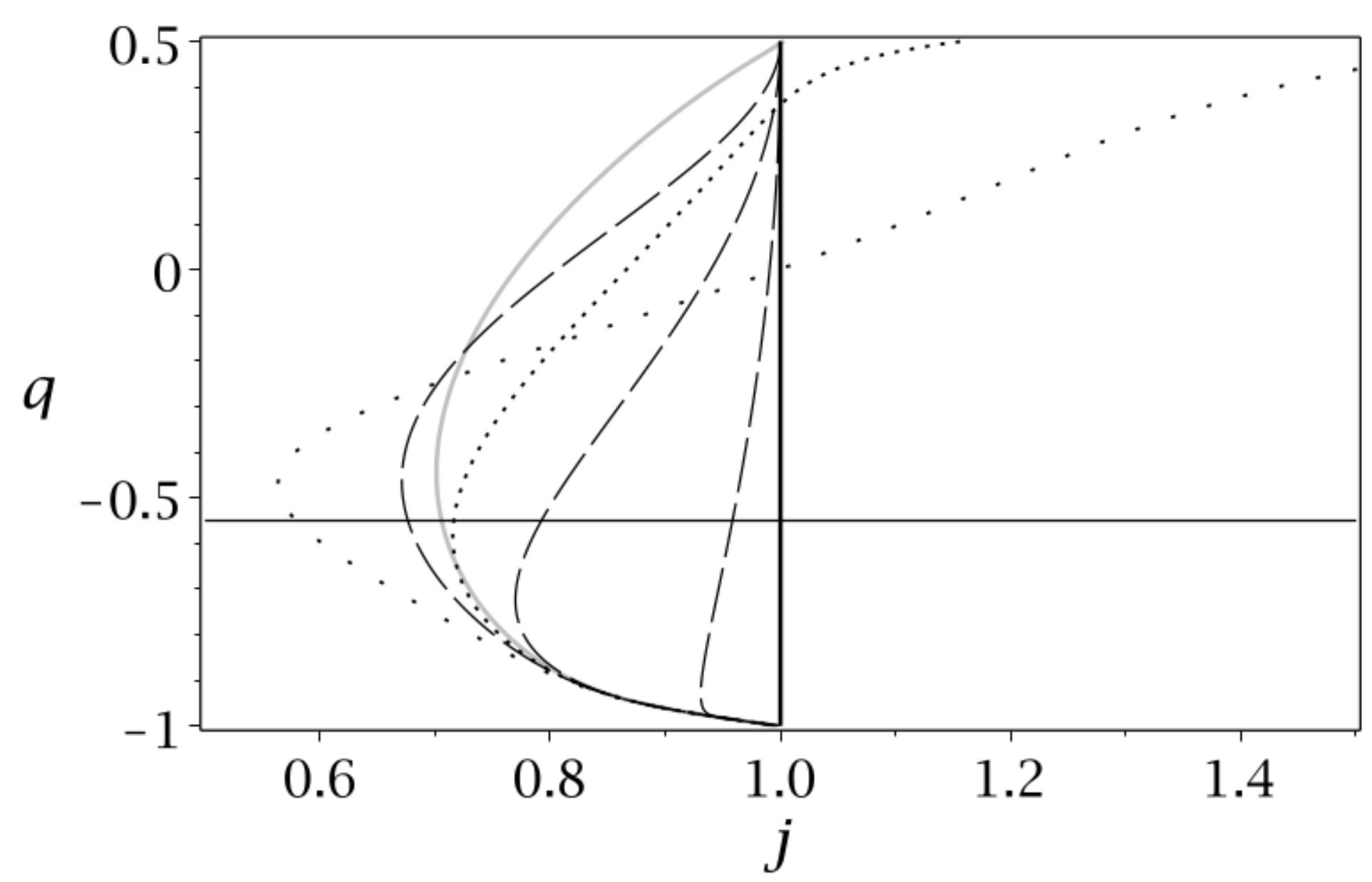}}
%\subfigure[$h-q-j$.]{\label{fig:h_q_j_RP_alpha1_2}
%\includegraphics[width=0.45\textwidth, trim = 0cm 0cm 0cm 0cm]{h_q_j_RP1_2.pdf}}
\vspace{-0.5cm}
\end{center}
\caption{Observables for dust and an inverse power-law potential $V=V_0\phi^{-\alpha}$ with $\alpha=1/2$. 
The thick solid line represents $\Lambda$CDM cosmology; the horizontal lines represent now, i.e., zero redshift;
and the other lines correspond to the solutions in Fig.~\ref{fig:RP_alpha1_2}, where the grey line corresponds 
to the tracker solution.}
\label{fig:Cosm_RP_alpha1_2}
\end{figure}

The solution structure for $\alpha=1/2$, and especially that of the subset
that originates from $\mathrm{FL}_Z$, is depicted in
Fig.~\ref{fig:RP_alpha1_2}.
The topological structure is the same as for $\alpha=6$, but the subset that
originates from $\mathrm{FL}_Z$ is now located closer to the $x=0$-plane and
as a consequence the dynamics of solutions with $\Omega_m$ close to 1
initially more closely resembles that of the $\Lambda$CDM models. These
results are therefore in line with the conclusion in~\cite{tsu13}, that it is only
for quite small $\alpha$ that the tracker solution is observationally viable,
which restricts the observational relevance of these models. The deviations
from $\Lambda$CDM cosmology can be explicitly seen in the diagrams for the
observables in Fig.~\ref{fig:Cosm_RP_alpha1_2}.
%As for
%$\alpha=6$, the tracker solution and the nearby solutions are those with the
%worst evolution in this set of solutions.
%, in the sense that they all deviate
%substantially from $\Lambda$CDM evolution, in spite of a quite small value of
%$\alpha$.
%

Note that Figs.~\ref{fig:RP_alpha6} and~\ref{fig:RP_alpha1_2} also show that it is not only the
tracker solution that attracts nearby solutions, but the whole
$\mathrm{FL}_Z$ saddle subset. The underlying reasons for this are that (i)
the whole line of fixed points $\mathrm{FL}_Z$ acts as a line of saddle
points, and (ii) all solutions in this subset are attracted to the
center manifold of $\mathrm{dS}_1$ at late times before asymptotically
approaching the global future attractor $\mathrm{dS}_1$.

Finally, it is worth noting that the underlying reason for why the early
behavior of the tracker solution exhibits a ``tracking property'' is because
the present case can be viewed as a bifurcation where
$\mathrm{EM}_0$ merges with $\mathrm{FL}_0$ in the limit $\lambda \rightarrow
\infty$. The property that $\gamma_\phi=\gamma_m=1$ for $\mathrm{EM}$,
irrespective of $\lambda$ (as long as $\mathrm{EM}$ exists until it merges
with $\mathrm{PL}$), can subsequently be exploited by comparing the reduced
two-dimensional dynamics of a sequence of reduced exponential scalar field
state spaces, as done in~\cite{tam14,ngetal01,ure12}, to yield an approximate
heuristic description of the behavior of the tracking solution. Nevertheless,
it should be clear that the dynamics of an inverse power-law potential model
is globally very different from that of a given exponential potential, which
is what can be expected from a model that does not exhibit scaling
symmetries.

%, although lately its observational relevance has been
%questioned~\cite{tsu13}. That this is the case should really not come as a
%surprise if one considers the situation from a dynamical systems perspective.

The observational viability difficulties of the dust and inverse power-law
potential models, including those of tracker solutions, arise because
$\lambda\rightarrow \infty$ when $\phi\rightarrow \infty$, since a large
$\lambda$ makes the dynamics deviate from $\Lambda$CDM dynamics. This
motivates us to consider other types of scalar field potentials for which
$\lambda$ is globally and asymptotically regularized. For this reason, we
therefore phenomenologically regularize $\lambda$ of the inverse power-law
potential and use~\eqref{Vfromlambda} to derive $V$, in arguably the
simplest possible way, so that $\lambda$ becomes finite and one obtains
continuous deformations of $\Lambda$CDM cosmology.

%-----------------------------------------------------------------
\subsection{$\lambda$-regularization of inverse power-law potentials}
%-----------------------------------------------------------------

\begin{figure}[ht!]
\begin{center}
\subfigure[Potential diagram]{\label{fig:Va6C10}
\includegraphics[width=0.48\textwidth, trim = 0cm 0cm 0cm 0cm]{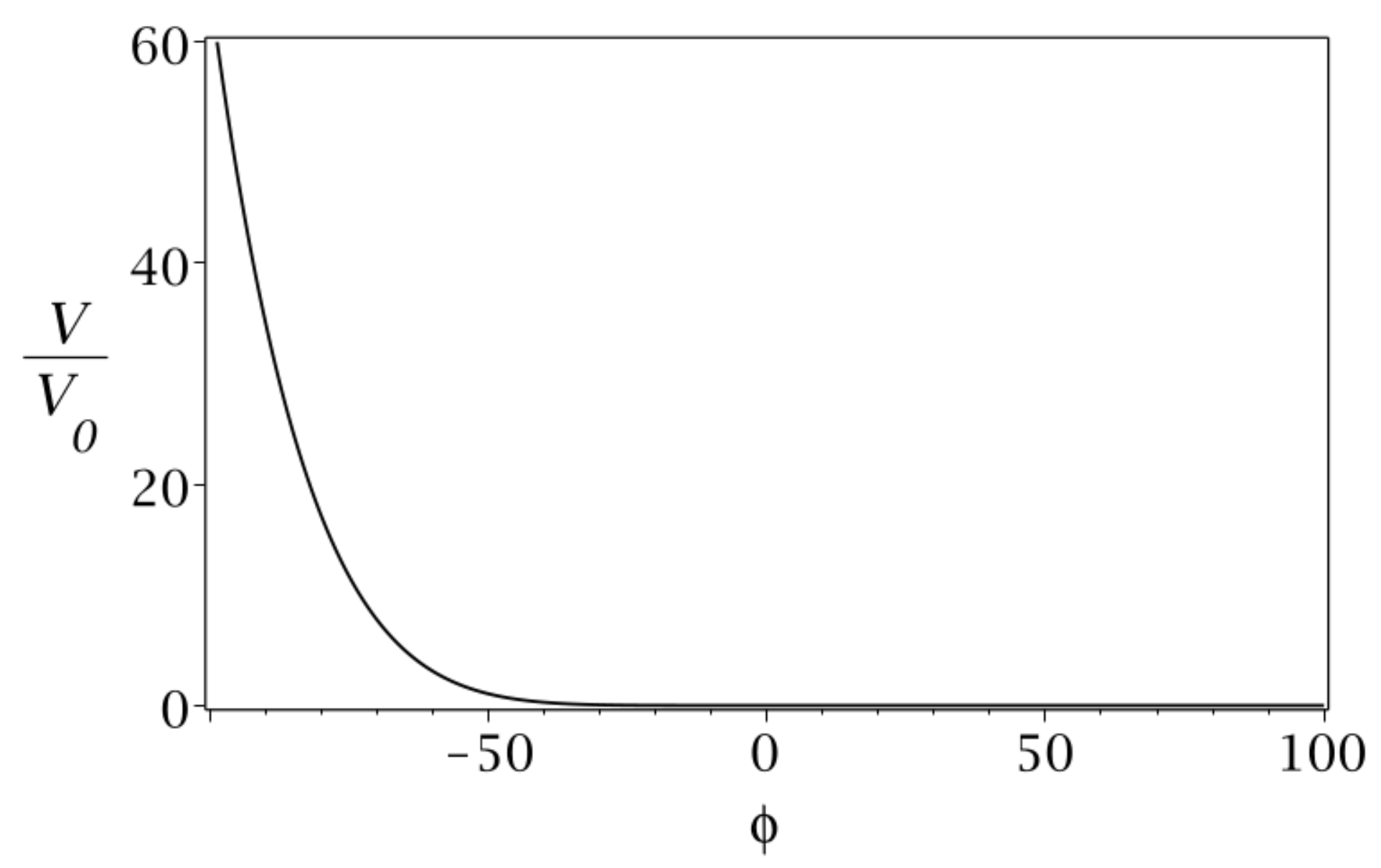}}
\subfigure[$\lambda$ diagram]{\label{fig:La6C10}
\includegraphics[width=0.48\textwidth, trim = 0cm 0cm 0cm 0cm]{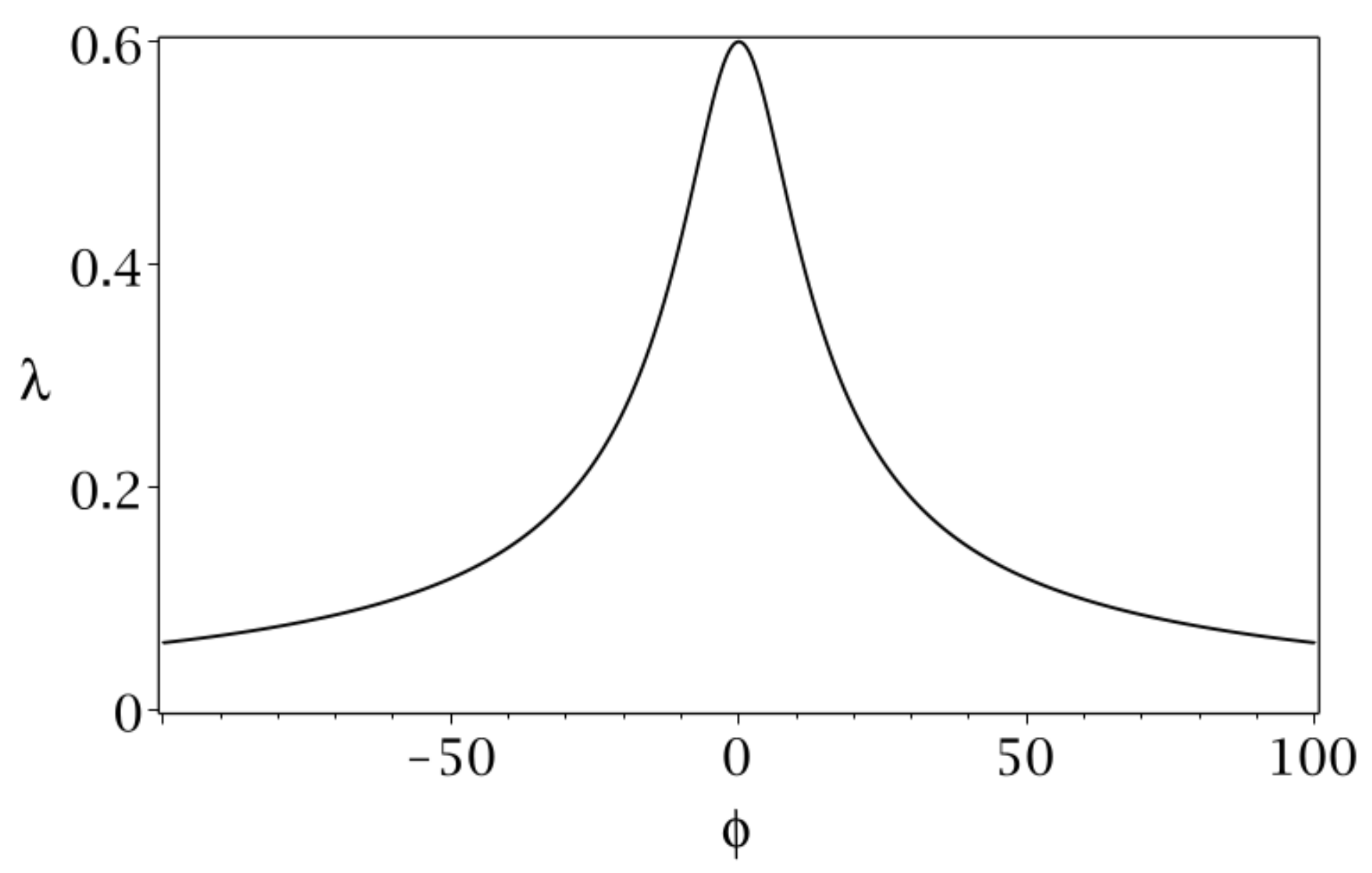}}
%\subfigure[$h-q-j$.]{\label{fig:h_q_j_RP_alpha1_2}
%\includegraphics[width=0.45\textwidth, trim = 0cm 0cm 0cm 0cm]{h_q_j_RP1_2.pdf}}
\vspace{-0.5cm}
\end{center}
\caption{The potential $V = V_0(\phi +\sqrt{C^2+\phi^2})^{-\alpha}$ and associated $\lambda= \alpha/\sqrt{C^2 + \phi^2}$
for $\alpha=6$ and $C=10$.}
\label{fig:Va10}
\end{figure}

To obtain a type of potential that has a regular $\lambda$ and still behaves
like an inverse power-law potential for large $\phi$, we generalize the
inverse power-law potential by choosing
\begin{equation}
\lambda = \frac{\alpha}{\sqrt{C^2 + \phi^2}} = \frac{\lambda_\mathrm{max}}{\sqrt{1 + (\phi/C)^2}}, \quad C>0.
\end{equation}
Here $\lambda_\mathrm{max} = \alpha/C$ is the maximum value of
$\lambda(\phi)$ while $C$ describes the peak width of the present
$\lambda(\phi)$. Using~\eqref{Vfromlambda}, this ``regularized'' inverse
power-law $\lambda(\phi)$ leads to the potential
\begin{equation}
V = V_0(\phi +\sqrt{C^2+\phi^2})^{-\alpha}.
\end{equation}
It follows that $\lim_{\phi \rightarrow \pm\infty}(\phi^{\pm\alpha} V) =
\mathrm{const}$. Note that this class of potentials contains the inverse
power-law potential by setting $C=0$ (restricting dynamics to $\phi>0$) and a
cosmological constant by setting $\alpha=0$. As an example, we give the
potential and $\lambda$ for $\alpha=6$ and $C=10$, depicted in
Fig.~\ref{fig:Va10}.

The above is (so far) a purely phenomenologically motivated potential, but so
was the original motivation for the inverse power-law
potential~\cite{peerat88,ratpee88}. The point here is to give a simple
specific example of continuous $\Lambda$CDM scalar field deformations,
illustrating features that \emph{any} observationally viable model arising from some
fundamental theory must have.

\begin{figure}[ht!]
\begin{center}
\subfigure[$(x,\Omega_{m},Z)$ state space.]{\label{fig:SS_RegAlpha6}
\includegraphics[width=0.45\textwidth, trim = 1cm 2.2cm 2cm 0cm]{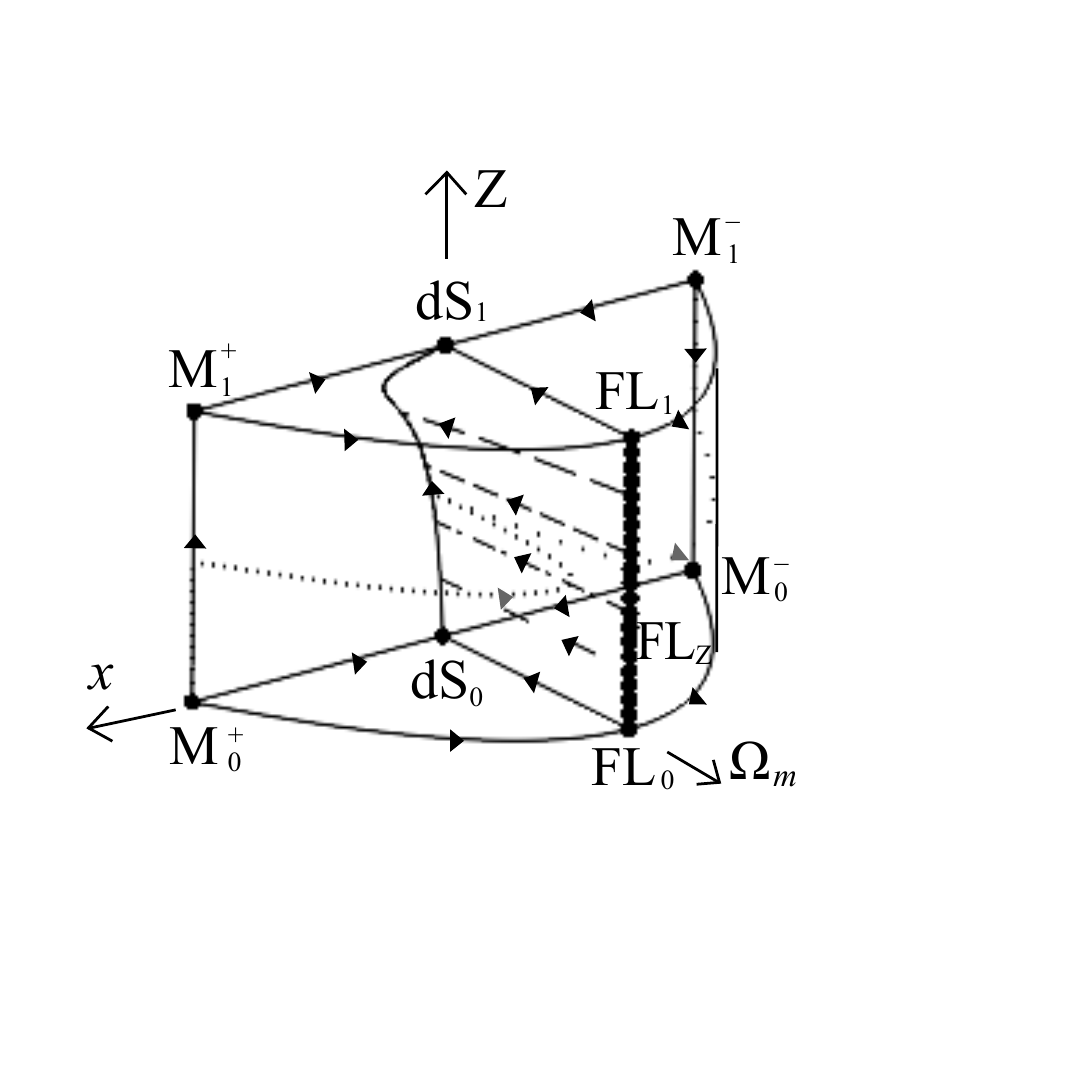}}
\subfigure[Projection  of $(x,\Omega_{m},Z)$ state space onto  $(x,\Omega_{m})$.]{\label{fig:SS_RegAlpha6_Top}
\includegraphics[width=0.45\textwidth, trim = 0.5cm 2cm 1cm 0cm]{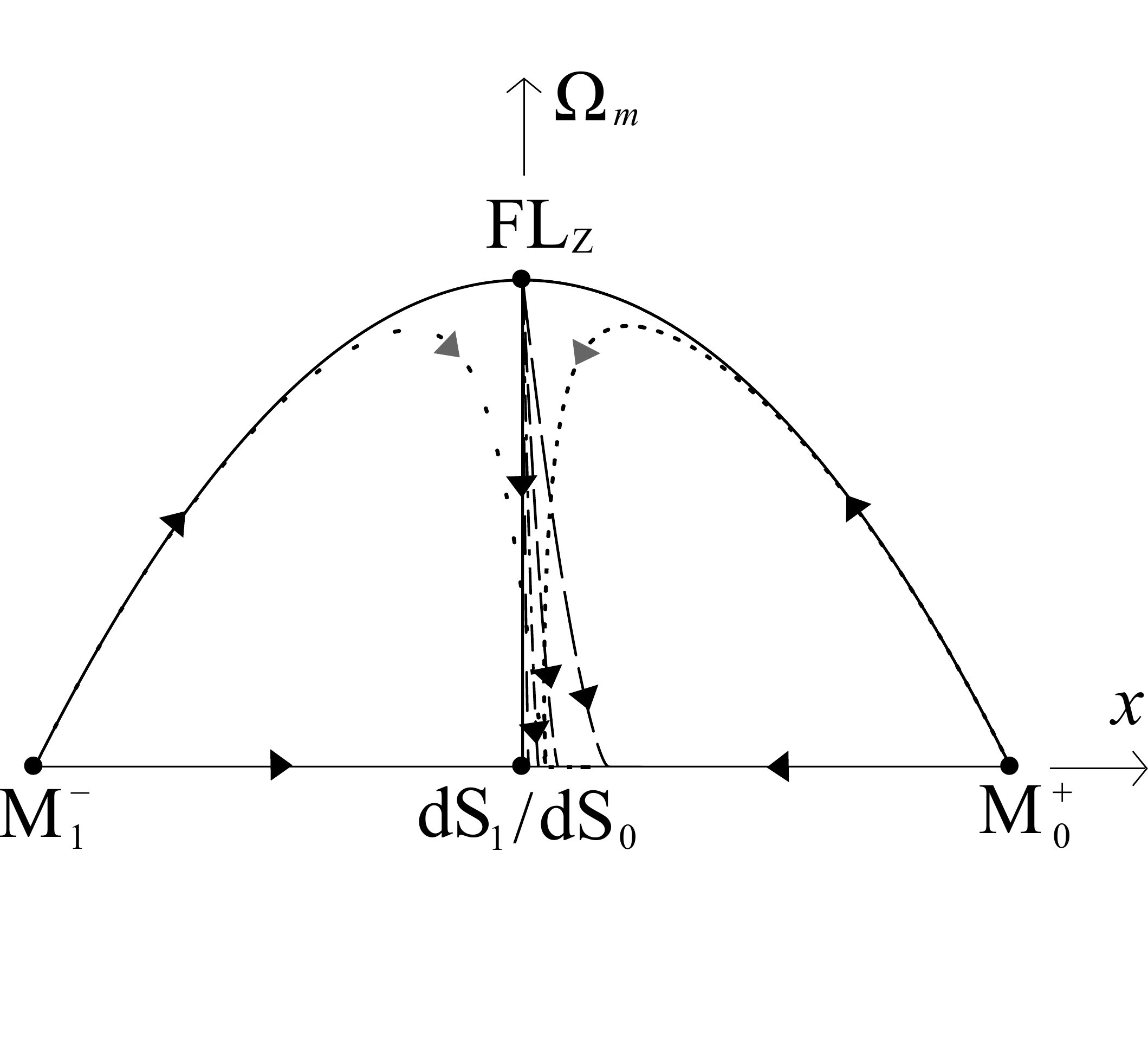}}
%\subfigure[$h-q-j$.]{\label{fig:h_q_j_RP_alpha1_2}
%\includegraphics[width=0.45\textwidth, trim = 0cm 0cm 0cm 0cm]{h_q_j_RP1_2.pdf}}
\vspace{-0.5cm}
\end{center}
\caption{Depiction of the solution space of flat FLRW models with dust and a scalar field with a potential
$V = V_0(\phi +\sqrt{C^2+\phi^2})^{-\alpha}$ and associated $\lambda= \alpha/\sqrt{C^2 + \phi^2}$
for $\alpha=6$ and $C=10$.}
\label{fig:sol_a6_C10}
\end{figure}

\begin{figure}[ht!]
\begin{center}
\subfigure[$\Omega_{m}-q$ diagram]{\label{fig:q_M_RegAlpha6}
\includegraphics[width=0.48\textwidth, trim = 0cm 0cm 0cm 0cm]{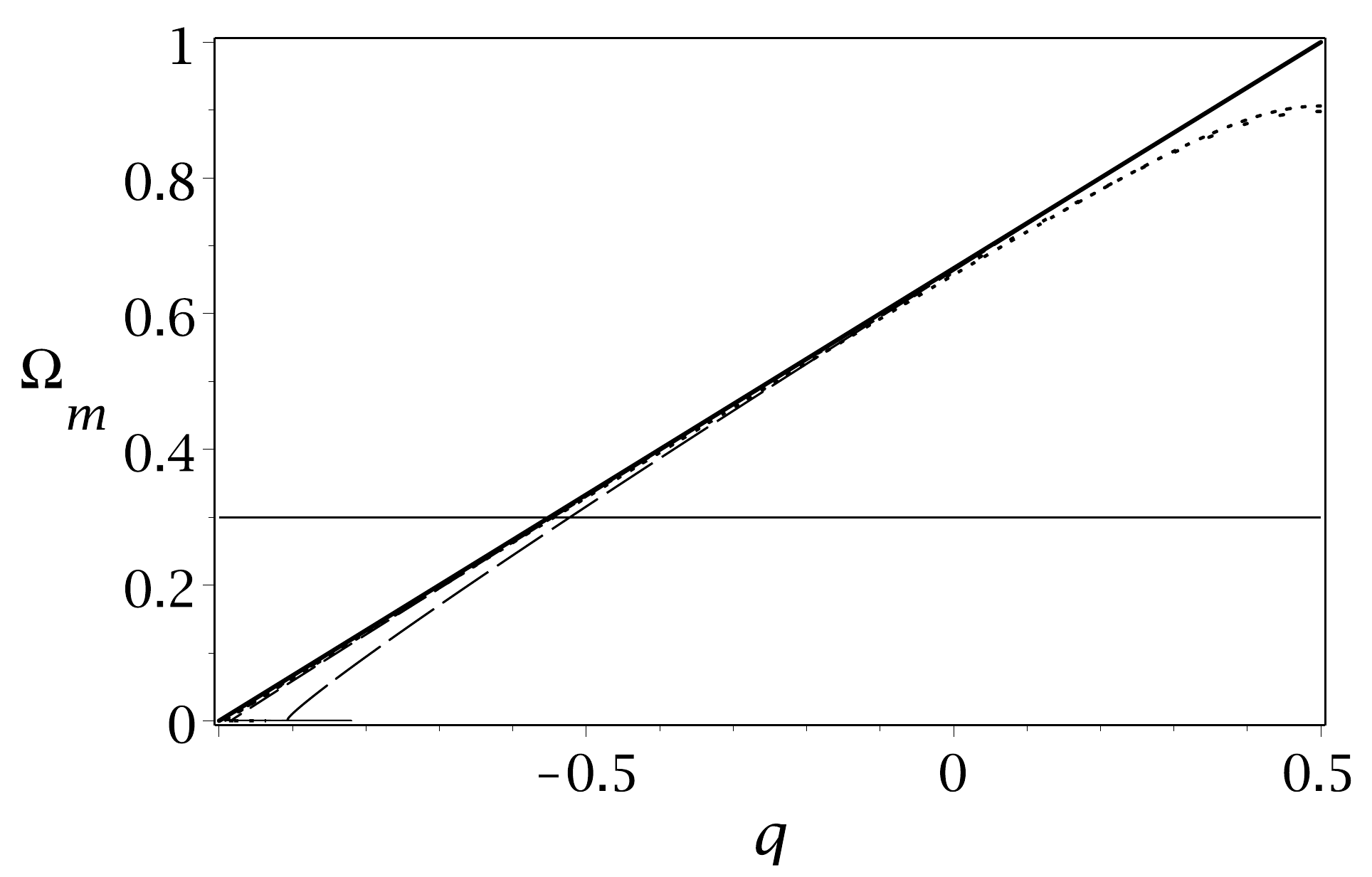}}
\subfigure[$h-q$ diagram]{\label{fig:h_q_RegAlpha6}
\includegraphics[width=0.48\textwidth, trim = 0cm 0cm 0cm 0cm]{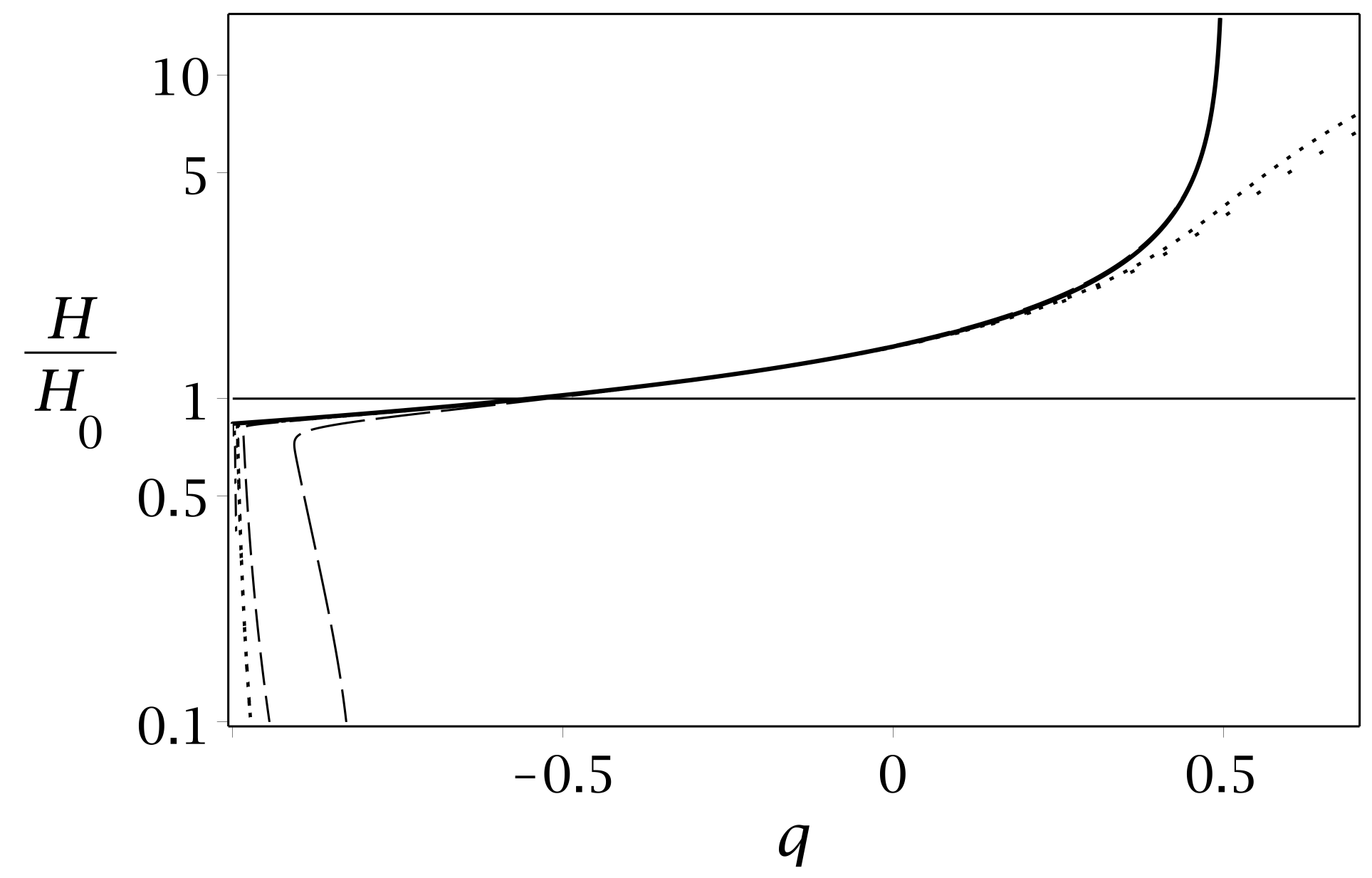}}
\subfigure[$h-j$ diagram]{\label{fig:h_j_RegAlpha6}
\includegraphics[width=0.48\textwidth, trim = 0cm 0cm 0cm 0cm]{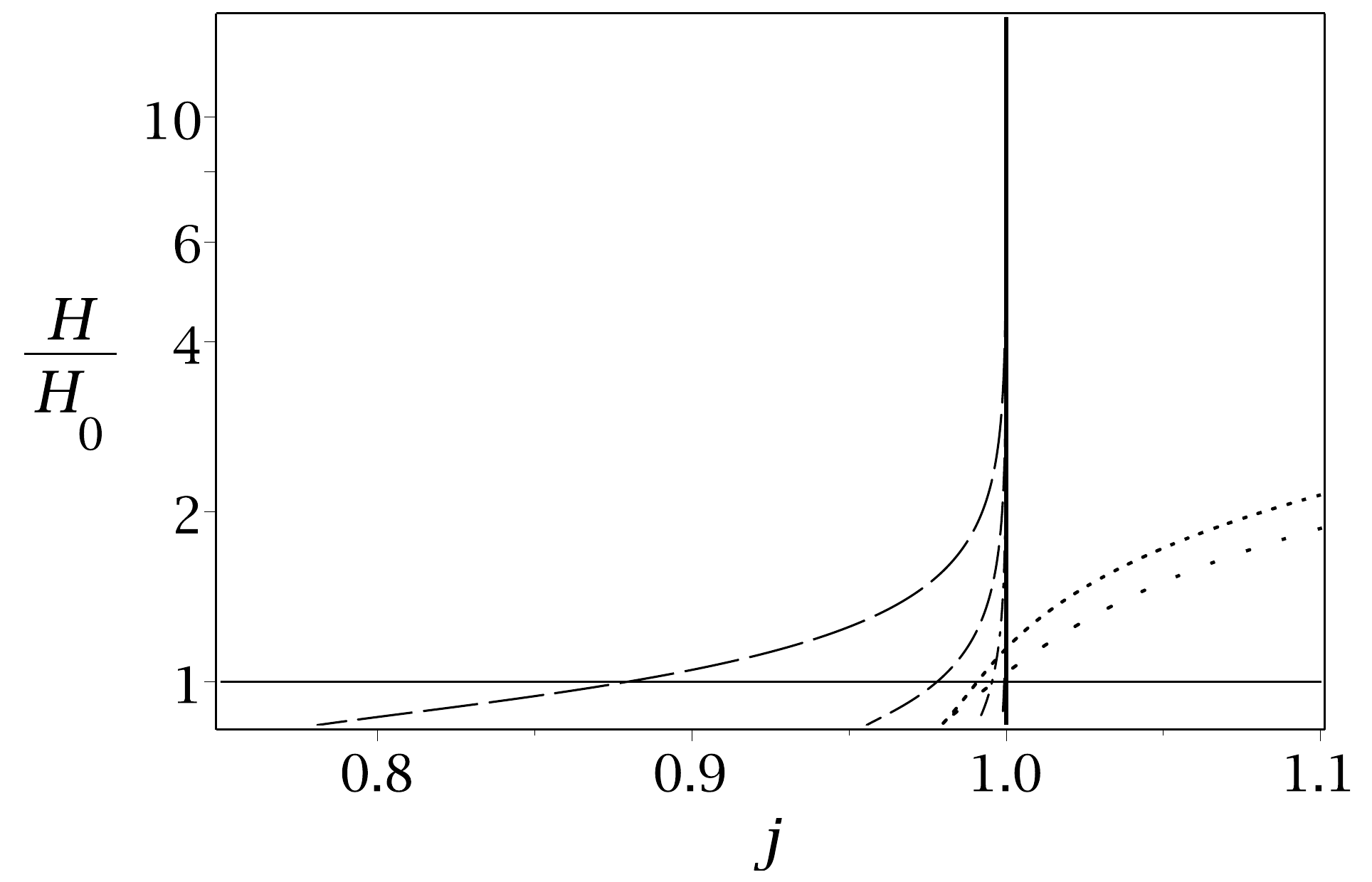}}\,\,
\subfigure[$q-j$ diagram]{\label{fig:j_q_RegAlpha6}
\includegraphics[width=0.48\textwidth, trim = 0cm 0cm 0cm 0cm]{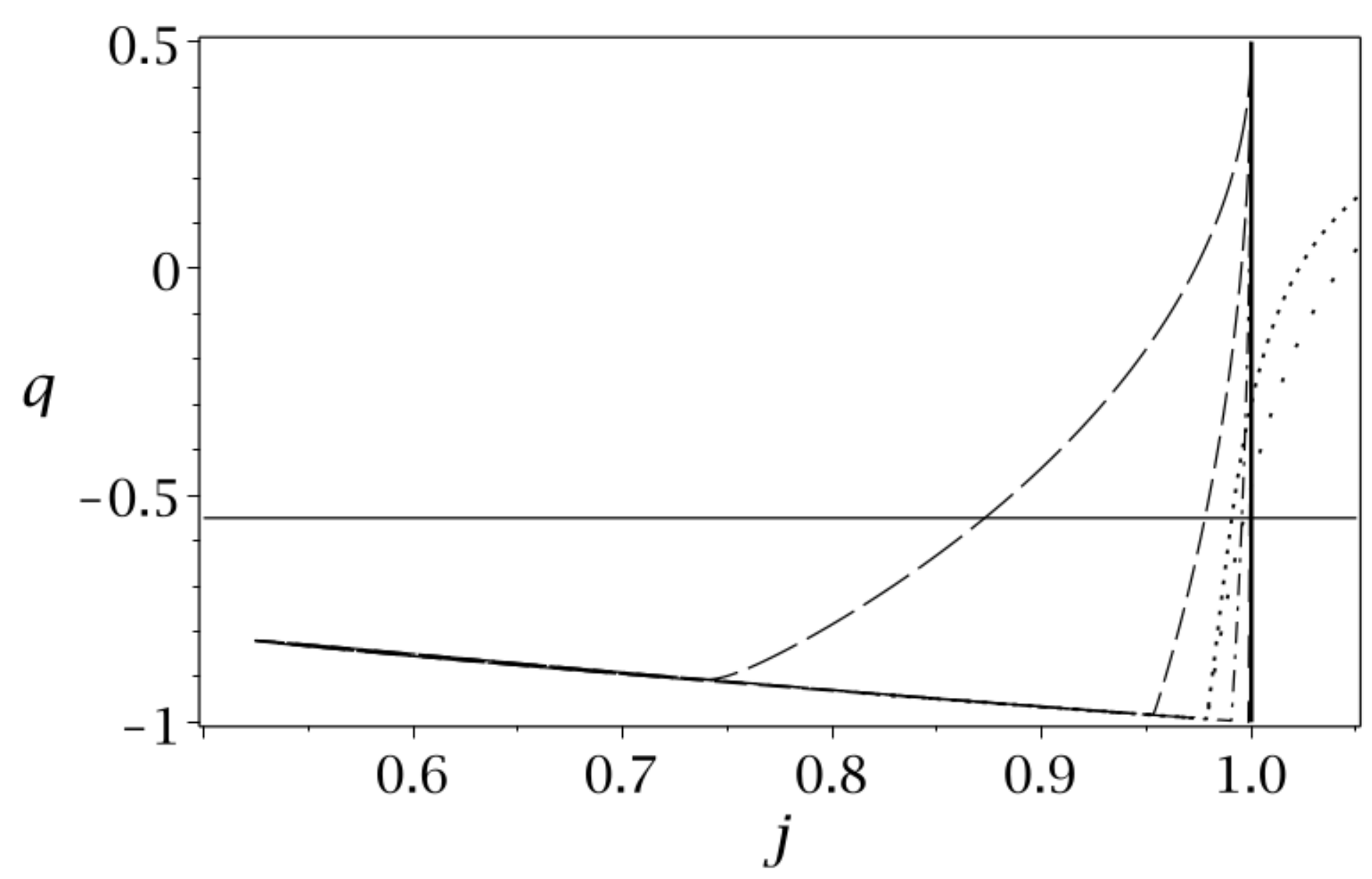}}
%\subfigure[$h-q-j$.]{\label{fig:h_q_j_RP_alpha1_2}
%\includegraphics[width=0.45\textwidth, trim = 0cm 0cm 0cm 0cm]{h_q_j_RP1_2.pdf}}
\vspace{-0.5cm}
\end{center}
\caption{Observables for flat FLRW models with dust and a scalar field with the
potential $V = V_0(\phi +\sqrt{C^2+\phi^2})^{-\alpha}$ and associated $\lambda= \alpha/\sqrt{C^2 + \phi^2}$
for $\alpha=6$ and $C=10$. The thick solid line represents $\Lambda$CDM cosmology; 
the horizontal lines represent now, i.e., zero redshift;
and the other lines correspond to the solutions in Fig.~\ref{fig:sol_a6_C10}.}
\label{fig:reglamdacosmkin}
\end{figure}

To continue we choose
\begin{equation}
Z= \frac{z}{1+z}, \qquad z = \phi + \sqrt{C^2 + \phi^2}, \qquad \phi = \frac{z^2-C^2}{2z} 
= \frac{Z^2-C^2(1-Z)^2}{2Z(1-Z)}
\end{equation}
as the scalar field variable (here $z$ should not be confused with the redshift). 
This leads to
\begin{equation}
\lambda = \frac{2\alpha Z(1-Z)}{Z^2+C^2(1-Z)^2}
\end{equation}
and the regular dynamical system
\begin{subequations}\label{dynsysreg}
\begin{align}
x^\prime &= -(2-q)x + \sqrt{\frac{3}{2}}\lambda(Z)(1 - x^2 - \Omega_m),\label{xeqreg}\\
\Omega_m^\prime &= 3\left[2x^2 - \gamma_m(1 - \Omega_m)\right]\Omega_m,\label{Omeqreg}\\
Z^\prime &= \sqrt{6} \alpha^{-1}\lambda Z(1-Z)\,x = \sqrt{6}\frac{2Z^2(1-Z)^2}{Z^2+C^2(1-Z)^2}\,x\label{Zeqreg}
\end{align}
\end{subequations}
%
%where
%
%\begin{equation}
%\frac{dZ}{d\phi} = \alpha^{-1}\lambda Z(1-Z) = \frac{2Z^2(1-Z)^2}{Z^2+C^2(1-Z)^2}
%\end{equation}
%
where $2-q=3(1-x^2) - \frac32 \Omega_m$. The qualitative properties of this
system follow from the general discussion in Sec.~\ref{sec:dynsys}. We
give an explicit representation of the solution space for the case $C=10$ and
$\alpha=6$ in Fig.~\ref{fig:sol_a6_C10}.
The associated diagrams for physical observables are given in
Fig.~\ref{fig:reglamdacosmkin}.
As can be seen the $\lambda$-regularized models give much better
results than the corresponding models with an inverse power-law potential with 
the same $\alpha$, when $C$ is chosen so that it is of the same order of 
magnitude as $\alpha$, and even better if $C$ is larger (cf. 
Figs.~\ref{fig:Cosm_RP_alpha6} and~\ref{fig:reglamdacosmkin}, which both have 
$\alpha=6$). The results basically linearly improve with increasing $C$ for a
fixed $\alpha$, which should not come as a surprise since
$\lambda_\mathrm{max} = \alpha/C$; for $\alpha =6$ and $C=100$, and therefore
$\lambda_\mathrm{max}=0.06$, the dynamics for models with $\Omega_m$ close to
1 initially is almost indistinguishable from $\Lambda$CDM dynamics. Toward
smaller $C$, the models increasingly yield similar results as the inverse
power-law potential, which again is to be expected.

%-----------------------------------------------------------------
\subsection{Observational $\lambda$ conditions for continuous $\Lambda$CDM deformation}
%-----------------------------------------------------------------

The structure of the flat FLRW models with dust and inverse power-law
potential strongly suggests that it does not suffice to just have a
monotonically decreasing potential with a shallow potential tail, since even
if $\lambda\rightarrow 0$ when $\phi \rightarrow +\infty$ a large
$\lambda(\phi)$ for some $\phi$
%for models with $\Omega_m$ close to one,
generally results in considerable deviations from $\Lambda$CDM cosmology.
There are significant differences in the solution structure if $\lambda$
is globally and asymptotically bounded or not.

This suggests that observational viability implies \emph{global and
asymptotic boundedness conditions} on $\lambda$, 
so that the whole subset that originates from $\mathrm{FL}_Z$ into
${\bf S}$ only deviates moderately from the $x=0$ $\Lambda$CDM surface. It
should be pointed out that this not only produces an observationally viable
set of models, but also alleviates the coincidence and possibly other
fine-tuning problems since \emph{all} orbits with initial values of
$\Omega_m$ close to 1 behave similarly because the invariant subset
that originates from $\mathrm{FL}_Z$ then constitutes an attracting separatrix
surface, eliminating the need for any measure.\footnote{Because open sets of 
solutions do not follow the tracker solution closely,
it was argued in~\cite{tam14} that solutions that were close
to the tracker solution were in some sense generic due to the center manifold
structures. However, such structures can be eliminated or changed by a change
of variables, while the fact that there are open sets of solutions that
behave quite differently than the tracker solution in terms of physical
observables cannot. It is therefore a question of measures if trackerlike
behavior is generic or not, and a consensus about measures seems rather
unlikely. On the other hand, if center manifold structures reflect behavior
in physical observables, they might very well play an important role in
cosmology.}

\subsection*{Acknowledgments}
A. A. is supported by the projects CERN/FP/123609/2011, EXCL/MAT-GEO/0222/2012,
and CAMGSD, Instituto Superior T{\'e}cnico by FCT/Portugal through
UID/MAT/04459/2013, and the FCT Grant No. SFRH/BPD/85194/2012. Furthermore, A. A.
also thanks the Department of Engineering and Physics at Karlstad University,
Sweden, for kind hospitality.

\begin{appendix}
%%%%%%%%%%%%%%%%%%%%%%%%%%%%%%%%%%%%%%%%%%%%%%%%%%%%%%%%%%%%%%%%%%
\section*{A scalar field and two fluids}\label{app:twofluid}
%%%%%%%%%%%%%%%%%%%%%%%%%%%%%%%%%%%%%%%%%%%%%%%%%%%%%%%%%%%%%%%%%%

In this work we have considered a matter content that consists of a fluid
without pressure. However, the early Universe after inflation is radiation
dominated and therefore somewhat more sophisticated models have both
radiation and dust as matter content. In a dynamical systems context this
situation can be treated as follows. Consider two noninteracting fluids with
energy densities $\rho_1$ and $\rho_2$, respectively. It is then convenient
to introduce
\begin{equation}\label{chi}
\rho_m = \rho_1 + \rho_2,\qquad \chi = \frac{\rho_1}{\rho_m},
\end{equation}
as the two perfect fluid variables instead of $\rho_1$ and
$\rho_2$.\footnote{For alternative ways of handling several fluids, see,
e.g.,~\cite{colwai92,heietal05}.} Assume further that the fluids obey linear
equations of state characterized by the constant equation of state parameters
$\gamma_1$ and $\gamma_2$, where we without loss of generality set
$\gamma_2>\gamma_1$ (if $\gamma_2=\gamma_1$ we treat the problem as a single
fluid). The system~\eqref{dynsys0} holds for fluids with general barotropic
equations of state, which includes the present case. However, having two
fluids instead of one with linear equations of state makes it convenient to
augment the system~\eqref{dynsys0} with the additional variable $\chi$, which
leads to
\begin{subequations}\label{dynsys2}
\begin{align}
x^\prime &= -(2-q)x + \sqrt{\frac{3}{2}}\lambda(Z)\,\Omega_V,\label{xeq2}\\
\Omega_m^\prime &= 3\left[2x^2 - \gamma_m(1 - \Omega_m)\right]\Omega_m,\label{Omeq2}\\
Z^\prime &= \sqrt{6}\frac{dZ}{d\phi}\,x,\label{Zeq2}\\
\chi^\prime &= 3(\gamma_2-\gamma_1)\chi(1-\chi),\label{chi2}
\end{align}
\end{subequations}
where, as before, $\Omega_V = 1 - x^2 - \Omega_m$ and $\Omega_m=
\rho_m/3H^2$. The deceleration parameter is given by
\begin{equation}\label{qchi}
q = - 1 + 3x^2 + \frac32\gamma_m\Omega_m = - 1 + 3x^2 + \frac32\left(\gamma_1\chi + \gamma_2(1-\chi)\right)\Omega_m.
\end{equation}
since $\gamma_m = \frac{\rho_1 + \rho_2 + p_1+p_2}{\rho_1+\rho_2} =
\gamma_1\chi + \gamma_2(1-\chi)$. Due to the similarity with the single fluid
case we choose to include the same boundaries, but with the addition of
$\chi$ we also consider $\chi \in [0,1]$, where $\chi=0$ ($\chi=1$)
corresponds to the single fluid case for $\rho_2$ ($\rho_1$).

Since $\gamma_2>\gamma_1$ it follows that $\chi$ is strictly monotonically
increasing when $0<\chi<1$. Furthermore, $\chi$ can be solved in terms of the
scale factor $a$ by using the conservation equation $d\rho_i/d\ln a = -
3\gamma_i\rho_i$, which yields $\rho_i \propto a^{-3\gamma_i}$ and therefore
\begin{equation}\label{chia}
\chi = \frac{1}{1 + k(a/a_0)^{\gamma_1-\gamma_2}} = \frac{1}{1 + k\exp[(\gamma_1-\gamma_2)\tau]},
\end{equation}
where $k= \rho_{20}/\rho_{10} = \Omega_{20}/\Omega_{10}$. It follows that
$\chi \rightarrow 0$ ($\chi \rightarrow 1$) when $a\rightarrow 0$, i.e.,
$\tau \rightarrow - \infty$ ($a\rightarrow \infty$, i.e. $\tau \rightarrow +
\infty$). Thus $\chi =0$ in the asymptotic past and $\chi=1$ in the
asymptotic future, which tells us that the fluid with the most soft (stiff)
equation of state dominates asymptotically over the other fluid at late
(early) times. As a consequence of the above features it follows that
asymptotic future (past) dynamics reside on the $\chi =1$ ($\chi =0$) subset.

Because it is possible to solve the conservation equations for each of
the two fluids in terms of the scale factor, this gives rise to a constant of
motion, which also reflects that the problem is really a
three-dimensional one. To obtain this conserved quantity we first note that
\begin{equation}
\Omega_1 = \chi\Omega_m, \qquad \Omega_2 = (1-\chi)\Omega_m.
\end{equation}
Then, since $\rho_1 \propto a^{-3\gamma_1}$ and $\rho_2 \propto
a^{-3\gamma_2}$, it follows that
\begin{equation}
\mathrm{const.} = \frac{\rho_1^{\gamma_2}}{\rho_2^{\gamma_1}} \propto
\frac{\Omega_1^{\gamma_2}}{\Omega_2^{\gamma_1}}(H^2)^{\gamma_2-\gamma_1}
= \chi^{\gamma_2}(1-\chi)^{-\gamma_1}\Omega_m^{\gamma_2-\gamma_1}(H^2)^{\gamma_2-\gamma_1},
\end{equation}
where $H^2$ is given in terms of the state space variables in Eq.~\eqref{H2}.

%Thus, since $\gamma_2 > \gamma_1$, if $\chi \rightarrow 0$ or $\Omega_m
%\rightarrow 0$ then $T \rightarrow 0$, and vice versa, or if $\chi
%\rightarrow 1$ then $T\rightarrow 1$, and vice versa.
%This, of course,
%assumes that $\rho_1\rho_2>0$ and say nothing about the special cases with
%only one or no fluids.

The matter subset $\Omega_m=1$, i.e., $x=0$, $\Omega_V=0$, now corresponds to
the two-dimensional system
\begin{subequations}\label{dynsysOm1}
\begin{align}
Z^\prime &= 0,\label{ZeqOm1}\\
\chi^\prime &= 3(\gamma_2-\gamma_1)\chi(1-\chi),\label{chiOm1}
\end{align}
\end{subequations}
which has the solution $Z=\mathrm{const}.$, $\chi = 1/(1 +
k\exp[(\gamma_1-\gamma_2)\tau])$. The analysis of the single fluid case is
directly translatable to the present context, and the line of fixed points
$\mathrm{FL}_Z$ is now replaced with the above two-dimensional subset. It
follows that the inclusion of radiation changes little of the previous
discussion as regards qualitative dynamical features.

It should perhaps be pointed out that instead of $\chi$ and $\Omega_m $ it is
possible to use $\Omega_1=\rho_1/3H^2$ and $\Omega_2=\rho_2/3H^2$  as
variables, where
\begin{equation}
\frac{d\Omega_i}{d\tau} = [2q - (3\gamma_i-2)]\Omega_i,\qquad q = - 1 + 3x^2 +
\frac32(\gamma_1\Omega_1 + \gamma_2\Omega_2),
\end{equation}
and where $\Omega_V = 1 - x^2 - \Omega_1 - \Omega_2$.

\end{appendix}

\end{document}